\documentclass[aps,prd,amsmath,amssymb,preprintnumbers,nofootinbib]{revtex4-1}
\usepackage[utf8]{inputenc} 
\usepackage{graphicx}

\usepackage{dcolumn}
\graphicspath{{./figures/}}
\usepackage{url}
\usepackage[bookmarks, pagebackref=false]{hyperref}
\usepackage[usenames,dvipsnames]{xcolor}
\definecolor{orange}{cmyk}{0,0.5,1,0}
\definecolor{rossoCP3}{cmyk}{0,.88,.77,.40}
\definecolor{graa}{rgb}{0.8,0.8,0.8}
\definecolor{blaa}{rgb}{0.2,0.2,0.6}
		\hypersetup{
			colorlinks,
			bookmarksopen,
			bookmarksnumbered,
			citecolor=blaa, 		
			linkcolor=rossoCP3,	
			urlcolor=rossoCP3,			
			}
\usepackage{amsthm}
\usepackage{bm}
\usepackage{bbm}
\usepackage{pxfonts}

\usepackage{amsmath,amssymb,amsfonts}
\usepackage{color}
\usepackage{float}
\usepackage{hyperref}
\usepackage[Symbolsmallscale]{upgreek}
\usepackage{amsmath}
\usepackage{amsfonts}
\usepackage{amssymb,dsfont}
\usepackage{graphicx}
\usepackage{amssymb}
\usepackage[vcentermath]{youngtab}
\usepackage[all]{xy}
\usepackage{pstricks}
\usepackage{dsfont}%
\usepackage{bbold}

\usepackage{placeins}
\usepackage{xspace}
\usepackage{cancel} 

\usepackage{slashed}

\usepackage{natbib}

\usepackage{feynmf}

\usepackage{braket}

\newcommand{\beq}{\begin{eqnarray}}
\newcommand{\eeq}{\end{eqnarray}}

\newcommand{\bmp}{\noindent\begin{minipage}{16cm}}
\newcommand{\emp}{\end{minipage}\vskip 7mm} 

\newcommand   \cO {\mathcal{O}}

\newcommand   \cA {\mathcal{A}}

\newcommand{\Tr}{\text{Tr}}
\newcommand*{\del}{\mathop{\mathrm{{}\partial}}\mathopen{}}

\def\lsim{\mathrel{\rlap{\lower4pt\hbox{\hskip1pt$\sim$}}
    \raise1pt\hbox{$<$}}}                
\def\gsim{\mathrel{\rlap{\lower4pt\hbox{\hskip1pt$\sim$}}
    \raise1pt\hbox{$>$}}}                

\newcommand{\mG}{\mathcal{G}}
\newcommand{\mO}{\mathcal{O}}
\newcommand{\mL}{\mathcal{L}}
\newcommand{\mQ}{\mathcal{Q}}
\newcommand{\mLt}{\widetilde{\mathcal{L}}}
\newcommand{\mH}{\mathcal{H}}
\newcommand{\Bbar}{\overline{B}}
\newcommand{\mQbar}{\overline{\mathcal{Q}}}
\newcommand{\fh}{\mathfrak{h}}
\newcommand{\slnc}{\text{sl}(N;\mathbb{C})}
\newcommand{\adjl}{\mathbf{Ad}{{\mathbf{j}}_{L}}}
\newcommand{\adjr}{\mathbf{Ad}{{\mathbf{j}}_{R}}}
\begin{document}

\title{ Untangling scaling dimensions of fixed charge operators in Higgs Theories }
\author{Oleg {\sc Antipin}$^{\color{rossoCP3}{\clubsuit}}$}
\email{oantipin@irb.hr}
\author{Jahmall {\sc Bersini}
$^{\color{rossoCP3}{\clubsuit}}$}
\email{jbersini@irb.hr}
\author{Francesco {\sc Sannino} $^{\color{rossoCP3}{\diamondsuit},\color{rossoCP3}{\heartsuit}}$}
\email{sannino@cp3.sdu.dk}
\author{Zhi-Wei Wang $^{\color{rossoCP3}{\diamondsuit}}$}
\email{wang@cp3.sdu.dk}
\author{Chen Zhang $^{\color{rossoCP3}{\spadesuit}}$}
\email{czhang@cts.nthu.edu.tw}
\affiliation{{ $^{\color{rossoCP3}{\clubsuit}}$ Rudjer Boskovic Institute, Division of Theoretical Physics, Bijeni\v cka 54, 10000 Zagreb, Croatia}\\{ $^{\color{rossoCP3}{\diamondsuit}}$\color{rossoCP3} {CP}$^{ \bf 3}${-Origins}} \& the Danish Institute for Advanced Study {\color{rossoCP3}\rm{Danish IAS}},  University of Southern Denmark, Campusvej 55, DK-5230 Odense M, Denmark. \\\mbox{ $^{\color{rossoCP3}{\heartsuit}}$Dipartimento di Fisica “E. Pancini”, Università di Napoli Federico II | INFN sezione di Napoli}\\ \mbox{Complesso Universitario di Monte S. Angelo Edificio 6, via Cintia, 80126 Napoli, Italy.}\\
{$^{\color{rossoCP3}{\spadesuit}}$ Physics Division, National Center for Theoretical Sciences, Hsinchu, Taiwan 300}}

\begin{abstract}
We go beyond a  systematic review of the semiclassical approaches for determining the scaling dimensions of fixed-charge operators in  $U(1)$ and $O(N)$ models by introducing a general strategy apt at determining the relation between a given charge configuration and the associated operators for more involved symmetry groups such as the $U(N) \times U(M)$. We show how, varying the charge configuration, it is possible to access anomalous dimensions of different operators transforming according to a variety of irreducible representations of the non-abelian symmetry group  without the aid of diagrammatical computations.  We illustrate our computational strategy by determining the anomalous dimensions of several composite operators to the next-to-leading order in the semiclassical expansion for the  $U(N) \times U(M)$ conformal field theory (CFT) in $4-\epsilon$ dimensions. Thanks to the powerful interplay between semiclassical methods and group theory we can, for the first time, extract scaling dimensions for a wide range of operators.
\\
  {\footnotesize  \it Preprint: RBI-ThPhys-2021-8, CP$^3$-Origins-2021-01 DNRF90} 

\end{abstract}

\maketitle

\tableofcontents

  \section{Introduction}

Recently there has been a flurry of interest in studying conformal field theories with continuous global symmetries in the limit of a large conserved charge $\bar Q$ in order to access non-perturbative corners of Quantum Field Theories (QFT)s.  By identifying the emergence of effective field theories (EFT)s stemming from the large charge dynamics \cite{Hellerman:2015nra, Alvarez-Gaume:2016vff, Jafferis:2017zna, Hellerman:2017sur} one can use them to extract relevant data in inverse powers of the charge as reviewed in  \cite{Gaume:2020bmp}.  Typically one is led to determine the scaling dimension of certain fixed-charge operators for a subgroup of QFTs that display conformal invariance and denoted as CFTs. One can, however, go beyond the CFT limit  \cite{Orlando:2019skh, Orlando:2020yii} which is relevant for establishing the spectrum and dynamics of near conformal dynamics emerging from quantum phase transitions \cite{Sannino:2009za, Cacciapaglia:2020kgq}.  The approach is often referred to as semiclassical in the sense that the path-integral is typically dominated by trajectories near the solution of the classical equation of motion.

The hurdle in the strongly coupled regime is that the identification and construction of the specific operators associated to a given charge configuration are impossible before solving the theory.  This is the reason why such an identification, except perhaps for some symmetry-protected operators, is left unspecified in the literature.

There is, however, another relevant limit in which the semiclassical approach is useful. This is the one in which the CFT is perturbative and controlled by a small parameter $\epsilon$ emerging because there is a non-trivial interacting fixed point near the loss of asymptotic freedom of either perturbatively safe \cite{Litim:2014uca} or infrared nature of the Banks-Zaks type \cite{Banks:1981nn}. For the safe case this was investigated first in \cite{Orlando:2019hte}. Another way to introduce a small parameter is to slightly modify the number of space-time dimensions typically injecting, for UV free theories, perturbative infrared fixed points in lower than four dimensions. Here, the charge expansion captures higher orders in the ordinary perturbative coupling corrections \cite{Badel:2019oxl, Arias-Tamargo:2019xld, Antipin:2020abu, Antipin:2020rdw,Jack:2020wvs}. The reason being that the presence of a small parameter allows studying the fixed-charge sectors of a CFT by defining a 't Hooft-like coupling $\mathcal{A} = \epsilon \bar Q$ in which one can take the limit $\epsilon \to 0$ while maintaining $\mathcal{A}$ fixed. In fact, one can now resum ordinary perturbation series by providing all-order results in the $\mathcal{A}$ coupling. In particular, much attention has been paid to the time-honoured $O(N)$ model, first investigated for any $N$  in $4-\epsilon$ dimensions in \cite{ Antipin:2020abu}. The results were recently successfully tested against ordinary perturbation theory in \cite{Jack:2021ypd} to four loops. Later the O(N) model was investigated via the  semiclassical approach at large $N$ in various dimensions in \cite{Arias-Tamargo:2020fow,Giombi:2020enj}.

Another, yet unexplored possibility, is to introduce two independent small parameters, one that takes into account the deformation of the number of space-time dimensions and the other that controls the original fixed point \cite{Codello:2016muj}. This last case has not yet been investigated in the literature and will be considered elsewhere.

Compared to conventional perturbation theory according to which one chooses a specific composite operator and then diagrammatically determines its scaling dimension, in the semiclassical fixed-charged framework one needs to reverse engineer the given charge configuration to determine the irreducible representation of the related composite operator. This has, so far, restricted the semiclassical method to the highest
weight representation operators where there is no ambiguity with respect to the chosen charge configuration. In this work we show how to access different operators belonging to distinct irreducible representations.  This is achieved by means of group-theoretical considerations applied to the semiclassical approach. The resulting efficient procedure will entail:
\begin{itemize}
 \item[1.] {Establishing the mathematical connection between classical operator dimensions and group-theoretical weights; }
 \item[2.] {Varying the charge configuration and using the first point to arrive at different operators transforming in a variety of  irreducible representations and determine their scaling dimensions;}
 \item[3.] {Developing the strategy to deal with charge configurations that give rise to non-trivial chemical potentials.}
\end{itemize}





To test the power of our strategy we investigate several non-abelian global symmetries in various space-time dimensions culminating in the general $U(N)\times U(M)$ global symmetry case.
%
%
%

The work is organized as follows. In \autoref{SOA}, we introduce the semiclassical methods at fixed charge focusing on their applications and limitations. In doing this, we review a series of results obtained in the literature for $U(1)$ and $O(N)$ invariant theories. We then move to \autoref{themap} in which we provide the map between operators and their group structure. we show how to identify the fixed charge operators in the $U(1)$, $O(N)$ and we then generalise it to the case of the $U(N) \times U(M)$ symmetry group.
In \autoref{computations} we study various charge configurations in the $U(N) \times U(M)$ model, compute the associated scaling dimensions, and identify the corresponding fixed charge operators. The results are used to establish the connection between charge configuration and fixed charge operators.
We  offer our conclusions in \autoref{conclusion}. The appendix contains details related to the scaling dimensions of the $U(N) \times U(M)$ model. Readers already familiar with the basics of fixed-charge semiclassical methods may start reading from \autoref{themap}, while readers who wish to quickly get to the main results of this work may directly start from \autoref{computations}.

\section{Review of semiclassical methods at fixed charge in weakly coupled theories} \label{SOA}

\subsection{The $U(1)$ model in $4-\epsilon$ and $3-\epsilon$ dimensions} \label{sub1}

We start this section with a brief introduction to semiclassical methods at fixed charge in QFT by considering the Abelian $U(1)$ theory in both $4-\epsilon$ and $3-\epsilon$ spacetime dimensions. These two cases have been investigated in \cite{Badel:2019oxl, Arias-Tamargo:2019xld, Watanabe:2019pdh, Arias-Tamargo:2019kfr} and \cite{Badel:2019khk, Jack:2020wvs}, respectively.
The Lagrangian reads
\begin{equation}
    \mathcal{L} = \partial \bar \phi \partial \phi +V\left ( \bar \phi \phi \right ) \,,
\label{eq:uonel}
\end{equation}
where $V_{d = 4-\epsilon}= \mathcal{N}_4 \lambda_0 \left( \bar \phi \phi \right )^2$ and $V_{d = 3-\epsilon}= \mathcal{N}_3 \lambda_0^2 \left( \bar \phi \phi \right )^3$, with $\lambda_0$ the bare coupling and $\mathcal{N}_{3,4}$ the normalizations.

By virtue of the Noether theorem, the $U(1)$ symmetry implies the existence of a conserved charge $\bar Q$ given by
\begin{equation} \label{U1charge}
    Q = \int d^{d-1} x \mathbb{j}_0  \ , \qquad {\rm with} \qquad
        \mathbb{j}_\mu  = \bar \phi \partial_\mu  \phi- \phi  \partial_\mu \bar \phi \,.
\end{equation}
We adopt conventions such that $\phi$ and $\bar \phi$ have charge $\bar Q = +1$ and $\bar Q =-1$, respectively. This model exhibits an infrared Wilson-Fisher (WF) fixed point (FP) $\lambda^* = \lambda^* (\epsilon)$ in both $d = 4 - \epsilon$ and $d = 3 - \epsilon$ dimensions \cite{Wilson}. The corresponding fixed point theory is scale-invariant, and we assume it to be invariant under the full set of conformal transformations. Furthermore, for $\epsilon \ll 1$ the theory is weakly coupled. Our goal is to compute the fixed point scaling dimension  $\Delta_{\phi^{\bar Q}}$ of the $\phi^{\bar Q}$ operators, which we \emph{define} to be the charge $\bar Q$ operators with the smallest scaling dimension \footnote{Since in the free theory limit derivatives increase the scaling dimension, in the perturbative regime $\phi^{\bar Q}$ is the lowest-lying operator with charge $\bar Q$. On the other hand, at large coupling, level crossing with other operators can, in principle, occur.}.

The CFT associated with the WF fixed point defined in a flat spacetime can be mapped to a QFT defined on a cylinder geometry in a Weyl-invariant manner. Weyl invariance then dictates a correspondence between correlation functions of the two theories~\footnote{We refer the reader to Refs.~\cite{Rychkov:2016iqz,Simmons-Duffin:2016gjk} for introductory accounts of the Weyl map, and especially Ref.~\cite{Farnsworth:2017tbz} for conceptual clarification between conformal invariance and Weyl invariance including the implication for correlation functions.}. Considering polar coordinates $(r, \Omega_{d-1})$ for $\mathbb{R}^d$, the map reads
 \begin{equation}
\mathbb{R}^d \to \mathbb{R} \times S^{d-1}\,, \qquad (r, \Omega_{d-1}) \to  (\tau, \Omega_{d-1}) \,, \qquad r= R e^{\tau/R},\, \qquad
ds^2_{\rm cyl}=d\tau^2+R^2d\Omega_{d-1}^2=\frac{R^2}{r^2}ds^2_{\rm flat}
\end{equation}
with R the radius of the sphere and $\tau$ the time coordinate on the cylinder.
According to the state-operator correspondence \cite{Cardy:1984rp, Cardy:1985lth}, the action of an operator $\mathcal{O}$ at $\tau = -\infty$ creates a state with the same quantum numbers and with energy given by $E_{\mathcal{O}} \equiv \frac{\Delta_{\mathcal{O}}}{R}$. Since we are looking for the smallest scaling dimension, our goal turned to the computation of the ground state energy (at fixed charge $\bar Q$) on the cylinder, $E_{\phi^{\bar Q}}$. This can be calculated by considering the expectation value of the evolution operator $e^{-H T}$ (with $H$ the Hamiltonian and $T = \tau_f - \tau_i$) in an arbitrary state $\ket{\bar Q}$ with fixed charge $\bar Q$ and then taking the limit $T \to \infty$ to project out the ground state from it. That is
\begin{equation} \label{evolution}
    \bra{\bar Q} e^{-HT}\ket{\bar Q}  \underset{T\to \infty}{=} \tilde{ \mathcal{N}} e^{-E_{\phi^{\bar Q}} T} \,.
\end{equation}
Notice that we only require $\ket{\bar Q}$ have a non-zero overlap with the lowest-lying state in the fixed charge sector.
Then when we insert a complete set of energy eigenstates in the left-hand side (LHS) of the above equation, only the contribution of the lowest energy state survives, with a prefactor $\tilde{ \mathcal{N}}$ that is independent of $T$ but depends on the overlap between the states. Therefore, we may always extract the ground state energy from the $T$-dependent part of the expectation value. We consider polar coordinates for the field:
\begin{equation}
\phi = \frac{\rho}{\sqrt{2}} \ e^{i \chi}, \qquad \qquad \qquad \bar\phi = \frac{\rho}{\sqrt{2}} \ e^{-i \chi}\,.
\end{equation}
Then a convenient choice for $\ket{\bar Q}$ yields \cite{Badel:2019oxl}~\footnote{Since $\ket{\bar Q}$ can be chosen arbitrarily as long as it has a nonzero overlap with the lowest lying fixed charge state, we do not impose any boundary condition on $\rho$. This allows us to compute the saddle point expansion for the path integral.}
\begin{equation} \label{eff}
\bra{\bar Q}e^{-HT}\ket{\bar Q} =\mathcal{Z}^{-1}\int \mathcal{D} \rho  \mathcal{D} \chi e^{-S_{eff}}\,,
\end{equation}
where
\begin{equation}
S_{eff} =  \int_{-T/2}^{T/2} d\tau\int d\Omega_{d-1} \left[ \frac{1}{2} (\partial \rho)^2 +  \frac{1}{2} \rho^2 ( \partial \chi )^2 + \tilde{V}(\rho^2) +i\frac{\bar Q}{R^{d-1}\Omega_{d-1}}\,\dot{\chi}\right]\,, \qquad \qquad  \tilde{V}(\rho^2) =  V(\rho^2)+\frac{m^2}{2} \rho^2 \,.
\label{eq:effActCyl}
\end{equation}
The prefactor $\mathcal{Z}$ is a $T$-independent constant that does not affect the determination of the scaling dimension. The mass term $m^2 =\frac{1}{4} \frac{d-2}{d-1} \mathcal{R}$ in \eqref{eq:effActCyl} arises from the conformal coupling to the Ricci scalar $\mathcal{R}$ of $S^{d-1}$ \cite{Brown:1980qq}. On a $d-1$-dimensional sphere of radius $R$, we have $\mathcal{R} = \frac{(d-1)(d-2)}{R^2}$ and thus $m^2 =\left(\frac{d-2}{2 R}\right)^2$. Rescaling the field as $\rho \to \rho/\lambda_0^{1/2}$ and collecting an overall $\lambda_0^{-1}$ as loop counting parameter, we see that, at small coupling, this path integral can be computed via a saddle point expansion, resulting in
\begin{equation}
 \label{renorm}
\begin{split}
   R E_{\phi^{\bar Q}} & =  \sum_{k=-1} \lambda_0^k e_k (\mathcal{A}_0, d) =  \sum_{k=-1} \lambda^k \bar e_k (\mathcal{A}, R M, d) \,.
\end{split}
\end{equation}
where $M$ is the renormalization group (RG) scale, $\bar e_k (\mathcal{A}, R M, d)$ the renormalized coefficients of the semiclassical expansion and we introduced the 't Hooft coupling $\cA_0 \equiv \lambda_0 \bar Q$ ($\cA \equiv \lambda \bar Q$ as the renormalized one). In the last equality, we have renormalized the result by separating the divergent part in every term of the expansion and absorbing it in the $\tilde{ \mathcal{N}}$ coefficient in \eqref{evolution}.
At the fixed point, the dependence on $R M$ drops and we obtain
\begin{equation}
    \Delta_{\phi^{\bar Q}} =\frac{1}{\lambda^*}\Delta_{-1}\left(\cA^* \right)+\Delta_{0}\left(\cA^*\right)+\lambda^*\Delta_{1}\left(\cA^*\right)+\cdots \ \,,
\end{equation}
where the star notation $``*"$ denotes the value of the coupling at the FP, and the $\Delta_k$ are the $(k+1)$-loop corrections in the saddle point expansion. Note that the equation above can also be written in an equivalent “dual” form\footnote{These two dual forms are equivalent to each other. It can be easily shown that the 't Hooft coupling $\mathcal{A}$ plays the role as a standard ``ruler". For any chosen 't Hooft coupling (no matter in the perturbative regime $\mathcal{A}\ll1$ or super-fluid regime $\mathcal{A}\gg1$), to make the semiclassical expansion order by order under control, we can either deduce the upper bound of $\lambda$ or the equivalent lower bound of $\bar Q$ simply through $\bar Q=\mathcal{A}/\lambda$.} as a large charge expansion in $1/Q$ in the limit $Q \to \infty$ and $\cA$ fixed, i.e.
\begin{equation} \label{expans}
    \Delta_{\phi^{\bar Q}} =\sum_{k=-1} \frac{1}{\bar Q^k} \Tilde{\Delta}_k (\mathcal{A}^*) \,, \qquad \qquad \Tilde{\Delta}_k \equiv \Delta_k \mathcal{A}^k \,,
\end{equation}
which is akin to the large number of flavor expansions in gauge theories\footnote{A key difference of $1/N_f$ expansion in gauge theories is that it has a finite radius convergence in the 't Hooft coupling which is determined by the pole structure. For example in QED, the radius convergence is $\mathcal{A}=\lambda N_f=15/2$, and if we fix $\mathcal{A}\le15/2$ we can obtain a lower bound on $N_f$ to make the $1/N_f$ expansion under control \cite{Holdom:2010qs}. However, less is known about the pole structure of the charge expansion.} \cite{PalanquesMestre:1983zy,Gracey:1996he,Holdom:2010qs,Antipin:2018zdg,Mann:2017wzh,Pelaggi:2017abg,Kowalska:2017pkt,Molinaro:2018kjz,Alanne:2018csn,Wang:2018yer,Sannino:2019sch,Dondi:2020qfj,Huang:2020bbe,Cacciapaglia:2020tzd}.

To compute the leading order (LO) contributions $\Delta_{-1}(\mathcal{A}^*)$, we need to solve the classical system and evaluate $S_{eff}$ on the solution. The solution of the EOM with the lowest energy at fixed $\bar Q$ is spatially homogeneous and reads
\begin{equation} \label{U(1)sol}
    \rho = f \,, \qquad \qquad \chi = - i \mu \tau + \text{const.} \,,
\end{equation}
where
\begin{equation} \label{total}
    \mu^2 =\frac{1}{f} \frac{\partial \Tilde{V}(\rho^2)}{\partial \rho} \bigg\rvert _{\rho = f}  \qquad \quad \frac{\bar Q}{ R^{d-1} \Omega_{d-1}}=\mu f^2  \,.
\end{equation}
Fixing the charge produces spontaneous symmetry breaking and the fields take a non-zero vev with $\mu$ playing the role of a fixed chemical potential.  $\Delta_{-1}(\mathcal{A}^*)$ is then given by Eq.\eqref{eq:effActCyl} evaluated on the solution \eqref{U(1)sol} at the fixed point
\begin{equation} \label{firstorder}
  \frac{1}{\lambda^*}\frac{\Delta_{-1} (\mathcal{A}^*)}{R}= \frac{S_{eff}}{T}=\frac{1}{2}\mu Q +\Tilde{V}(f^2) \,.
\end{equation}

Then, by plugging the second equation in \eqref{total} into the first one, we have
\begin{equation} \label{generalmu}
(R \mu)^{2\frac{d-1}{d-2}} -\left(\frac{d-2}{2}\right)^2 (R \mu)^{\frac{2}{d-2}} =  \mathcal{C} \mathcal{A}^{*\frac{2}{d-2}}    \ , \qquad \quad \mathcal{C} =\frac{\mathcal{N}_{3,4} \mathcal{M}^{\frac{d}{2-d}} 16^{\frac{1}{2-d}} d \pi^{-\frac{d}{d-2}}}{\Gamma\left(\frac{d}{2}\right)^{-\frac{2}{d-2}}(d-2)} \,,
\end{equation}
where $\mathcal{N}_{3,4}$ and $\mathcal{M}$ are the normalizations of the potential and the kinetic term, respectively.
Inserting the solution of Eq.\eqref{generalmu} into Eq.\eqref{firstorder} one obtains  $\Delta_{-1}$ as a function of the 't Hooft coupling $\mathcal{A}^*$.

The case $d=4-\epsilon$ (with $\mathcal{N}_4 =1/4$) has been considered in \cite{Badel:2019oxl}. The results read
\begin{align}
 \label{classic}
  \frac{\Delta_{-1}}{\mathcal{A}^*} = \frac{1}{4} F_{4d}(x) \,, \qquad \quad F_{4d}(x) \equiv \frac{3^\frac{2}{3}x^{\frac{1}{3}}}{3^\frac{1}{3}+x^{\frac{2}{3}}}  + \frac{3^\frac{1}{3}\left(3^\frac{1}{3}+x^{\frac{2}{3}}\right)}{x^{\frac{1}{3}}},\,\qquad x= 9 \frac{\mathcal{A}^*}{(4 \pi)^2}  + \sqrt{-3+81  \frac{\mathcal{A}^{*2}}{(4 \pi)^4} }\,.    \end{align}
Notice that this classical result resums an infinite series of Feynman diagrams in the conventional perturbative expansion. In particular, it resums the leading power of $\bar Q$ at every perturbative order.

The next leading order (NLO) contributions $\Delta_0$ is given by the functional determinant of the fluctuation around the classical solution. Its bare form can be written in terms of the dispersion relations of the fluctuations, $w_+$ and $w_-$, as
\begin{equation}\label{eq:E1loop}
e_0(\mathcal{A}_0 ,d)=\frac{R}{2}\sum_{\ell=0}^{\infty}
 n_{\ell}\left[\omega_+(\ell)+\omega_-(\ell)\right].
\end{equation}
where $\ell$ labels the eigenvalues of the Laplacian on the sphere $J^2_\ell= \ell (\ell+d-2)/R^2$ and $n_\ell=\frac{(2\ell+d-2) \Gamma(\ell+d-2)}{\Gamma(\ell+1) \Gamma(d-1)}$ is the Laplacian multiplicity on $S^{d-1}$.
By expanding around the classical solution as
\begin{equation}
    \rho(x) = f +r(x), \qquad \chi(x) = -i\mu \tau + \frac{1}{f \sqrt{2}}\pi(x) \,,
\end{equation}
we obtain the action at the quadratic order in the fluctuations
\begin{eqnarray}
   S^{(2)}   =  \int_{-T/2}^{T/2} d\tau\int d\Omega_{d-1}\left[\frac{1}{2}(\partial \pi)^2+\frac{1}{2}(\partial r)^2 +\frac{2}{d-2} (\mu^2-m^2)r^2-2 \ i \ \mu \ r \ \dot \pi \right] \ .\end{eqnarray}
where $\dot \pi = \frac{\partial \pi}{\partial \tau}$. From the quadratic action we can then easily obtain the dispersion relations of the spectrum, which read
\begin{equation} \label{dispersion}
    \omega_{\pm}(l) = \sqrt{\frac{(d-2)J^2_\ell+ 2(d-1)\mu^2-2 m^2 \pm 2\sqrt{m^4+ \mu^2 \left( (d-2)^2 J^2_\ell -2 (d-1) m^2 \right) + (d-1)^2 \mu^4}}{d-2}} \,.
\end{equation}

The  spectrum  contains  one  relativistic  Goldstone boson (the conformal mode) and one massive state with mass $2\sqrt{\frac{(d-1) \mu^2- m^2}{(d-2)}}$.

We proceed by fixing $d=4-\epsilon$ and $\mathcal{N}_4 =1/4$. Eq.\eqref{eq:E1loop} needs to be renormalized. This is achieved by expanding $\lambda_0 = M^\epsilon \lambda Z_{\lambda}$ in Eq.\eqref{renorm} and keeping the terms of order $\lambda^0$. Then, to obtain $\Delta_0$ we set $d = 4$ in $\bar{e}_0(\mathcal{A},RM,d)$ and add the expansion of the LO term $\bar e_{-1}/\lambda$ to first order in $\epsilon$. Notice that the procedure mixes different orders of the bare expansion. In particular, $\bar{e}_0(\mathcal{A},RM,d)$ contains two $\frac{1}{\epsilon}$ terms that have to cancel each other in order to be able to take $d=4$ and obtain $\Delta_0$. The first comes from renormalizing $e_{-1}$ while the second can be isolated by regularizing the sum over $\ell$ in $e_0$ which formally diverges. Thus, since these two terms come from different orders of the bare expansion, their cancellation can be used as a non-trivial self-consistency check of the correctness of the calculations which can be particularly useful when dealing with more complicated models.
We have:
\begin{eqnarray}
\Delta_0(\mathcal{A}^*)= -\frac{15 \mu^4  R^4+6 \mu^2  R^2-5}{16}
+\frac{1}{2} \sum_{\ell=1}^\infty\sigma(\ell)
+\frac{\sqrt{3\mu^2R^2-1}}{\sqrt{2}} \,,
\label{Delta0}
\end{eqnarray}
where
\begin{align}
\sigma(\ell) &=  R (1+\ell)^2\left[\omega_+(\ell)+\omega_-(\ell)\right]- 2 \ell^3- 6 \ell^2 - \left(2 \mu^2 R^2 +4 \right)\ell -2 \mu^2 R^2 + \frac{1}{\ell} \frac{5}{4}\left( \mu^2 R^2 -1\right)^2 \,,
\end{align}
with $R \mu$ given by Eq.\eqref{generalmu} in $d=4$. Summing $\Delta_{-1}$ and $\Delta_0$, expanding the result for small $\mathcal{A}^*$, and using the FP value $\lambda^*= \frac{1}{5}\epsilon +\frac{3}{25}\epsilon^2 + \mathcal{O}\left(\epsilon^3 \right)$,  we have
\begin{equation}
    \Delta_{\phi^{\bar Q}} =\bar Q \left ( \frac{d}{2} -1 \right )+\frac{\epsilon}{10} \bar Q (\bar Q-1)- \frac{\epsilon^2}{50} \bar Q({\bar Q}^2-4 \bar Q)+\mathcal{O}\left(\epsilon^2 \bar Q,\epsilon^3 {\bar Q}^4\right).
\end{equation}
  This result has been obtained in \cite{Badel:2019oxl} and checked via a Feynman diagram calculation. Similarly, the expansion for large $\mathcal{A}$ is
  \begin{equation} \label{eq:Delta1LoopLarge2}
\Delta_{\phi^{\bar Q}}=\frac{1}{\epsilon}\left(\frac{2}{5}\epsilon \bar Q\right)^{\frac{4-\epsilon}{3-\epsilon}}\left[ \frac{15}{8}+\epsilon \left(-0.575331+\frac{3}{8}\right)+\mathcal{O}\left(\epsilon^2\right)\right]+\frac{1}{\epsilon}\left(\frac{2}{5}\epsilon \bar Q\right)^{\frac{2-\epsilon}{3-\epsilon}}\left[
\frac{5}{4}+\epsilon \left(-0.9371-\frac{1}{4}\right)+\mathcal{O}\left(\epsilon^2\right)\right] +\mathcal{O}\left((\epsilon \bar Q)^0\right)  \,, \end{equation}
The above expression is of the form predicted by the large charge EFT approach, which in arbitrary dimensions reads \cite{Gaume:2020bmp}
\begin{align}
\Delta_{\cO_{\bar Q}}= {\bar Q}^{\frac{d}{d-1}}\left[\alpha_{1}+ \alpha_{2} {\bar Q}^{\frac{-2}{d-1}}+\alpha_3 {\bar Q}^{\frac{-4}{d-1}}+\ldots\right] +\bar Q^0\left[\beta_0+ \beta_{1} {\bar Q}^{\frac{-2}{d-1}}+\ldots\right] + \ldots \ \ .
\label{largecharge}
\end{align}
In \cite{{Badel:2019oxl}}, the authors have compared Eq.\eqref{eq:Delta1LoopLarge2} with the results of lattice studies of the $3$-dimensional $U(1)$ model in the large charge limit \cite{Banerjee:2017fcx}, with mild results compatible with the limitations related to taking $\epsilon = 1$. A similar comparison has been performed in \cite{Watanabe:2019pdh} via a slightly different approach and reaching a similar conclusion.

 \bigskip
We now move to consider the $d=3-\epsilon$ case with $\mathcal{N}_3 = 1/36$. Unlike the previous case, the beta function of the model starts at two loops, and thus the $1$-loop theory is conformal invariant in exactly three dimensions. This allows a more direct comparison to the predictions of the large charge  EFT  of  three-dimensional CFT. In particular, we can compare the coefficient of the ${\bar Q}^0$ term in \eqref{largecharge}, which in three dimensions is calculable and a theory-independent number related to the sound speed and the $1$-loop Casimir energy on the sphere \cite{Hellerman:2015nra}. The value of this coefficient as computed in the EFT approach reads: $\beta_0 \vline_{d=3} = - 0.0937256$ \cite{Monin:2016jmo, Alvarez-Gaume:2016vff} and agrees with the result of Montecarlo simulations \cite{Banerjee:2017fcx}. Reproducing this number via semiclassical methods provides a nontrivial consistency check of the large charge expansion framework.

The leading order energy follows again from Eqs.\eqref{firstorder} and \eqref{generalmu}, which give
\begin{equation}
\frac{\Delta_{-1}(\mathcal{A}^*)}{\mathcal{A}^*}= F_{3d}\left(\frac{\mathcal{A}^{*2}}{12\pi^2}\right), \qquad \qquad F_{3d}(x) \equiv \frac{1+\sqrt{1+x}+x/3}{\sqrt{2}(1+\sqrt{1+x})^{3/2}} \,,
\label{DelmQ}
\end{equation}
while $\Delta_0$ is obtained by regularizing Eq.\eqref{eq:E1loop}. Since the $1$-loop beta function vanishes, the result is finite in $d = 3$ and reads
\begin{align}
\Delta_{0}(\mathcal{A}^*)=\frac{1}{4}-3(R\mu)^2+\tfrac{1}{2}\sqrt{8R^2\mu^2-1} + \frac{1}{2}  \sum_{\ell=1}^{\infty}\sigma(\ell) \,,
\end{align}
where
 \begin{align}
\sigma(\ell)=(1+2 \ell)R\left[\omega_+(\ell)+\omega_-(\ell)\right] -4 \ell( \ell+1)-6(R\mu)^2 + \frac{1}{2} \,.
\end{align}
Combining the previous results and expanding in the perturbative regime, we obtain
\begin{align}
\Delta_{\phi^{\bar Q}}=\frac{{\bar Q}}{2} +\kappa\left[\frac{{\bar Q}^3-3{\bar Q}^2}{9}+{\mathcal{O}}({\bar Q})\right] -\kappa^2\left[\frac{{\bar Q}^5}{9}-\frac{{\bar Q}^4(64-9\pi^2)}{72}+{\mathcal{O}}(\bar Q^3)\right]+\kappa^3\left[\frac{2 {\bar Q}^7}{9}+\frac29\left\{-13+\frac{10}{9}\pi^2+\frac{3}{32}\pi^4\right\}{\bar Q}^6+\mathcal{ O}({\bar Q}^5)\right]+\mathcal{O}\left(\kappa^4\right) \,,
\label{Delexp}
\end{align}
where $\kappa = \left(\frac{\lambda}{8\pi}\right)^2$. This $6$-loops result has been verified via diagrammatic computations in \cite{Jack:2020wvs}.

We now proceed by analyzing the large $\mathcal{A}^*$ limit which is captured by the large charge EFT. $\Delta_{-1}$ can be expanded analytically while $\Delta_0$ can be computed numerically and then fitted to the expected functional form \eqref{largecharge}.
The value of the coefficients can be found in \cite{Badel:2019khk, Jack:2020wvs}. Here, we just report the result for $\beta_0$ which reads
\begin{equation}
   \beta_0\vline_{d=3}=-0.0937255 \,,
\end{equation}
with an error of $3$ on the last digit. The universal coefficient $\beta_0$ agrees to high accuracy with the value obtained in the EFT approach and Montecarlo simulations.

It has been recently pointed out that in $4$ dimensions the large charge EFT predicts the existence of a universal $\text{log} \bar Q$ term with calculable coefficient $\delta_0\vline_{d=3} = -\frac{1}{48 \sqrt{3}}$ \cite{Cuomo:2020rgt} \footnote{This term arises in the renormalization of $\beta_0$, which features a pole for $d=4$.}.
It would be interesting to test this prediction in the semiclassical framework as done for the three-dimensional case.


\subsection{The $O(N)$ model in $4-\epsilon$ and $3-\epsilon$ dimensions} \label{ONsec}

In this section, we analyse the large charge expansion in the non-abelian $O(N)$ vector model, which constitutes the natural generalization of the $U(1)$ model investigated in the previous section. For $N = 1, 2, 3$, it defines respectively the \emph{Ising}, \emph{XY}, and \emph{Heisenberg} universality classes while for $N=4$, it describes the standard model Higgs. In euclidean spacetime, the $O(N)$ theory is defined by the action
\begin{equation}
        \mathcal{S} = \int d^d x \frac{1}{2} \partial^{\mu}  \phi_a  \partial^{\mu} \phi_a +  V\left( \phi_a \phi_a \right)  \qquad \qquad a = 1,\dots, N\ .
\label{eq:onla}
\end{equation}
As before, we consider the massless theory in $d=4- \epsilon$ and $d=3 -\epsilon$ with potentials $V_{d = 4-\epsilon}= \mathcal{N}_4 g_0 \left( \bar \phi_a \phi_a \right )^2$ and $V_{d = 3-\epsilon}= \mathcal{N}_3 g_0^2 \left( \bar \phi_a \phi_a \right )^3$ with $g_0$ the bare coupling and $\mathcal{N}_{3,4}$ the normalization. The conserved Noether current associated with the global $O(N)$ symmetry transforms in the adjoint representation of $O(N)$ and it is given by
\begin{equation} \label{current}
        \left( \mathbb{j}^\mu \right)^{a b} = \left(\phi^a \partial^\mu \phi^b - \phi^b \partial^\mu \phi^a \right) \,.
\end{equation}
The corresponding conserved charge is matrix-valued and can be decomposed in terms of the generators of the algebra $T^A$
\begin{equation}
   \mathcal{Q}^{a b}= \int d^{d-1} x \ \left( \mathbb{j}^0 \right)^{a b} = \sum_A Q_A \left(T^A\right)^{a b} \,.
\end{equation}
The $O(N)$ group with even or odd $N$ has rank $\frac{N}{2}$ or $\frac{N-1}{2}$ respectively, which corresponds to the number of ``charges" $Q_A$ we can fix. Without loss of generality, we focus on the even-$N$ case and we fix $k < N/2$ charges via $k$ constraints $Q_i = \bar Q_i$, where $\{\bar Q_i\}$ is a set of fixed constants and  $i = 1, \dots , k$. Using the fact that the $O(N)$ model has a $SU(N/2) \times U(1)$ subalgebra, it is useful to introduce $N/2$ complex field variables as
\begin{align}\label{complex}
  \varphi_1 &= \frac{1}{\sqrt 2} \left(\phi_1 + i \phi_ 2 \right) = \frac{1}{\sqrt 2} \sigma_1 \ e^{i \chi_1}\,, \qquad \qquad \varphi_2 = \frac{1}{\sqrt 2} \left( \phi_3 + i \phi_4\right) = \frac{1}{\sqrt 2} \sigma_2 \ e^{i \chi_2}\,,  \qquad \qquad \varphi_3 =\dots \ \ .
\end{align}
such that $\varphi_i$ or $\bar{\varphi_i}$ has charge $\bar{Q_i} = +1$ or $-1$ respectively.
As before, we will consider the $O(N)$ theory on a cylinder $S^{d-1} \times \mathbb{R}$ where the action reads
\begin{equation}
        \mathcal{S}_{cyl} = \int d^d x \sqrt{g} \left[\frac{1}{2}g_{\mu\nu}\partial^{\mu} \sigma_i  \partial^{\nu} \sigma_i  + \frac{1}{2}  \sigma_i \sigma_i g_{\mu\nu} (\partial^{\mu} \chi_i \partial^{\nu} \chi_i)+ \Tilde{V} (\sigma_i \sigma_i ) \right] \ ,  \quad \quad  \Tilde{V}\left(\sigma_i \sigma_i \right) = V \left(\sigma_i \sigma_i\right) + \frac{m^2}{2} \sigma_i \sigma_i \ , \quad \quad i = 1, \dots, N/2\ .
\end{equation}
In analogy with the abelian case, we look for a spatially homogeneous solution of the EOM. This has the lowest energy at fixed charge and reads
\begin{equation} \label{SolEOM}
  \begin{cases}
\sigma_i =  A_i\, \ , \ \ \chi_i =- i \mu t \, & i=1,\dots,k \,,\\
\varphi_{k+j} = 0, & j=1,\dots,N/2-k \, .
\end{cases}
\end{equation}
The chemical potential $\mu$ is the same for all the $\chi_i$ even if the charges $\bar Q_i$ are all different. For $i=1,\dots,k$ this ground state describes circular motion in the plane spanned by the real and imaginary parts of $\varphi_i$. The motions in different planes are synchronous with the same angular velocity $\mu$ and different radii of the circles $A_i$. These parameters are fixed by the EOM and Eq.\eqref{current} as
\begin{equation}
    \mu^2 =F(v^2) \,  \qquad \quad \frac{\bar Q}{ R^{d-1} \Omega_{d-1}}=\mu v^2  \,,
\end{equation}
where we have defined
\begin{equation}
  v^2 \equiv \sum_{i=1}^k A_i^2 \,, \quad \qquad  \bar Q \equiv \sum_{i=1}^k \bar Q_i \,, \qquad \qquad F (v^2)=\frac{1}{A_j} \frac{\partial \Tilde{V}(\sigma_i \sigma_i)}{\partial \sigma_j} \bigg\rvert _{\sigma_j = A_j} \,,
\end{equation}
with $\bar Q$ the sum of the charges.

It can be shown that the symmetry breaking pattern induced by fixing the charges can be seen as an explicit symmetry breaking  $O(N) \longrightarrow O(N-2k)\times U(k)$ followed by a spontaneous symmetry breaking (SSB) $U(k) \longrightarrow U(k-1)$ \cite{Alvarez-Gaume:2016vff, Antipin:2020abu}.
The latter can be understood by noting that we can always use an $O(N)$ rotation to rotate the ground state to
\begin{equation}\label{rot}
\frac{1}{\sqrt 2} (A_1,..., A_k, 0,...,0) \longrightarrow \big(  \underbrace{0, ...,0}_{k-1}\,,\, \tfrac{v}{\sqrt 2} \,,\, \underbrace{0,...,0}_{N/2-k} \big)
\,.
\end{equation}

This analysis shows that we can organize the saddle point computation as a single coupling 't Hooft expansion in $\mathcal{A} = g \bar Q$ in full analogy with the $U(1)$ case. The sum of the charges acts as a single $U(1)$ charge while the charge configuration plays no role at all. In order to access more general charge configurations is necessary to consider non-homogeneous ground states as done for the $O(4)$ critical model in \cite{Banerjee:2019jpw, Hellerman:2017efx, Hellerman:2018sjf}. The above considerations lead directly to the path integral expression for the ground state energy, which reads
\begin{equation} \label{finale}
    \bra{\bar Q}e^{-HT}\ket{\bar Q} =   \frac{1}{\mathcal{Z}} \int D^{k} \sigma \ D^k \chi \ e^{- \mathcal{S}_{eff}} \,,
    \end{equation}
where
\begin{eqnarray}
 \label{O2n_Lagrangian}
\mathcal{S}_{eff} =&& \int^{T/2}_{-T/2} d t \ \int d\Omega_{d-1} \left(\frac{1}{2}\partial \sigma_i \partial \sigma_i + \frac{1}{2} \sigma_i^2 (\partial \chi_i \partial \chi_i)+ \Tilde{V} (\sigma_i \sigma_i ) + \frac{i} { R^{d-1} \Omega_{d-1}} \  \bar Q  \ \dot \chi_{N/2} \right) \ .
    \end{eqnarray}
The sums over $i$ run from $1$ to $N/2$, i.e. we fixed the all the $N/2$ charges.
Computing this path integral via a saddle point expansion, the scaling dimension at the fixed point $\mathcal{A}^* \equiv g^* \bar Q $ of the lowest-lying operator carrying a total charge $\bar Q$ takes the form
\begin{equation} \label{TQ}
\Delta_{T_{\bar Q}} = E_{T_{\bar Q}}R = \sum_{j=-1}^{\infty} g^{*j}\Delta_j(\mathcal{A}^*)  = \sum_{j=-1}^{\infty} \frac{1}{\bar Q^j} \Tilde{\Delta}_j(\mathcal{A}^*) \ .
\end{equation}
As in the $U(1)$ case, the leading term $\Delta_{-1}(\mathcal{A}^*)$ is  given by Eq.\eqref{firstorder} with $R \mu$ and $\mathcal{A}$ related by Eq.\eqref{generalmu}.

To compute the leading quantum corrections $\Delta_0$ we consider the ground state in \eqref{rot} and parametrize the fluctuations around it as
\begin{equation}
\label{Parametrization}
  \begin{cases}
\chi_i =- i \mu t \ +\frac{1}{v}p_i(x) \,, & i=1,\dots,N/2-1 \,,\\
\chi_{N/2} =- i \mu t \ +\frac{1}{v}\pi(x) \,,\\
\sigma_i =  s_i(x)\,, & i=1,\dots, N/2-1 \,,\\
\sigma_{N/2} = v + r(x)\
\end{cases}
 \,,
\end{equation}
The Lagrangian at the quadratic order in the fluctuations reads
\begin{eqnarray}
    \mathcal{L}_2   =\frac{1}{2}(\partial \pi)^2+\frac{1}{2}(\partial r)^2 +\frac{2}{d-2} (\mu^2-m^2)r^2-2 \ i \ \mu \ r \ \dot \pi  + \frac{1}{2} \partial s_i \partial s_i + \frac{1}{2} \partial p_i \partial p_i -2 \ i \ \mu \ s_i \ \dot p_i \ .\end{eqnarray}

The spectrum contains states that are already present in the $U(1)$ case, i.e. the conformal mode $\chi_{N/2}$ and one massive state $\sigma_{N/2}$ with dispersion relations given by Eq.\eqref{dispersion}.

Additionally, we now have also $\frac{N}{2}-1$ non-relativistic (Type II) Goldstone bosons $\chi_i$ and as many massive states $\sigma_i$ with mass $2\mu$ and dispersion relations
\begin{equation} \label{type2}
\omega_{\pm\pm}(\ell) = \sqrt{J^2_\ell + \mu^2} \pm \mu \,.
\end{equation}

According to the Nielsen-Chada theorem \cite{Nielsen:1975hm}, Type II Goldstone bosons count double with respect to the number of broken generators. Thus we have
\begin{equation}
  1 + 2\times \left( \frac{N}{2}-1 \right) = N - 1 = \dim \left(U\left(\frac{N}{2}\right)/U\left( \frac{N}{2}-1 \right)\right) \,.
\end{equation}

$\Delta_0$ is again given by the fluctuation functional determinant. It can be easily shown that the generalization of Eq.\eqref{eq:E1loop} to general non-Abelian scalar theories is
\begin{equation}
\label{eq:one-loop-det1}
\Delta_0 = \frac{R}{2}\sum_{\ell=0}^\infty n_{\ell} \sum_{i} g_i \omega_i(\ell)\,,
\end{equation}
where the sum over $i$ runs over all the fluctuations' dispersion relations $\omega_i$, each counted with its multiplicity $g_i$. In the $O(N)$ case, we have

\begin{equation}
\label{eq:one-loop-ON}
\Delta_0 = \frac{R}{2}\sum_{\ell=0}^\infty n_{\ell}\left[\omega_+(\ell)+\omega_-(\ell)+(\tfrac{N}{2}-1)(\omega_{++}(\ell)+\omega_{--}(\ell))\right]\,.
\end{equation}

It is instructive to analyze what happens to our computation if we don't fix all the $N/2$ charges $Q_i$ but only $k < N/2$ out of them. It is easy to show that in such a case, the number of Type II Goldstone bosons and massive particles with dispersion relation in Eq.\eqref{type2} becomes $k-1$, whereas the spectrum is completed by $2 \times [(N/2-1) - (k-1)]= N -2k$ new massive states with mass $\mu$ and dispersion relation
\begin{equation}
\omega_{*}(\ell) = \sqrt{J^2_\ell + \mu^2} \,.
\end{equation}
Accordingly, the expression for $\Delta_0$ becomes
\begin{equation}
\Delta_0 = \frac{R}{2}\sum_{\ell=0}^\infty n_{\ell}\left[\omega_+(\ell)+\omega_-(\ell)+(k-1)(\omega_{++}(\ell)+\omega_{--}(\ell)) + (N - 2k) \omega_* \right]\,.
\end{equation}
Since $\omega_{++}(\ell)+\omega_{--}(\ell)= 2 \omega_* (\ell)$, $\Delta_0$ does not depend on the number of charges that are fixed. This result is consistent with the scaling dimension not being sensitive to the charge configuration but only to the sum of the charges.

In parallel with the previous section, we now proceed by providing explicit results starting from the case $d=4-\epsilon$ and $\mathcal{N}_4 =\frac{(4 \pi)^2}{4!}$ which has been considered by us in \cite{Antipin:2020abu}. As his Abelian relative, this theory features an infrared WF FP, which for small $\epsilon$ can be expressed as a power series in $\epsilon$.

The computation of the leading order $\Delta_{-1}$ is analogous to the $U(1)$ case and leads to the same result

\begin{align}
 \label{classicON}
  \frac{\Delta_{-1}}{\mathcal{A}^*} = \frac{1}{4} F_{4d}(x) \,, \qquad \quad x \equiv 6 \mathcal{A}^*  + \sqrt{-3+36 \mathcal{A}^{*2}} \,,
   \end{align}
where $F_{4d}$  has been defined in Eq.~\eqref{classic}.
To compute the leading quantum correction, we start from Eq.\eqref{eq:one-loop-ON} and we follow the procedure of Sec.\ref{sub1} in order to regularize and renormalize the fluctuation determinant. As a result, we obtain
\begin{eqnarray}
\Delta_0(\cA^*)= -\frac{15 \mu^4  R^4+6 \mu^2  R^2-5}{16}
+\frac{1}{2} \sum_{\ell=1}^\infty\sigma(\ell)
+\frac{\sqrt{3\mu^2R^2-1}}{\sqrt{2}}  -\frac{1}{16}\left(\frac{N}{2}-1\right)\left[7 + R \mu \left(-16+6R\mu+3R^3 \mu^3\right)\right] \ .
\label{Delta0ON}
\end{eqnarray}
where
\begin{align}
\sigma(\ell) &=  R (1+\ell)^2\left[\omega_+(\ell)+\omega_-(\ell)+(\tfrac{N}{2}-1)(\omega_{++}(\ell)+\omega_{--}(\ell))\right]+\frac{1}{8}(N+8)\left( R^2 \mu ^2 -1\right)^2\frac{1}{\ell} \nonumber \\
& +\frac{1}{2}\left(2-N-(N+2) R^2 \mu^2 \right)+\frac{1}{2}\left(2-5N-(2+N) R^2 \mu^2 \right) \ \ell -3 N \ \ell^2- N \ \ell^3 \ .
\end{align}
Again all the quantities are evaluated in $d=4$ dimensions.

To check this result in the perturbative regime, we sum classical contribution \eqref{classicON} and leading quantum correction \eqref{Delta0ON}, and we perform an expansion for small $\mathcal{A}^* = g^* Q$, obtaining
 \begin{eqnarray}
\label{3loopcomplete}
    && \Delta_{T_{\bar Q}}=\textcolor{red}{\bar Q}+\textcolor{red}{\left(-\frac{\bar Q}{2}+\frac{\bar Q(\bar Q-1)}{8+N}\right)}\epsilon-\left[\frac{184+N(14-3N)}{4(8+N)^3}\bar Q+\textcolor{red}{\frac{(N-22)(N+6)}{2(8+N)^3}\bar Q^2+\frac{2}{(8+N)^2}\bar Q^3} \right]\epsilon^2 + \left[\textcolor{red}{\frac{8}{(8+N)^3}\bar Q^4} \right. \nonumber\\
    &&\textcolor{red}{ +\frac{-456-64N+N^2+2(8+N)(14+N)\zeta(3)}{(8+N)^4}\bar Q^3} +\frac{-N^4-57 N^3+258 N^2-24 (N+6) (N+8) (N+26) \zeta (3)+8176 N+31008}{4 (N+8)^5}\bar Q^2 \nonumber \\
    &&\left. +\frac{-69504 + 3 N [-5216+ N (184+ N(86+N))]+64 (8+N)(178+ N (37 + N))\zeta(3)}{16
   (N+8)^5}\bar Q\right]\epsilon^3 + \left[\textcolor{red}{-\frac{42}{(8+N)^4} \bar Q^5} \right. \nonumber \\
    &&\left. \textcolor{red}{+\frac{-4 N^2-5 (N+8) (N+30) \zeta(5) -2 (N+8) (6 N+65) \zeta(3) +476
   N+3344}{(8+N)^5} \bar Q^4} + \frac{1}{60 (8+ N)^6} \left(\pi^4 N^4+60 N^4+38 \pi^4 N^3+4020 N^3 \right. \right. \nonumber \\
    &&\left. \left.+528 \pi^4 N^2-88800 N^2-4200 (N-2) (N+8)^2 \zeta (5)-60 (N+8) (N (N (3
   N-44)-1720)-7464) \zeta(3) +3200 \pi^4 N-1577280 N \right. \right. \nonumber \\
    &&\left. \left. +7168 \pi^4-5662560 \right) \bar Q^3 - \frac{1}{80 (N+8)^7} \left( 10 N^6+4 \pi ^4 N^5+915 N^5+224 \pi^4 N^4+34120 N^4+4464 \pi^4
   N^3+86600 N^3+41600 \pi ^4 N^2 \right. \right. \nonumber \\
    &&\left. \left. -3928440 N^2-400 (N+8)^2 (N (65
   N+958)+2496) \zeta (5)-20 (N+8) (N (N (N (N
   (N+52)+904)-12224)-181184)-514112) \zeta (3) \right. \right. \nonumber \\
    &&\left. \left. +185344 \pi^4  N-35161600 N+319488 \pi^4 - 87127680\right) \bar Q^2 +\frac{1}{960 (8+ N)^7} \left(45 N^6+32 \pi ^4 N^5+5820 N^5+1952 \pi^4 N^4+322440 N^4  \right. \right. \nonumber \\
    &&\left. \left. +40256
   \pi^4 N^3+1972440 N^3+380416 \pi ^4 N^2-16196640 N^2-9600
   (N+8)^2 (N (25 N+418)+1240) \zeta (5)  \right. \right. \nonumber \\
    &&\left. \left. -240 (N+8) (N (N (N (N
   (N+40)+1056)-3496)-100480)-300096) \zeta (3)+ 1699840 \pi ^4 N -191091840 N+2916352 \pi^4  \right. \right. \nonumber \\
    &&\left. \left. -494461440  \big) \bar Q \Bigg]\epsilon^4  +  \mathcal{O}\left(\epsilon^5 \right)  \right. \right. \,,
 \end{eqnarray}
 where the terms highlighted in red stem from the semiclassical computation.
 To two loops, they agree with the known $2$-loop anomalous dimension of the $\bar Q$-index traceless symmetric $O(N)$ tensor with classical dimension $\bar Q$ \cite{Kehrein:1995ia}, which can be depicted as a $\bar Q$-boxes Young tableau with one row. In \cite{Antipin:2020abu}, we also obtained all the black terms at three and four loops by combining the knowledge of the red ones with the known perturbative results for the $\bar Q=1$ \cite{Kleinert:1991rg}, $\bar Q = 2$ \cite{Braun:2013tva, Kompaniets:2019zes} and $\bar Q=4$ \cite{ Calabrese:2002bm} cases. This result corrects and extends the one obtained long ago in \cite{Wallace:1974nu}. This is an example of how the semiclassical expansion at fixed charge can be helpful in improving and checking perturbative results. Conversely, the diagrammatic check of the semiclassical computation has been recently further extended to the $4$-loop level in \cite{Jack:2021ypd}.
 Notice that for $\bar Q = 1$ the relevant operator is the $\phi$ field, while for $\bar Q = 2$, it is the bilinear traceless symmetric $O(N)$ tensor $\phi_a \phi_b -\frac{1}{N} \phi_c \phi_c$, which is of interest to many critical phenomena being responsible for crossover behaviour in the $O(N)$ theory. Its anomalous dimension defines a so-called \emph{crossover exponent} describing the instability of the theory against anisotropy \cite{Calabrese:2002qi}.  As we will see in the next section, in the perturbative regime the operator identification can be proven via group-theoretical arguments. For sake of completeness, we report here also the large $\cA$ expansion of $\Delta_{T_{\bar Q}}$ \cite{Jack:2021ypd}
\begin{align}
\Delta_{T_{\bar Q}}&=\frac{1}{\epsilon}\left(\frac{4\epsilon \bar Q}{N+8}\right)^{\frac{d}{d-1}}\left[
\frac{3(N+8)}{16}+\epsilon\left(-1.5559-0.2293 N+\frac{3(3N+14)}{16(N+8)}\right)+\cO(\epsilon^2)\right] \nonumber \\
&+\frac{1}{\epsilon}\left(\frac{4\epsilon \bar Q}{N+8}\right)^{\frac{d-2}{d-1}}\left[
\frac{N+8}{8}+\epsilon\left(-0.05413+0.0383N-\frac{3N+14}{8(N+8)}\right)+\cO(\epsilon^2)\right]+\cO[(\epsilon \bar Q)^0] \,.
\end{align}

We end this review section with the $O(N)$ sextic theory in $d=3-\epsilon$ dimension with $\mathcal{N}_3 = \frac{1}{48}$ which has been studied in \cite{Jack:2020wvs}. The leading order energy is once again given by Eqs.\eqref{firstorder} and \eqref{generalmu} and reads
\begin{equation}
\frac{\Delta_{-1}(\mathcal{A}^*)}{\mathcal{A}^*}= F_{3d}\left(\frac{\mathcal{A}^{*2}}{2\pi^2}\right) \,,
\label{final3d}
\end{equation}
where $F_{3d}$ has been defined in Eq.~\eqref{DelmQ}.

The regularized version of Eq.\eqref{eq:one-loop-ON} provides the $1$-loop correction in the semiclassical expansion as
\begin{align}
\Delta_{0}(\mathcal{A}^*)=\frac{1}{4}-3(R\mu)^2+\tfrac{1}{2}\sqrt{8R^2\mu^2-1} -\left(\frac{N}{2}-1\right)\left(\frac{1}{4} + (R\mu)^2 - R\mu\right)+ \frac{1}{2} \sum_{\ell=1}^{\infty}\sigma(\ell) \,,
\label{Delab}
\end{align}
 where
\begin{align}
\sigma(\ell)=(1+2 \ell)R\left[\omega_+(\ell)+\omega_-(\ell)+\left(\frac{N}{2}-1\right) \omega_{++}(\ell)+\omega_{--}(\ell)\right] -4\ell(\ell+1)-\left(6(R\mu)^2-\frac{1}{2}\right)- \left (\frac{N}{2}-1\right)\left( 4\ell(\ell+1) +\ 2(R\mu)^2+\frac{1}{2} \right) \,,
\label{sigab}
\end{align}
which is again constructed so that the sum is convergent in $d=3$.
In \cite{Jack:2020wvs}, this result has been  verified via conventional diagrammatic techniques at the $6$-loop level. The large $\mathcal{A}^*$ expansion can be studied numerically, as in the $U(1)$ case. The outcome is corrections (proportional to $N-2$) to the $U(1)$ values of the $\alpha$'s coefficients in \eqref{largecharge}, while the $\beta$'s receives no new contributions.

\section{From group theory to operators: The map}\label{themap}

\subsection{The power of symmetries}\label{Symmetry}

Our knowledge of a quantum field theory (QFT) is generally encoded in the correlation functions of local operators. If the QFT under consideration has some internal global compact symmetry group $\mG$ which is neither explicitly broken nor spontaneously broken, then without loss of generality we may restrict ourselves to local operators that transform under definite unitary irreducible representations of $\mG$, since any other local operator should be able to be expressed as a linear combination of those local operators with definite transformation properties.

Therefore let us consider a set of local operators $\mO^p_1,\mO^p_2,...,\mO^p_{d_p}$ that transform under a $d_p$-dimensional unitary irreducible representation
$\Gamma^p$ of $\mG$.~\footnote{Much of the basic group theory introduced in this section is based on the textbooks by J. F. Cornwell ~\cite{Cornwell:1985xs,Cornwell:1985xt} and by B. C. Hall ~\cite{Hall:2015tb}, which leads us to the proofs of several important results needed for application in the fixed-charge semiclassical approach to CFT.} This implies they have implicitly the same spacetime (Lorentz) transformation properties, and form a basis of a carrier
space $V_p$ for $\Gamma^p$, that is, for $i=1,2,...,d_p$ and all $T\in\mG$
\begin{align}
\Phi(T)\mO^p_i=\sum_{j=1}^{j=d_p}\Gamma^p(T)_{ji}\mO^p_j
\label{eq:transg}
\end{align}
where $\Phi(T)$ denotes the linear transformation operator corresponding to $T\in\mG$ that acts on $V_p$, and $\Gamma^p(T)$ denotes the representation matrix corresponding to $T\in\mG$. It is important to note that the complete symmetry property of an operator is encoded
in two indices. For the set of operators $\mO^p_i$, one index is $p$, which refers to the irreducible representation the operator
belongs to, up to equivalence. The other index is $i$, referring to which row of $\Gamma^p$ the operator $\mO^p_i$ transforms according to.
There is a counterpart of Eq.~\eqref{eq:transg} in Lie algebra representation theory. Suppose $\mLt$ is the complexification of the
real Lie algebra of $\mG$, then for $i=1,2,...,d_p$ and all $a\in\mLt$
\begin{align}
\Phi(a)\mO^p_i=\sum_{j=1}^{j=d_p}\Gamma^p(a)_{ji}\mO^p_j
\label{eq:transl}
\end{align}
where now $\Phi(a)$ denotes the linear transformation operator corresponding to $a\in\mLt$ that acts on $V_p$ and $\Gamma^p(a)$ denotes the
representation matrix corresponding to $a\in\mLt$. Let $\mH$ be a Cartan subalgebra of $\mLt$. Without loss of generality we may assume that, the set of operators $\mO^p_i$ are chosen such that $\Gamma^p(h)$ is diagonal for all $h\in\mH$. This fact can be represented by the following equation:
\begin{align}
\Phi(h)\mO^p_i=\lambda_i(h)\mO^p_i
\label{eq:weight}
\end{align}
for $i=1,2,...,d_p$. This defines $d_p$ linear functionals $\lambda_i(h)$ that act on $\mH$, which are the \emph{weights} of the irreducible representation $\Gamma^p$ in mathematical terms. Therefore, the index $i$ plays the role of labeling the weights of $\Gamma^p$. From a physical point of view, a weight when acting on a set of elements in $\mH$, gives the Cartan charges associated with the elements, and thus
specifies a charge configuration.

We state two important consequences of the intact (i.e. neither explicitly nor spontaneously broken) symmetry $\mG$.\\

\noindent
\textbf{Consequence 1:} Operators of different symmetry properties (i.e. belonging to inequivalent irreducible representations of $\mG$, or belonging to
equivalent irreducible representations of $\mG$ but correspond to different weights) do not mix under the renormalization group.\\

\noindent
\textbf{Consequence 2:} In a conformal field theory (CFT), operators that transform \emph{in} the same irreducible representations of $\mG$ but correspond to different weights have identical
scaling dimensions, if they do not mix with operators that do not belong to their carrier space under renormalization.

\vskip .3cm
Here and hereafter, we always assume that for equivalent irreducible representations an appropriate similarity transformation has
been applied to make them identical. The first {\it consequence} above is simply the requirement that renormalization of the theory preserves its
global symmetry. The second {\it consequence} can be verified by examining the two-point correlator of the operators in question and making use
of the Wigner-Eckart theorem, which states that matrix elements of irreducible tensor operators can be factorized into two parts, with
the first part solely determined by the corresponding Clebsch-Gordan coefficients, and the remaining part called reduced matrix elements
which are independent of the magnetic quantum numbers (weights). When we consider two-point correlators like $\langle\Omega|\mO^p_i(x)\mO^q_j(y)|\Omega\rangle$ ($|\Omega\rangle$ being the vacuum), we may view $\mO^q_j(y)|\Omega\rangle,\mO^p_i(x)|\Omega\rangle$  as a whole, and the identity operator as the irreducible tensor operator, in order to apply the Wigner-Eckart theorem. The Clebsch-Gordan coefficients are trivial for $p=q$ and the Wigner-Eckart theorem tells us for $i=j$ the two-point correlator is independent of the weight label $i$. With the further assumption that this set of operators do not mix with other operators, we deduce that their scaling dimensions must be the same (since in a CFT scaling dimension $\Delta$ of an operator $\mO$ can be completely determined from its two-point correlator as $\langle\Omega|\mO(x)\mO(y)|\Omega\rangle=|x-y|^{-2\Delta}$). Note that it is important to require the two operators to transform \emph{in} the same irreducible representation, i.e. they live in the same irreducible carrier space. If they both merely transform according to some irreducible representation $\Gamma^p$, but are not in the same irreducible carrier space, then we cannot claim anything about their scaling dimensions.

\subsection{The nature of charge fixing}

The fixed-charge approach has proven to be very powerful in probing the dynamics of a QFT with global symmetries in regimes that are difficult to access by conventional methods. In most applications so far a CFT is considered since one can employ the Weyl
invariance of the theory to map the CFT to a cylinder, with the computation of scaling dimensions of fixed-charge operators turned into the computation of the ground state energies in the corresponding fixed-charge sectors of the cylinder theory. Obtaining results for non-CFTs may also be possible in various cases~\cite{Antipin:2020rdw}, however for the moment we will restrict our presentation to the case of CFTs for simplicity. Conventional perturbation theory can probe the small-charge regime, up to a certain power in the coupling expansion,
limited by computational capabilities, while the large-charge regime is beyond its validity range. In the fixed-charge approach,
however, both the small-charge and large-charge regimes are dealt with by a semiclassical expansion around a nontrivial fixed-charge trajectory in the path integral.

An important feature of the fixed-charge approach to scaling dimension computation is that  \emph{a priori} it does \emph{not} fix the full symmetry properties of the fixed-charge operator under consideration. This can be inferred from the general discussion of the fixed-charge path integral (see Sec.~\ref{SOA}), in which only the eigenvalues corresponding to a set of Cartan charges are required. Put it simpler, only weights are known and fixed, and we do not know which irreducible representation the operator belongs to. Multiple irreducible representations can share the same weight, while a given irreducible representation can be realized by different sets of local operators. This is where the Lie algebraic theory cannot tell us more and we need some dynamical information.

The dynamical information is hidden in the starting point of the derivation of a fixed-charge path integral. In a Euclidean field theory,
one considers the expectation of the evolution operator $e^{-HT}$ in an \emph{arbitrary} state $|\psi\rangle$ with a given fixed charge (i.e. weight). In the limit $T\rightarrow\infty$ the expectation gets saturated by the lowest energy state contained in $|\psi\rangle$ which typically has a nonzero overlap with the lowest-lying energy state corresponding to the given charge. Therefore by using state-operator correspondence, the scaling dimension one obtains for the fixed-charge operator should correspond to
the scaling dimension of the lowest-lying operators with the given fixed charge.

Unfortunately, for a given fixed charge, we still do not know \emph{a priori} which operator in which irreducible representation is
lowest-lying. Nevertheless, progress can be made by
\begin{enumerate}
\item Assuming that the lowest-lying operator has the minimal classical scaling dimension (MCSD) with which an operator can be constructed
corresponding to a given charge configuration. This will be called the MCSD assumption. If the MCSD assumption is valid with a unique operator $\mO_{MCSD}$ saturating the MCSD for a given charge configuration, then $\mO_{MCSD}$ must have a definite scaling dimension (i.e. it does not mix with other operators). However, in more general cases multiple operators may saturate the MCSD for a given charge configuration, and some appropriate linear combination of them will become the genuine lowest-lying operator and have a definite scaling dimension. For spin-0 fixed-charge operators~\footnote{In this work we are only concerned with spin-0 fixed-charge operators. Operators with nonzero spin would correspond to inhomogeneous ground states on the cylinder~\cite{Banerjee:2019jpw}.} (corresponding to homogeneous ground states in the cylinder theory) the MCSD assumption obviously requires we consider non-derivative operators only, as extra spacetime derivatives necessarily increase the classical scaling dimension.

\item Carrying out semiclassical computations for various weights of a given irreducible representation. Then, for each weight, list all the irreducible representations that contain it. If semiclassical computation gives different results of scaling dimensions for different weights, and the MCSD assumption and Consequence 2 are used, it might be possible to pin down the correspondence between weights and representations in the semiclassical computation.
\end{enumerate}

To summarize, an important feature of the fixed-charge semiclassical computation is that a priori it only fixes the weight, while the correspondence between the weight and the irreducible representation is hidden in the fact that only the lowest-lying state is projected
out. Further progress in disentangling weights and representations can be made by making the MCSD assumption and carrying out semiclassical
computations for multiple weights in question. To illustrate the main idea, in the following we will first review the simpler case of
$U(1)$ and $O(N)$ vector models, and then turn to the more complicated $U(N)\times U(M)$ linear sigma model which entails a sophisticated
group-theoretic analysis.

Here, we would like to highlight a few important motivations to figure out the correspondence between the weight and the irreducible representation.
First, logically one should always prove the \emph{existence} of the fixed-charge operators. Simply obtaining the result from a fixed-charge semiclassical computation does not guarantee that the results obtained are correct. Second, when comparing the results of the fixed-charge semiclassical computation with
results obtained by other methods (e.g. conventional perturbation theory), it is relevant to know the correspondence between the weight and the irreducible representation, or the explicit form of the fixed-charge operator. Third, for theory and application purposes we might just wish to know the scaling dimension of certain operators that transform according to given irreducible representations.

\subsection{Group-theoretic analysis: $U(1)$ and $O(N)$ vector models}

The simplest QFT with an internal continuous global symmetry is the theory of a complex scalar field $\phi$ that has the $U(1)$ symmetry transformation $\phi\rightarrow e^{-i\alpha}\phi $ with $\alpha$ being an arbitrary constant real phase. To allow for a nontrivial fixed point we consider the theory in $d=4-\epsilon$ Euclidean spacetime dimensions. The Lagrangian of the theory and the Noether charge $Q$ associated
with the global symmetry can be read off from in Eq.~\eqref{eq:uonel} and Eq.~\eqref{U1charge}.
One can derive the commutation relation
\begin{align}
[Q,\phi]=\phi
\label{eq:phicharge}
\end{align}
from canonical commutation relations for the field operator $\phi$. The integrated form of Eq.~\eqref{eq:phicharge} reads
\begin{align}
e^{-i\alpha Q}\phi e^{i\alpha Q}=e^{-i\alpha}\phi
\label{eq:phichargeint}
\end{align}
for an arbitrary real phase constant $\alpha$. It is from Eq.~\eqref{eq:phicharge} and Eq.~\eqref{eq:phichargeint} that we deduce
the $U(1)$ charge of $\phi$ to be $+1$. Then the operator $\phi^n$ ($n$ is a positive integer) has $U(1)$ charge $n$. The normalization
of the $U(1)$ charge changes by multiplying via a nonzero real number. Nevertheless, one can always compute
the $U(1)$ charge of a local operator $\mO$ by computing the parameter $q$ in the commutator equation
\begin{align}
[Q,\mO]=q\mO
\label{eq:ce}
\end{align}
If $\mO$ carries a definite $U(1)$ charge, then there should exist a real number $q$ such that Eq.~\eqref{eq:ce} holds. If the charge normalization is such that $\phi$ carries the charge $+1$, then $q$ must be an integer, as long as $\mO$ can be written as a linear
combination of products of $\phi$ and its derivatives (non-integer powers operators are ill-defined). This simple $U(1)$ example illustrates the well known fact that the charge is
discretized for well-defined local operators of the theory, regardless of the normalization convention while, after a Weyl map to the cylinder, the charge density
can be adjusted continuously by changing the compactification volume.

For a given $U(1)$ charge $n>0$, the operator with MCSD is obviously $\phi^n$. Any additional $(\bar{\phi}\phi)$ factor or derivative
would necessarily increase the classical scaling dimension. Therefore, with the MCSD assumption we expect a semiclassical computation
in the charge-$n$ sector with a homogeneous ground state to deliver the scaling dimension of the operator $\phi^n$.

From a group-theoretic point of view, the next-to-simplest case turns out to be the critical $O(N)$ vector model in $d=4-\epsilon$
dimensions (c.f. Section~\ref{ONsec}). This model features a $N$-component real scalar field $\phi=(\phi_1,\phi_2,...,\phi_N)$.
Its Lagrangian density in Euclidean spacetime is given in Eq.~\eqref{eq:onla}. For definiteness, let us consider the case where $N$ is even. Extension to the case of odd $N$ is
straightforward. The maximal commuting set of charges we can fix corresponds to the maximal set of Cartan generators, which can be made explicit by defining the complex fields
$\varphi_1=\frac{1}{\sqrt{2}}\left(\phi_1+i\phi_2\right),\varphi_2=\frac{1}{\sqrt{2}}\left(\phi_3+i\phi_4\right),...
\varphi_{N/2}=\frac{1}{\sqrt{2}}\left(\phi_{N-1}+i\phi_N\right)$. For each $j=1,2,...,N/2$, there exists an independent phase rotation $\varphi_j\rightarrow\varphi_j e^{-i\alpha_j}$ as a symmetry
transformation of the theory corresponding to a Cartan generator, with $\alpha_j$ being an arbitrary real phase. A generic charge
configuration (i.e. weight) can thus be characterized by $[m]\equiv(m_1,m_2,...,m_{N/2})$, with $m_i$ representing the
charge associated with the $i$th Cartan generator. The normalization of the Cartan charges can be chosen such that
$\varphi_i$ corresponds to $(0,0,...,m_i=+1,0,...,0)$ for $i=1,2,...,N/2$, which we adopt. This implies for a generic charge configuration
$[m]=(m_1,m_2,...,m_{N/2})$,$m_i$'s are all integers. Without loss of generality we may consider only the case in which all $m_i$'s are
nonnegative since the sign of the Cartan charge is a matter of convention. The operator with MCSD that corresponds to
$[m]=(m_1,m_2,...,m_{N/2})$ is then easily constructed:
\begin{align}
\mO_{[m]}\equiv\prod_{i=1}^{i=N/2}(\varphi_i)^{m_i}
\end{align}
If some $m_i$ is negative, we may simply use $\varphi_i^*$ instead of $\varphi_i$ for the corresponding factor.
As in the $U(1)$ case, any additional factor of $(\bar{\varphi_i}\varphi_i)$ or derivative would necessarily increase the classical scaling dimension.

Let us note $\mO_{[m]}$ thus constructed live in the traceless fully symmetric subspace of $O(N)$ transformations. It is fully symmetric
because it is a product of commuting scalar fields. It is traceless because otherwise it would contain some factor like $\phi^2$ which
would violate the MCSD assumption. Therefore, $\mO_{[m]}$ corresponds to an irreducible representation of $O(N)$ and has a definite scaling
dimension. The argument also shows that operators that have the same value of $\sum_{i=1}^{i=N/2}|m_i|$ and MCSD all belong to the same irreducible $O(N)$ representation and thus have the same scaling dimension, in agreement with the expectation that by an $O(N)$ rotation we can associate all charges to a single Cartan generator.


\subsection{Group-theoretic analysis: the $U(N)\times U(M)$ linear sigma model} \label{groupanalysis}
\subsubsection{Introduction}
The $O(N)$ vector model is simple in the fixed-charge semiclassical approach as charge fixing can always be associated with a single Cartan charge by virtue of a symmetry rotation. Thus all charge configurations are similar and solely characterized by $\sum_{i=1}^{i=N/2}|m_i|$.

To allow for more variations in the charge configuration, here we consider the $U(N)\times U(M)$ linear sigma model in $d=4-\epsilon$
dimensions, with $N>1$ and $M>1$ being integers. The Lagrangian density in Euclidean spacetime is given by
\begin{align}
\mL=\Tr(\del_\mu H^\dagger \del^\mu H ) + u_0\Tr(H^\dagger H)^2 + v_0(\Tr H^\dagger H )^2
\end{align}
Here $H$ denotes an $N\times M$ complex matrix scalar field. Without loss of generality we may assume $N\leq M$. The model has the global symmetry
\begin{align}
\mathcal{G} \equiv SU(N)_L\times SU(M)_R\times U(1)_A
\label{eq:gsym}
\end{align}
in which $U(1)_A$ is the universal phase rotation of the $H$ field~\footnote{It does not matter whether this universal $U(1)$ rotation acts from the left or from the right. Therefore precisely speaking the global symmetry should be written as Eq.~\eqref{eq:gsym}. Writing it as $U(N)\times U(M)$ is less rigorous but more convenient, see ~\cite{Calabrese:2004uk} for example.}. Under $SU(N)_L\times SU(M)_R$, the $H$ and $H^\dagger$ fields transform
as
\begin{align}
H\rightarrow LHR^\dagger,\quad H^\dagger\rightarrow RH^\dagger L^\dagger
\end{align}
with $L$ being an arbitrary $N\times N$ constant special unitary matrix, and $R$ being an arbitrary $M\times M$ constant special unitary matrix.

Depending on the value of $N$ and $M$, the model may feature fully-interacting real or complex fixed points. At such a fixed point,
we perform a Weyl map to a cylinder of radius $R$ (i.e. $\mathbb{R}^d \to \mathbb{R} \times S^{d-1}$), with the cylinder action given by
\begin{align} \label{Scyl}
\mathcal{S}_{cyl}=\int d^d x \sqrt{g}\Big[ \Tr(\del_\mu H^\dagger \del^\mu H ) + u_0\Tr(H^\dagger H)^2 + v_0(\Tr H^\dagger H )^2
+ m^2 \Tr( H^\dagger H)\Big].
\end{align}
Here $g$ denotes the metric determinant and $m^2=\left(\frac{d-2}{2R}\right)^2$ is the coefficient of the conformal coupling required by Weyl invariance.

We consider a homogeneous ground state with the ansatz
\begin{equation}
H_0\left(\tau\right)=e^{2iM_E \tau}\Bbar\ ,
\label{eq:ansatz}
\end{equation}
where $\tau$ denotes the cylinder time and $M_E$ is an $N\times N$ diagonal matrix. $\Bbar$ is an $N\times M$ matrix in the form
\begin{align}
{{\Bbar}_{N\times M}}=\left( \begin{matrix}
   {{B}_{N\times N}} & {{\mathbf{0}}_{N\times (M-N)}}  \\
\end{matrix} \right)
\end{align}
in which $B$ is an $N\times N$ diagonal matrix. The Noether charges associated with Cartan generators are encoded
in the following charge configuration matrices:
\begin{align}
{{\mathcal{Q}}_{L}}=-V{{\dot{H}}_{0}}H_{0}^{\dagger }, \qquad {\mQbar_{R}}=VH_{0}^{\dagger }{{\dot{H}}_{0}}
\label{eq:qdef}
\end{align}
with $V= R^{d-1} \Omega_{d-1}$ being the volume of $S^{d-1}$. Plugging in the ansatz Eq.~\eqref{eq:ansatz}, it is straightforward to show
\begin{align}
{{\mathcal{Q}}_{L}}=-2iVM_E{{B}^{\dagger }}B,\qquad {\mQbar_{R}}=2iVM_E\left( \begin{matrix}
   {{B}^{\dagger }}B & {{\mathbf{0}}_{N\times (M-N)}}  \\
   {{\mathbf{0}}_{(M-N)\times N}} & {{\mathbf{0}}_{(M-N)\times (M-N)}}  \\
\end{matrix} \right)
\label{eq:qlqr}
\end{align}
If we parametrize $\mQbar_R$ as
\begin{align}
{{\overline{\mathcal{Q}}}_{R}}=\left( \begin{matrix}
   {{\mathcal{Q}}_{R}} & {{\mathbf{0}}_{N\times (M-N)}}  \\
   {{\mathbf{0}}_{(M-N)\times N}} & {{\mathbf{0}}_{(M-N)\times (M-N)}}  \\
\end{matrix} \right)
\end{align}
Then from Eq.~\eqref{eq:qlqr} we find the constraint
\begin{align} \label{constraint}
\mQ_L+\mQ_R=0
\end{align}
As $B$ is diagonal, $\mQ_L,\mQ_R$ are also diagonal. We will restrict our attention to the sector neutral under $U(1)_A$,
which implies
\begin{align}
\Tr\mQ_L=\Tr\mQ_R=0
\end{align}
To simplify the notation, in the following we use $\mQ$ to denote $\mQ_L$, that is
\begin{align}
\mQ\equiv\mQ_L=-\mQ_R
\end{align}

In the following, we will first determine what are the admissible charge configuration matrices $\mQ$ and then we will disentangle
which irreducible representations they correspond to. Although we work in $d=4-\epsilon$ dimensions we indicate the classical scaling dimensions (CSD) with the corresponding one in $4$ dimensions. For example, the CSD of the field $H$ is $1$ and the one of the operator $\Tr(H^\dagger H)$ is $2$. The transition to $d=4-\epsilon$ dimensions is straightforward.
All the discussion will be restricted to the homogeneous ground state ansatz in Eq.~\eqref{eq:ansatz} and the associated traceless charge
configuration of $\mQ$.

We first introduce and prove several propositions that underlay the determination of the irreducible representation associated with a given charge configuration from Lie algebraic considerations.

\vskip .3cm
\noindent
\textbf{Proposition 1:}$\quad$ Suppose $\mO$ is a fixed-charge operator that corresponds to a traceless charge configuration with MCSD. Let us denote the CSD of $\mO$ by $D$. Let us also suppose $\mO$ belongs to some irreducible representation $(\Gamma_L,\Gamma_R)$ of $SU(N)_L\times SU(M)_R$ in the $U(1)_A$-neutral sector. Then $\Gamma_L$ must appear in ${{(\mathbf{Adj}_L)}^{D/2}}$ , with $\mathbf{Adj}_L$  being the adjoint representation of $SU(N)_L$; $\Gamma_R$ must appear in ${{(\mathbf{Adj}_R)}^{D/2}}$ , with $\mathbf{Adj}_R$  being the adjoint representation of $SU(M)_R$.

\vskip .2cm
\noindent
{\emph{Proof of proposition 1:}}$\quad$ Since we are considering a homogeneous ground state, the corresponding fixed-charge operator $\mO$ must be a Lorentz scalar. Within the MCSD assumption, this implies no derivative can appear in the construction of $\mO$, and thus the operator $\mO$ with CSD $D$ must be built out of the product of $D/2$ $H$ fields and $D/2$ $H^\dagger$ fields (so that $\mO$ is also neutral under $U(1)_A$). Now, under $SU(N)_L\times SU(M)_R$
\begin{align}
H\sim({{\mathbf{F}}_{L}},{{\mathbf{\bar{F}}}_{R}}),\text{ }{{H}^{\dagger }}\sim({{\mathbf{\bar{F}}}_{L}},{{\mathbf{F}}_{R}})
\end{align}
Here $\mathbf{F}_{L}$ denotes the fundamental representation of $SU(N)_L$, and ${\mathbf{\bar{F}}}_{L}$ denotes the anti-fundamental representation of $SU(N)_L$. The notation for representations of $SU(M)_R$ is self-explanatory. Therefore $\mO$ must transform as an irreducible component inside the reducible representation
\begin{align}
({{\Gamma }_{L0}},{{\Gamma }_{R0}}),\text{ with }{{\Gamma }_{L0}}\equiv {{({{\mathbf{F}}_{L}}\otimes {{\mathbf{\bar{F}}}_{L}})}^{D/2}},{{\Gamma }_{R0}}\equiv {{({{\mathbf{F}}_{R}}\otimes {{\mathbf{\bar{F}}}_{R}})}^{D/2}}
\end{align}
Now for representations of special unitary groups we know that
\begin{align}
{{\mathbf{F}}_{L}}\otimes {{\mathbf{\bar{F}}}_{L}}={{\mathbf{1}}_{L}}\oplus \mathbf{Ad}{{\mathbf{j}}_{L}},\text{  }{{\mathbf{F}}_{R}}\otimes {{\mathbf{\bar{F}}}_{R}}={{\mathbf{1}}_{R}}\oplus \mathbf{Ad}{{\mathbf{j}}_{R}}
\end{align}
and thus
\begin{align}
{{\Gamma }_{L0}}={{({{\mathbf{1}}_{L}}\oplus \mathbf{Ad}{{\mathbf{j}}_{L}})}^{D/2}},\text{  }{{\Gamma }_{R0}}={{({{\mathbf{1}}_{R}}\oplus \mathbf{Ad}{{\mathbf{j}}_{R}})}^{D/2}}
\end{align}
All singlet components in ${{\mathbf{1}}_{L}}\oplus \mathbf{Ad}{{\mathbf{j}}_{L}}$ and ${{\mathbf{1}}_{R}}\oplus \mathbf{Ad}{{\mathbf{j}}_{R}}$ can actually be dropped because $\mO$ corresponds to an MCSD operator. If a singlet component contributes then one would be able to construct another operator that corresponds to the same charge configuration with less number of $H$ and $H^\dagger$ fields, in contradiction to the MCSD requirement. Therefore we conclude that the operator $\mO$ belongs to $(\Gamma_L,\Gamma_R)$ , where $\Gamma_L$  and $ \Gamma_R$  must appear respectively in ${{(\mathbf{Adj}_L)}^{D/2}}$ and  ${{(\mathbf{Adj}_R)}^{D/2}}$.

\vskip .3cm
\noindent
\textbf{Proposition 2:}$\quad$ Suppose that the CDS of $\mO$  is D  and the MCDS fixed-charge operator corresponds to a  traceless charge configuration. Let us also suppose $\mO$ belongs to some irreducible representation $(\Gamma_L,\Gamma_R)$ of $SU(N)_L\times SU(M)_R$ in the $U(1)_A$-neutral sector. Then $(\Gamma_L,\Gamma_R)$ must appear in the $U(1)_A$-neutral sector of the decomposition of the $D$-index traceless fully symmetric tensor of $O(2NM)$ under the branching
\begin{align}
O(2NM)\supset SU(NM)\times U(1)_A\supset SU(N)_L\times SU(M)_R\times U(1)_A
\label{eq:o2nmbranching}
\end{align}

\vskip .2cm
\noindent
{\emph{Proof of proposition 2:}}$\quad$ $H$ is a complex $N\times M$ matrix field with $2NM$ real components. As $\mO$ is constructed with MCSD, it cannot contain derivatives and therefore if its CSD is $D$, it must live in the carrier space of a $D$-index fully symmetric tensor of $O(2NM)$.
On the other hand the real symmetry of the theory is $SU(N)_L\times SU(M)_R\times U(1)_A\subset SU(NM)\times U(1)_A\subset O(2NM)$,
therefore $(\Gamma_L,\Gamma_R)$ must appear in the decomposition of a $D$-index fully symmetric tensor of $O(2NM)$ under the branching
in Eq.~\eqref{eq:o2nmbranching}. In fact the $D$-index fully symmetric tensor must be traceless, because we are considering operators constructed with MCSD. If the tensor contains a trace part, then it would be possible to factor out the trace and build a new operator with the same symmetry properties but with a smaller CSD, in contradiction to the MCSD assumption.
\subsubsection{The correspondence between weight and charge configuration}
In the above propositions, no explicit reference is made yet about the charge configuration matrix $\mQ$ which we will consider now.  Let us start with the explicit form of $\mQ$  and determine the precise correspondence between the charge configuration matrix and the weight of an irreducible representation. The matrix $\mQ$ belongs to the
special linear algebra $\text{sl}(N;\mathbb{C})$, which is the space of all $N\times N$ complex matrices $X$ for which $\Tr X=0$. $\text{sl}(N;\mathbb{C})$ is exactly the complexification of the real Lie algebra of the $SU(N)$ group ~\cite{Hall:2015tb}. The Cartan subalgebra $\fh$ of
$\text{sl}(N;\mathbb{C})$ can be characterized by
\begin{align}
\mathfrak{h}=\left\{ \left. \left( \begin{matrix}
   {{\lambda }_{1}} & {} & {}  \\
   {} & ... & {}  \\
   {} & {} & {{\lambda }_{N}}  \\
\end{matrix} \right) \right|{{\lambda }_{j}}\in \mathbb{C},\text{ }{{\lambda }_{1}}+...+{{\lambda }_{N}}=0 \right\}
\end{align}
A weight is a linear functional on $\fh$. Nevertheless, it is convenient to identify linear functionals on $\fh$ with elements of $\fh$
itself, by virtue of an inner product on $\fh$. Suppose $K$ and $K'$ are two elements of $\fh$, we define their inner product by
\begin{align}
\langle K,K'\rangle=\Tr(K^*K')
\label{eq:ip}
\end{align}
If $\phi$ is a linear functional on $\fh$, there is a unique element $\lambda$ in $\fh$ such that
\begin{align}
\phi(K)=\langle\lambda,K\rangle
\label{eq:lf}
\end{align}
for all
$K\in\fh$. Therefore, a weight $\mu$ can be thought of as an element in $\fh$, by virtue of the inner product defined in Eq.~\eqref{eq:ip}.

The charge configuration matrix $\mQ$ should be proportional to some weight $\mu$ of a representation of $\slnc$. Let us now determine
the precise correspondence, assuming $\mQ=\mQ_L$ is normalized as in Eq.~\eqref{eq:qdef}. Suppose $\mQ$ can be decomposed as
\begin{align}
\mathcal{Q}=\sum\limits_{j=1}^{N-1}{{{x}_{j}}{{{\hat{h}}}_{j}}}
\label{eq:qdecom}
\end{align}
with $\hat{h}_j$ being a set of \emph{ortho-normal} basis elements of $\fh$, with the orthogonality defined by virtue of the inner product
in Eq.~\eqref{eq:ip}, and the normalization condition being
\begin{align}
\Tr(\hat{h}_{j}^{2})=\frac{1}{2},\text{  }j=1,2,...,N-1
\label{eq:hnorm}
\end{align}
For example, the following choice of one element is normalized
\begin{align}
{{\hat{h}}_{1}}=\frac{1}{2}({{E}_{11}}-{{E}_{22}})
\end{align}
where $E_{ij}$ denotes a $N\times N$ matrix with a "$1$" in the $(i,j)$ entry and "$0$" elsewhere. The normalization of basis elements is required as in Eq.~\eqref{eq:hnorm} because, for example, one can compute the commutation relation
\begin{align}
[{{\hat{h}}_{1}},{{E}_{12}}]={{E}_{12}}
\label{eq:excom}
\end{align}
which implies a raising operator constructed with a single $E_{12}$ will carry charge $+1$ corresponding to $\hat{h}_1$. One may wish to make this argument more precisely by rewriting Eq.~\eqref{eq:excom} as a commutation relation between the corresponding charge operator and the corresponding fixed-charge operator, computed with the help of canonical commutation relations of fundamental fields.

Then $x_j$ in Eq.~\eqref{eq:qdecom} gives the Cartan charge associated with $\hat{h}_j$, and can be computed as
\begin{align}
{{x}_{j}}=2\Tr(\mathcal{Q}{{\hat{h}}_{j}})
\label{eq:cc1}
\end{align}
On the other hand, for the weight $\mu$, the Cartan charge associated with $\hat{h}_j$ is given by
\begin{align}
\mu(\hat{h}_j)=\langle\mu,\hat{h}_j\rangle=\Tr(\mu^*\hat{h}_j)
\end{align}
Here we use the same symbol $\mu$ for the weight as a linear functional and as an element in $\fh$. We will only be concerned with
the case of real $\mu$ and therefore the Cartan charge reads
\begin{align}
\mu(\hat{h}_j)=\Tr(\mu\hat{h}_j)
\label{eq:cc2}
\end{align}
Comparing Eq.~\eqref{eq:cc1} and Eq.~\eqref{eq:cc2} we conclude the correspondence between $\mQ$ and $\mu$ is
\begin{align}
\mQ=\frac{1}{2}\mu
\label{eq:qwrelation}
\end{align}
This leads to the following proposition.

\vskip .3cm
\noindent
\textbf{Proposition 3:}$\quad$ Suppose $\mO$ is a fixed-charge operator that corresponds to a traceless charge configuration $\mQ$ with MCSD, with the CSD of $\mO$ being $D$. Let us also suppose $\mO$ belongs to some irreducible representation $(\Gamma_L,\Gamma_R)$ of $SU(N)_L\times SU(M)_R$ in the $U(1)_A$-neutral sector. Then $2\mQ$ must be a weight of $\Gamma_L$, and $-2\mQ$ must be a weight of $\Gamma_R$.

Since we know that the weights of Lie algebra representations sit on a discrete weight lattice, we then deduce from Eq.~\eqref{eq:qwrelation} that the charge configuration $\mQ$ is also quantized.

Because we want to consider fixed-charge operators corresponding to a traceless charge configuration $\mQ$ with MCSD, according to Proposition 1, the weights of our interest should belong to ${{(\mathbf{Ad}{{\mathbf{j}}_{L}})}^{D/2}}$, when we consider the $SU(N)_L$ factor. The nonzero weights of $\mathbf{Ad}{{\mathbf{j}}_{L}}$ are  nonzero roots of $\slnc$, which are given by ~\cite{Hall:2015tb}
\begin{align}
{{\alpha }_{jk}}={{e}_{j}}-{{e}_{k}},\text{  }j\ne k,\text{  }j,k=1,2,...,N
\label{eq:slncweight}
\end{align}
where $e_j$'s denote the standard basis elements of ${{\mathbb{C}}^{N}}$, that is
\begin{align}
{{e}_{j}}=\{\underbrace{0,...,0}_{j-1},1,\underbrace{0,...,0}_{N-j}\}
\end{align}
for $j=1,2,...,N$. Note that in this representation of roots we have identified $\fh$ with the subspace of ${{\mathbb{C}}^{N}}$ consisting of vectors whose components sum to zero ~\cite{Hall:2015tb}. Because all the weights of a tensor product representation are given by the sum of weights of
the component representations~\cite{Cornwell:1985xt}, we conclude that if $\mu$ is a weight of ${{(\mathbf{Ad}{{\mathbf{j}}_{L}})}^{D/2}}$, then $\mu$ must be able to be expressed as
\begin{align}
\mu =\sum\limits_{p=1}^{D/2}{{{s}_{p}}{{\alpha }_{{{j}_{p}}{{k}_{p}}}}},\text{  with }{{s}_{p}}=1\text{ or }0
\label{eq:mudecomp}
\end{align}
and for $p=1,2,...,D/2$, ${{\alpha }_{{{j}_{p}}{{k}_{p}}}}={{e}_{{{j}_{p}}}}-{{e}_{{{k}_{p}}}}$ with $j_p\neq k_p$ is one of the weights in
Eq.~\eqref{eq:slncweight}. From Eq.~\eqref{eq:mudecomp} we deduce that if we write $\mu =({{\mu }_{1}},{{\mu }_{2}},...,{{\mu }_{N}})$ , then ${{\mu }_{i}}\in \mathbb{Z},\text{ }\forall i=1,2,...,N$, and then Eq.~\eqref{eq:qwrelation} tells us the diagonal entries of $\mQ$  must be integers or half-integers.

Now for any vector $\nu \equiv ({{\nu }_{1}},{{\nu }_{2}},...,{{\nu }_{N}})\in {{\mathbb{C}}^{N}}$, define the A-length of $A[\nu]$ of
$\nu$ as
\begin{align}
A[\nu ]\equiv \sum\limits_{i=1}^{N}{|{{\nu }_{i}}|}
\end{align}
Suppose $\nu ,\rho \in {{\mathbb{C}}^{N}}$, then the following triangle inequality holds
\begin{align}
A[\nu +\rho ]\le A[\nu ]+A[\rho ]
\end{align}
This can be proved easily: Suppose $\nu \equiv ({{\nu }_{1}},{{\nu }_{2}},...,{{\nu }_{N}}),\rho =({{\rho }_{1}},{{\rho }_{2}},...,{{\rho }_{N}})$, then
\begin{align}
A[\nu +\rho ]=\sum\limits_{i=1}^{N}{|{{\nu }_{i}}+{{\rho }_{i}}|}\le \sum\limits_{i=1}^{N}{|{{\nu }_{i}}|}+\sum\limits_{i=1}^{N}{|{{\rho }_{i}}|}=A[\nu ]+A[\rho ]
\end{align}
It is also obvious that the A-length has a linearity property with respect to multiplication by a c-number
\begin{align}
A[c\nu ]=|c|A[\nu ],\text{  }\forall c\in \mathbb{C}
\end{align}
Then by using the triangle inequality and linearity property of the A-length, from Eq.~\eqref{eq:mudecomp} we can deduce
\begin{align}
A[\mu ]\le \sum\limits_{p=1}^{D/2}{A[{{s}_{p}}{{\alpha }_{{{j}_{p}}{{k}_{p}}}}]}=\sum\limits_{p=1}^{D/2}{|{{s}_{p}}|A[{{\alpha }_{{{j}_{p}}{{k}_{p}}}}]}
\end{align}
Now let us note that
\begin{align}
|{{s}_{p}}|\le 1,\text{  }A[{{\alpha }_{{{j}_{p}}{{k}_{p}}}}]=2
\end{align}
and thus
\begin{align}
A[\mu ]\le 2\times D/2=D
\label{eq:e34}
\end{align}
On the other hand, we see the charge configuration $\mQ$ corresponding to $\mu$ satisfies $\mQ=\frac{\mu}{2}$. To make a comparison we should also map $\mQ$ into ${{\mathbb{C}}^{N}}$ in the obvious manner, i.e.
\begin{align}
\mathcal{Q}=\text{diag}\{{{Q}_{1}},{{Q}_{2}},...,{{Q}_{N}}\}\to ({{Q}_{1}},{{Q}_{2}},...,{{Q}_{N}})\in {{\mathbb{C}}^{N}}
\end{align}
Then we can write
\begin{align}
A[\mathcal{Q}]=\frac{1}{2}A[\mu ]
\label{eq:e36}
\end{align}
Combining Eq.~\eqref{eq:e34} and Eq.~\eqref{eq:e36} we see immediately that
\begin{align}
D\ge 2A[\mathcal{Q}]
\label{eq:e37}
\end{align}
This leads to the following proposition.

\vskip .3cm
\noindent
\textbf{Proposition 4}$\quad\quad$ Suppose $\mO$ is a fixed-charge operator that corresponds to a traceless charge configuration $\mQ=\text{diag}\{{{Q}_{1}},{{Q}_{2}},...,{{Q}_{N}}\}$. Let us denote the CSD of $\mO$ by $D$. Then $D$ satisfies the inequality
\begin{align}
D\ge 2\sum_{i=1}^{i=N}|Q_i|
\label{eq:pp3}
\end{align}

\subsubsection{Scaling dimension and operator construction}
All the conclusions achieved up to now are deduced without the need of explicitly constructing the fixed-charge operators. On the other hand, one can show that the equality sign in Eq.~\eqref{eq:e37} and Eq.~\eqref{eq:pp3} can always be achieved by  constructing an  operator corresponding to a given charge configuration. To this end, we first consider building blocks that have simple
definite transformation properties under $SU(N)_L\times SU(M)_R\times U(1)_A$ and are $U(1)_A$-neutral. For example, we may consider
\begin{align}
\Tr(\tau_L H\tau_R^\dagger H^\dagger)
\label{eq:bblock}
\end{align}
where $\tau_L$ is an $N\times N$ matrix in some root subspace of the $\slnc$ Lie algebra, and $\tau_R$ is an $M\times M$ matrix related to
$\tau_L$ in the following manner
\begin{align}
{{\tau }_{R}}=\left( \begin{matrix}
   {{\tau }_{L}} & {{\mathbf{0}}_{N\times (M-N)}}  \\
   {{\mathbf{0}}_{(M-N)\times N}} & {{\mathbf{0}}_{(M-N)\times (M-N)}}  \\
\end{matrix} \right)
\end{align}
Obviously, the building block in Eq.~\eqref{eq:bblock} lives in the bi-adjoint representation of $SU(N)_L\times SU(M)_R$, i.e.
$(\adjl,\adjr)$. It is constructed in such a manner that $\mQ_L+\mQ_R=0$ is manifestly satisfied, with $\mQ_L$ corresponding to
a weight of $\adjl$. The explicit form of $\tau_L$ is given by
\begin{align}
\tau_L=E_{pq}
\end{align}
for some $p,q=1,2,...,N$ and $p\neq q$. This is because we have the commutation relation
\begin{align}
[\hat{h}_j,E_{pq}]=\frac{1}{2}(\delta_{jp}-\delta_{jq}-\delta_{j+1,p}+\delta_{j+1,q})E_{pq}
\label{eq:cca1}
\end{align}
Here $j=1,2,...,N-1$ and $\hat{h}_j$ is defined by
\begin{align}
\hat{h}_j\equiv\frac{1}{2}(E_{j,j}-E_{j+1,j+1})
\end{align}
which satisfy the normalization condition $\Tr(\hat{h}_j^2)=\frac{1}{2}$.

Let us first identify the charge configuration associated with Eq.~\eqref{eq:bblock}. Define a set of $N$ linear functionals
$\varepsilon_p,p=1,...,N$ acting on $\fh$ as follows
\begin{align}
\varepsilon_p(\hat{h}_j)=\frac{1}{2}(\delta_{jp}-\delta_{j+1,p})
\end{align}
Then Eq.~\eqref{eq:cca1} can be written as
\begin{align}
[\hat{h}_j,E_{pq}]=\{\varepsilon_p(\hat{h}_j)-\varepsilon_q(\hat{h}_j)\}E_{pq}
\end{align}
which means $E_{pq}$ corresponds to the root $\varepsilon_p-\varepsilon_q$, which when mapped into $\fh$ using the inner product
Eq.~\eqref{eq:ip} gives $\alpha_{pq}$ defined in Eq.~\eqref{eq:slncweight}~\footnote{This can be deduced from the results in Appendix G of the textbook by J. F. Cornwell ~\cite{Cornwell:1985xt}.}. This $\alpha_{pq}$ just corresponds to the weight of
$\adjl$ associated with Eq.~\eqref{eq:bblock} and the corresponding charge configuration is simply $\frac{1}{2}\alpha_{pq}$,
according to Eq.~\eqref{eq:qwrelation}.

To build operators in more general charge configurations, we may consider
\begin{align}
\Tr\Big[ \Pi_j (\tau_{Lj} H\tau_{Rj}^\dagger H^\dagger)^{y_j}\Big] \ .
\label{eq:ocon}
\end{align}
Here $y_j>0$ is a positive integer, and $\tau_{Lj}$ is an $N\times N$ matrix with the explicit form given by
$\tau_{Lj}=E_{p(j)q(j)}$ for some $p,q=1,2,...,N$ that depend on $j$. The way that Eq.~\eqref{eq:ocon} is constructed
implies that its charge configuration $\mQ$ is just the appropriate linear combination of the charge configuration $\mQ_j$ of its corresponding building blocks
\begin{align}
\mathcal{Q}=\sum_{j}y_j \mQ_j
\label{eq:qdec1}
\end{align}
where
\begin{align}
\mQ_j=\frac{1}{2}\alpha_{p(j)q(j)}
\end{align}
We can now reverse the logic and ask for a given $\mQ$ how one may choose $\tau_{Lj}$ and $y_j$ in order to construct a MCSD operator
in the form of Eq.~\eqref{eq:ocon}. To this end, we may rewrite Eq.~\eqref{eq:qdec1} as
\begin{align}
2\mathcal{Q}=\sum_{j}y_j \alpha_{p(j)q(j)}
\label{eq:qdec2}
\end{align}
Note $2\mathcal{Q}\in{\mathbb{C}}^{N}$, with all entries being integers and the sum of all entries is zero. We also have
$\alpha_{p(j)q(j)}\in{\mathbb{C}}^{N}$, which for a given $j$ there exists only two nonzero entries, filled by $+1$ and $-1$
respectively. MCSD requires the minimization of $\sum_{j}y_j$ for a given $\mQ$. We can rewrite Eq.~\eqref{eq:qdec2} as
\begin{align}
2\mathcal{Q}-\sum_{j}y_j \alpha_{p(j)q(j)}=0
\label{eq:qdec3}
\end{align}
where now the left-hand side indicates a process in which we subtract $\alpha_{p(j)q(j)}$'s from the given ${\mathbb{C}}^{N}$ vector $2\mQ$.
Suppose each time we are only allowed to subtract one $\alpha_{p(j)q(j)}$, which we call an elementary subtraction. (For a given $j$ we therefore eventually subtract it $y_j$ times). The sum $\sum_{j}y_j$ therefore equals the total number of times we need to perform such elementary subtractions to
make the resulting ${\mathbb{C}}^{N}$ vector vanish. To minimize $\sum_{j}y_j$ it is then obvious that during the subtraction process each entry
of the ${\mathbb{C}}^{N}$ vector should change in a \emph{monotonic} manner (or remain unchanged for some steps). As a concrete example, suppose $2\mQ=(2,-1,-1)$.
The following subtraction process is monotonic
\begin{align}
(2,-1,-1)\rightarrow (1,0,-1)\rightarrow (0,0,0)
\end{align}
while the following subtraction is not monotonic
\begin{align}
(2,-1,-1)\rightarrow (1,0,-1)\rightarrow (1,-1,0)\rightarrow (0,0,0)
\end{align}
It can be seen manifestly in this simple example that non-monotonic subtraction leads to an increase of the total number of times we need to subtract the vector
to zero, and this obviously generalize to general cases. Monotonic subtraction can always be realized, by subtracting from the positive entry with the largest absolute value and negative entry with the largest absolute value each time. In such a case, the total number of times we need to perform elementary subtractions
simply equals $A[2\mQ]/2=A[\mQ]$, that is
\begin{align}
A[\mQ]=\sum_{j}y_j
\end{align}
On the other hand, from it is obvious that the CSD $D$ of the operator in Eq.~\eqref{eq:ocon} is
\begin{align}
D=2\sum_{j}y_j
\end{align}
Therefore we conclude the MCSD can be achieved, with the relation
\begin{align} \label{MCSD}
D=2A[\mQ]
\end{align}
which is compatible with our previous finding Eq.~\eqref{eq:pp3} without explicit construction of the fixed-charge operator. Therefore we are led to the following proposition

\vskip .3cm
\noindent
\textbf{Proposition 5}$\quad\quad$ The equality sign in Eq.~\eqref{eq:pp3} can always be achieved.

\vskip .5cm

We emphasize that the method does not guarantee the unicity of the MCSD operator. In fact, one may choose to redistribute the trace operation (i.e. splitting one single trace to multiple traces),
change the order of matrix products for different $\tau_{Lj} H\tau_{Rj}^\dagger H^\dagger$ factors, or change the root basis, to obtain more operators associated with the same charge configuration. Even if we impose the MCSD requirement there can be multiple solutions. Algebraically they may lead to different or identical results. It is also not known whether the above method based on the $\tau_{Lj} H\tau_{Rj}^\dagger H^\dagger$ building blocks with appropriate application of the trace operation covers all fixed-charge MCSD operators. Nevertheless, for a special type of charge configuration matrix, there is a unique answer and we know the above way of explicit construction must lead to the unique answer. This charge configuration is
\begin{align} \label{special}
\mathcal{Q}_{L,J}=\rm{diag}\left\{-J,J,0,\cdots,0\right\}
\end{align}
with $J$ being an integer or half-integer. This charge configuration corresponds to the highest weight in the tensor product of
$\adjl$, which is in turn the sum of the highest weight of $\adjl$. The uniqueness results from the fact that the highest weight of
a representation is always simple. The irreducible representation associated with such a highest weight then has the Dynkin label
$(2J,0,...,0,2J)$.

The above five propositions we proved pave the way for a general identification of irreducible representations for a given charge configuration prescribed in a fixed-charge semiclassical computation. With the MCSD assumption the MCSD can be determined by virtue of Proposition 4 and 5 from the given charge configuration $\mQ$. Then the candidate irreducible representations must satisfy the requirements of Proposition 1-3.

\section{Semiclassics and anomalous dimensions in the $U(N) \times U(M)$ model} \label{computations}

In this section, we start the exploration of $U(N) \times U(M)$ model in $4-\epsilon$
dimensions with the fixed-charge semiclassical method. The necessary group-theoretic results (especially the 5 propositions in section \ref{themap}) will be used, and we refer the readers who are
interested in the detailed proofs to the previous section. In Euclidean spacetime, the Lagrangian of the theory
reads
\begin{equation}
\label{model}
  \begin{aligned}
    \mathcal{L} =
   \Tr(\del_\mu H^\dagger \del^\mu H ) + u_0\Tr(H^\dagger H)^2 + v_0(\Tr H^\dagger H )^2 \,,
  \end{aligned}
\end{equation}
where $H$ is a $N \times M$ complex matrix. For $N = M$ and $v_0 > 0$, it describes the finite-temperature phase transition in massless quantum chromodynamics \cite{Butti:2003nu} with $H$ the order parameter.
We work in the $\overline{MS}$ scheme. The couplings are renormalized as
\begin{equation} \label{rin}
u_0 M^{-\epsilon} = u + \sum_{n=0}^{\infty} \frac{a_u^{(n)}(u,v)} {\epsilon^n} \,,  \qquad v_0 M^{-\epsilon} = v + \sum_{n=0}^{\infty} \frac{a_v^{(n)}(u,v)}{\epsilon^n} \,,
\end{equation}
The beta functions of the couplings are given by
\begin{equation}
    \beta_u\equiv\frac{d u}{d \log M}|_{\epsilon=0}=-\epsilon u +u\frac{\partial a_u^{(1)}}{\partial u}+v\frac{\partial a_u^{(1)}}{\partial v}-a_u^{(1)} \,, \qquad \beta_v\equiv\frac{d v}{d \log M}|_{\epsilon=0}=-\epsilon v +u\frac{\partial a_v^{(1)}}{\partial u}+v\frac{\partial a_v^{(1)}}{\partial v}-a_v^{(1)} \ .
\end{equation}
and, at $1$-loop, read \cite{Pisarski:1980ix}
\begin{eqnarray}
\beta_{u} (u, v)&=&- \epsilon u + \frac{1}{4 \pi^2}\left( 6 u v + (N+ M) u^2 \right) \,, \\
\beta_{v} (u, v)&=& -\epsilon v +\frac{1}{4 \pi^2}\left(  (N M + 4) v^2 + 2 (N + M) u v + 3 u^2 \right)  \,.
\end{eqnarray}
At the $1$-loop level there are always a Gaussian FP ($u^* = v^* = 0$) and an $O(2 N M )$ one  ($u^* = 0$). Furthermore there are other two  FPs given by
\begin{equation}
   u^*_{\pm} = 4 \pi^2 \frac{A_{M N} \mp 3 \sqrt{R_{M N}}}{ D_{M N}} \epsilon \,, \qquad \qquad v^*_{\pm} = 4 \pi^2 \frac{B_{M N} \pm (M +N) \sqrt{R_{M N}}}{2 D_{M N}} \epsilon \,,
\end{equation}
where
\begin{eqnarray}
&& A_{MN} = N M^2 + M N^2 - 5 N - 5M \,,\qquad \qquad B_{M N} = 36 - (M + N)^2 \,, \nonumber \\
&& R_{MN} = 24 + M^2 + N^2 - 10 M N \,, \qquad \qquad D_{M N} =( M N - 8) (M + N)^2 + 108 \,.
\end{eqnarray}
 The beta functions to five loops have been derived in \cite{Calabrese:2004uk}, where the authors concluded that no stable FP exists for $N= M$ and $d = 3$, suggesting that the chiral phase transition in light QCD at finite temperature is first-order. When $R_{M N} < 0$ the fixed points are complex, and we expect regions of the parameters space in which the theory features near-conformal dynamics of the \emph{walking} type \cite{Kaplan:2009kr, Gorbenko:2018ncu}.

Since the $u$ coupling breaks $O(2 N M)$ symmetry to $SU(M)\times SU(N)$ subgroup, it is convenient to think about representations of this model as a decomposition of the $O(2 N M)$ multiplets with defining (vector) and the $2$-index traceless symmetric representations of $O(2 N M)$ as
\begin{eqnarray}
{\small \Yvcentermath1  \yng(1)}_{\ O(2 N M)}&=& {\bf 2 N M} = [N, \bar{M}] \oplus [\bar{N}, M] \,.  \\
 {\small \Yvcentermath1  \yng(2)}_{\ O(2 N M)} &=& (1,\text{Adj})  \oplus(\text{Adj},1)\oplus(\text{Adj},\text{Adj})   \oplus  \big[(\ {\tiny \Yvcentermath1  \yng(2)}\ ,{\tiny \Yvcentermath1  \yng(2)}\ ^* \ ) \oplus (\ {\tiny \Yvcentermath1  \yng(1,1)}\ ,{\tiny \Yvcentermath1  \yng(1,1)}\ ^* \ ) \oplus c.c. \big] \ .  \nonumber \\ {\bf 2 N^2 M^2+ N M-1} &=& {\bf N^2-1 \oplus  M^2-1 \oplus \left(N^2-1 \right)\left(M^2-1 \right) \oplus  2\left(\frac{N (N+1)}{2}\right)\left(\frac{M (M+1)}{2}\right)  \oplus 2 \left(\frac{N (N-1)}{2}\right)\left(\frac{M (M-1)}{2}\right) }  \,, \nonumber \\
 \end{eqnarray}
where in the last line we explicitly show the dimension of the representations appearing in the decomposition.

\subsection{Charging the system}

In this section, we analyze the symmetry breaking pattern induced by charge fixing and set up the semiclassical computation.
After a Weyl map to the cylinder, our starting points are the cylinder action \eqref{Scyl} and the spatially homogeneous ground state ansatz given by Eq.\eqref{eq:ansatz}. The Noether charges ${{\mathcal{Q}}_{L}}$ and ${{\mathcal{Q}}_{R}}$ associated with the $U(N) \times U(M)$ global symmetry are given by Eq.\eqref{eq:qdef} and satisfy the constraint  \eqref{constraint} $\mQ_L+\mQ_R=0$. The Euler-Lagrange equations read ($m^2=\left(\frac{d-2}{2R}\right)^2$)
 \begin{equation} \label{EL}
     \partial_0^2 H +\nabla^2 H +2 u_0 H \left( H^{\dagger}H\right) +2 v_0 \Tr \left( H^{\dagger}H\right) H +m^2 H= 0 \,.
 \end{equation}
and for our homogeneous ansatz Eq.~\eqref{eq:ansatz}, they reduce to:
\begin{equation}
   2 M_E^2 B = -u_0 B^{\dagger} B^2 -v_0 \Tr \left( B^{\dagger} B \right) B - \frac{m^2}{2} B \,.
\end{equation}

We label the entries on the diagonal of $M_{E,ii}$ with $M_{E,ii} = -i\mu_i$.  In this subsection $\mu_i$ is a chemical potential and should not be mistaken with the group theoretical weight matrix $\mu$ used elsewhere in the paper. For $B_{ii}$ we have $B_{ii} = b_i$, we can now rewrite the EOM as
\begin{equation}
    2 \mu_i^2  = u_0 b_i^2 +v_0 \sum_{k=1}^N b_k^2 +\frac{m^2}{2} \,,
\end{equation}
 while the corresponding "charges" read
 \begin{equation}
    J_i \equiv (\mathcal{Q}_L )_{ii} = -2 V b_i^2 \mu_i \,.
\end{equation}

The classical energy $E$ is given by evaluating the cylinder Lagrangian $\mathcal{L}_{\text{cyl}}$ in \eqref{Scyl} with an appropriate boundary term $ -\sum_{i=1}^N \mu_i \frac{\partial \mathcal{L}_{\text{cyl}}}{\partial \mu_i}$, which implements the charge fixing. We obtain
\begin{equation} \label{classE}
 \frac{E}{V} =\mathcal{L}_{\text{cyl}}-\sum_{i=1}^N \mu_i \frac{\partial \mathcal{L}_{\text{cyl}}}{\partial \mu_i}   = 4 \sum_{i=1}^N b_i^2 \mu_i^2 +u_0 \sum_{i=1}^N b_i^4 +v_0 \left(\sum_{i=1}^N b_i^2 \right)^2 + m^2 \sum_{i=1}^N b_i^2 \,.
\end{equation}
We proceed by considering a $2$-parameters family of charge configurations
\begin{equation} \label{complete}
    \mathcal{Q}_{L,J,s} = \text{diag}\big(\underbrace{J,J, \dots}_s , \underbrace{-J, -J, \dots}_s , \underbrace{0, 0, \dots}_{N-2s} \big) \,.
\end{equation}
For $N=M$ and varying $s$, this charge configuration interpolates between the ones considered in \cite{Antipin:2020rdw} ($s=1$; given in Eq.\eqref{special}) and \cite{Orlando:2019hte} ($s=N/2$).
As we will see later, at fixed CSD we can access the anomalous dimension of operators transforming in various irreducible representations by varying the parameters $s$ and $J$. This is the first time that the fixed-charge semiclassical methods are used to access the scaling dimension of operators with the same CSD and different irreducible representation by varying the charge configuration.


The classical energy for this charge configuration can be easily computed along the lines of Sec.\ref{SOA}. By parameterizing the $M_E$ and $B$ matrices as follow
\begin{equation}
     \mu_i =  \begin{cases}
 \mu  & i=1,\dots,s \,,\\
-\mu & i=s+1,\dots, 2s \,,\\
 0 & i=2s+1,\dots, N \,,
\end{cases} \qquad  \qquad    b_i =  \begin{cases}
 b  & i=1,\dots,2s \,,\\
 0 & i=2s+1,\dots, N \ ,
\end{cases}
\end{equation}
the charge condition and EOM become
\begin{equation}
    J = 2 V \mu b^2  \,, \qquad \qquad 2 \mu^2 =(u_0 + 2 s v_0)b^2 + \frac{m^2}{2} \ .
\end{equation}

From Eq.\eqref{MCSD} we have that the CSD in four dimensions is $\bar Q = 4 s J$. Since $J \ge 1/2$, this implies that the results obtained in \cite{Orlando:2019hte} for the case $M=N$, $s =N/2$ make sense only when $\bar Q \ge N$ and not for arbitrary values of $\bar Q$ and $N$.

It is useful to define rescaled (renormalized) 't Hooft couplings as
\begin{align}\label{eq:couplings}
\cA_h = J \frac{u N}{(4\pi)^2} \ , \qquad \cA_v = J \frac{ 2 s v N}{(4\pi)^2} \ ,
\end{align}
Then the above equations imply
\begin{equation} \label{xdef}
    2 \frac{\mu}{m} =\frac{3^\frac{1}{3}+x^\frac{2}{3}}{3^\frac{2}{3}x^\frac{1}{3}} \,, \qquad \qquad x=  \frac{72}{N}(\mathcal{A}_h + \mathcal{A}_v)  + \sqrt{-3+ \left(\frac{72}{N}(\mathcal{A}_h + \mathcal{A}_v) \right)^2 } \,,
\end{equation}
and our semiclassical expansion takes the form
\begin{equation} \label{forma}
    \Delta_{\cO_{J,s}}  =  \sum_{k=-1} \frac{1}{J^k} \Delta_k (\mathcal{A}_h^*, \mathcal{A}_v^*) \,.
\end{equation}
The leading order in the semiclassical expansion follows straightforwardly from the results above by setting $d=4$ and $\cA_{h,v} = \cA_{h,v}^*$, where the star denotes the value of the couplings at the FP. We have
\begin{equation}
    \Delta_{-1} (\mathcal{A}_h^*, \mathcal{A}_v^*) = \frac{s N}{144(\cA_h^* + \cA_V^* )} \frac{1}{x^{*\frac{4}{3}}}\left(\sqrt[3]{3} x^{*8/3}-3 x^{*4/3}+6 \sqrt[3]{3} x^{*2/3}+2\ 3^{2/3}
   x^{*2}+ 3^{5/3} \right) \,.
\end{equation}
The expansion for small $\cA_{h,v}^*$ reads
\begin{equation}
    J \Delta_{-1} (\mathcal{A}_h^*, \mathcal{A}_v^*) = \bar Q \left[1+ 4 \left( \frac{ \cA_h^* +\mathcal{A}_v^*}{N}\right) - 32 \left( \frac{ \cA_h^* +\mathcal{A}_v^*}{N}\right)^2+ 512 \left( \frac{ \cA_h^* +\mathcal{A}_v^*}{N}\right)^3 + \cO\left( \frac{ \cA_h^* +\mathcal{A}_v^*}{ N}\right)^4  \right] \,.
\end{equation}
Notice that the leading order depends neither on $M$ nor $N$ when rewritten in terms of the original couplings $u^*$ and $v^*$. This  is because at the classical level only the fields which take a non-zero vev contribute and whose number depends on $s$.

Before proceeding with the computation of $\Delta_0$, it is useful to study the induced symmetry breaking pattern.
The explicit breaking can in general be deduced by adding the charge-fixing boundary term to the Lagrangian and check which symmetries it preserves.
We obtain
\begin{equation}
    SU(N)_L \otimes SU(M)_R \otimes U(1)_A \underset{explicit}{\Longrightarrow} C(R)_L \otimes SU(M)_R \otimes U(1)_A \,,
\end{equation}
 where $ C(R)_L$ is the $ SU(N)_L$ subgroup that commutes with $
   P_1 = \text{diag}\big(\underbrace{1,1, \dots}_s , \underbrace{-1, -1, \dots}_s , \underbrace{0, 0, \dots}_{N-2s} \big)$ and it is explictily given by
\begin{equation}
    C(R)_L = SU(s)_{Lu} \otimes SU(s)_{Ld} \otimes SU(N-2s)_{Ld} \otimes U(1)_{L3} \otimes U(1)_{L5} \,,
\end{equation}
where $SU(s)_{Lu}$ and $SU(s)_{Ld}$ are rotations in the first and second upper $s \times s$ blocks of $SU(N)_{L}$ while $SU(N-2s)_{Ld}$ rotates the lower  $N-2s \times N-2s$ block. Finally $U(1)_{L3}$ and $U(1)_{L5}$ are generated, respectively, by $P_1$ and $P_2 = \text{diag}\big(\underbrace{1, 1, \dots}_{2s} , \underbrace{0,  0, \dots}_{N-2s} \big)$ and act on the left factor.

The spontaneous symmetry breaking is  determined by the vacuum configuration, which is proportional to the $P_2$ matrix defined above. We have
\begin{eqnarray}
  C(R)_L \otimes SU(M)_R \otimes U(1)_A \underset{SSB}{\Longrightarrow}
    SU(s)_{Lu} \otimes SU(s)_{Ld} \otimes SU(N-2s)_{Ld} \otimes U(1)_{D3} \otimes U(1)_{D5} \otimes SU(M-2s)_{Rd} \otimes U(1)_{A6} \,.
    \end{eqnarray}
Here, $U(1)_{D3,5}$ are the diagonal subgroup of  $U(1)_{L3,5} \otimes  U(1)_{R3,5}$ where $U(1)_{R3,5}$ are the counterparts of  $U(1)_{L3,5}$ acting on the right factor. Finally, $ SU(M-2s)_{Rd}$ is defined as the  $SU(M-2s)$ rotation in the lower $M-2s \times M-2s$ block of $SU(M)_R$. To define $U(1)_{A6}$, we first introduce
$U(1)_{L6}$ which is generated by $\text{diag}\big(\underbrace{0, 0, \dots}_{2s} , \underbrace{1,  1, \dots}_{N-2s} \big)$ and acts on the left, and $U(1)_{R6}$
which is generated by $\text{diag}\big(\underbrace{0, 0, \dots}_{2s} , \underbrace{1,  1, \dots}_{M-2s} \big)$ and acts on the right. $U(1)_{A6}$ is then defined as
the axial part of $U(1)_{L6}$ and $U(1)_{R6}$. Note the diagonal part of $U(1)_{L6}$ and $U(1)_{R6}$ is not independent from $U(1)_{D5}$ and is thus not counted.

Altogether, the number of broken generators is
\begin{equation} \label{broken}
    M^2 - 1 - \left[(M-2s)^2 -1 \right] = 4 s (M -s) \,.
\end{equation}

We parametrize the fluctuations as
\begin{equation}
H\left(\tau, \bf{x}\right)=e^{2iM_E \tau}(\Bbar + \Phi \left(\tau, \bf{x}\right) ) \,,
\label{flutt}
\end{equation}
where $\Phi\left(\tau, \bf{x}\right)$ is a $N \times M$ matrix. We have
\begin{align}
\mathcal{L}_{quad} = & \sum\limits_{i=1}^N \sum\limits_{j=1}^M \partial_{\mu} \Phi_{ij}  \partial^{\mu} \Phi_{ij}^* -2 \mu \left[ \sum\limits_{i=1}^s \sum\limits_{j=1}^M \left((\partial_{0} \Phi_{ij}) \Phi_{ij}^* - \Phi_{ij} \partial_{0} \Phi_{ij}^*\right) -\sum\limits_{i = s+1}^{2s} \sum\limits_{j=1}^M \left( (\partial_{0} \Phi_{ij}) \Phi_{ij}^* - \Phi_{ij}  \partial_{0} \Phi_{ij}^* \right) \right] \nonumber \\
& +2 u_0 b^2 \sum\limits_{i=1}^N \sum\limits_{j=1}^{2s} \Phi_{ij}^* \Phi_{ij} + u_0 b^2   \sum\limits_{i=1}^{2s} \sum\limits_{j=1}^{2s} \left(\Phi_{ij} \Phi_{ji}  + \Phi_{ij}^* \Phi_{ji}^*  \right) + v_0 b^2 \left [ \sum\limits_{i=1}^{2s} (\Phi_{ii}+ \Phi_{ii}^*  ) \right]^2 \nonumber \\
& + \left(4 s v_0 b^2 + m^2 \right) \sum\limits_{i = 2s+1}^{N} \sum\limits_{j=1}^M \Phi_{ij} \Phi_{ij}^* \,.
\end{align}
It is useful to write $\Phi$ in block form as
\begin{equation}
    \Phi = \left(\begin{array}{cc} \Phi_{2s \times 2s}^{(11)}  &\Phi_{2s \times (M- 2s)}^{(12)}  \\  \\
       \Phi_{(N - 2s) \times 2s}^{(21)}  &\Phi_{(N- 2s) \times( M- 2s)}^{(22)}  \\
    \end{array} \right) \,.
\end{equation}
The blocks decouple and we can decompose $\mathcal{L}_{quad}$ as  $\mathcal{L}_{quad} =\mathcal{L}_{quad}^{(11)}+\mathcal{L}_{quad}^{(12)}+\mathcal{L}_{quad}^{(21)}+\mathcal{L}_{quad}^{(22)}$.
The dispersion relations of the fluctuations read
\begin{equation}
\label{disp}
\begin{split}
\omega_1=&\sqrt{J^2_\ell+4\mu^2}\qquad 4s (N -2 s)~\rm{d.o.f.}\\
\omega_2=&\sqrt{J^2_\ell+ m_{2}^2}\qquad 2\left(N-2s \right)(M-2s)\,\rm{d.o.f}\\
\omega_{3,4}=&\sqrt{J^2_\ell+4\mu^2}\mp 2\mu \qquad 2s (2M - 3s)~\rm{d.o.f.}\\
\omega_{5,6}=&\sqrt{J^2_\ell+4\mu^2+m_1^2}\pm2\mu\qquad 2s^2~\rm{d.o.f.}\\
\omega_{7,8}=& \frac{1}{\sqrt{2}}\sqrt{2 J^2_\ell + m_1^2 +16 \mu^2 \pm \sqrt{\left(2 J^2_\ell + m_1^2 +16 \mu^2 \right)^2-4 J^2_\ell \left(J^2_\ell+m_1^2 \right) }} \qquad 4s^2 -2~\rm{d.o.f.}\\
\omega_{9,10}=& \frac{1}{\sqrt{2}}\sqrt{2 J^2_\ell + m_0^2 +16 \mu^2 \pm \sqrt{\left(2 J^2_\ell + m_0^2 +16 \mu^2 \right)^2-4 J^2_\ell \left(J^2_\ell+m_0^2 \right) }} \qquad 2~\rm{d.o.f.} \,,
\end{split}
\end{equation}
where
\begin{eqnarray}
  m_0^2 = 8 \mu^2 -2 m^2 \,,  \qquad \quad  m_1^2 = \left( 8 \mu^2 -2 m^2 \right) \frac{u_0}{u_0 + 2s v_0} \,,  \qquad \quad m_2^2 = 4s v_0 b^2+m^2 \,.
\end{eqnarray}
$\omega_3$ describes Type II Goldstone bosons while $\omega_8$ and $\omega_{10}$ correspond to relativistic Type I Goldstone bosons. The remaining dispersion relations describe gapped modes. It's easy to check that the number of real d.o.f. sums to $2 N M$ while the counting of Goldstone modes with respect to the number of broken generators is
\begin{equation}
    2 \times s(2 M - 3 s) + 2 s^2 -1 +1 = 4 s (M-s) \,,
\end{equation}
which agrees with Eq.\eqref{broken}, saturating the Nielsen-Chadha bound.
The NLO in the semiclassical expansion is given by the general formula Eq.\eqref{eq:one-loop-det1}. Regularization and renormalization are performed as explained in Sec.\ref{SOA}, yielding
\begin{equation} \label{NLO}
    \Delta_{0}(\cA_h^*, \cA_v^*) = \rho (x^*,M, N, s, \cA_h^*,  \cA_v^*)  + \frac{1}{2} \sum_{\ell=0}^\infty \left[ R (1+ \ell)^2
\left(\sum_i g_i(M,N,s) \omega_i(\ell, x^*, \cA_h^*, \cA_v^* )\right)_{d=4} + \sigma(\ell, x^*, M, N, s,  \cA_h^*,  \cA_v^*) \right] \,.
\end{equation}
where $x^*$ has been defined in Eq.\eqref{xdef} while $\rho (x^*,M, N, s, \cA_h^*,  \cA_v^*) $ and $\sigma(\ell, x^*, M, N, s,  \cA_h^*,  \cA_v^*)$ are given in App.\ref{funzioni}.

We checked the cancellation of the divergent terms between $\Delta_{-1}$ and $\Delta_{0}$ as explained above Eq.\eqref{Delta0}. For $\cA_h = 0$ the result reduces to its counterpart in the $O(2 N M)$ model. Finally, for $N=M$ and $s=2$ we obtain the results in \cite{Antipin:2020rdw}.

The perturbative expansion for small t' Hooft couplings reads
\begin{eqnarray}
 \label{NLO22}
   \Delta_{0}(\cA_h^*, \cA_v^*)  = && -\frac{  4 }{N  \left(\cA_h^*+\cA_v^*\right)} \left[2 s \cA_h^{*2} (M+N+7 s)+\cA_h^* \cA_v^* (2 s (2 (M+N)+s)+7)+(M N+5)\cA_v^{*2}\right] -\frac{16 }{N^2}  \left[2 s \cA_h^{*2} (M+N-s)
    \right. \nonumber \\ && \left.   +4 \cA_h^* \cA_v^* (s (M+N-2 s)-1)+(M N - 3) \cA_v^{*2}\right] +\frac{128}{N^3} \left[ 2 \cA_h^{*2} \cA_v^* \left(6 \zeta (3) (s (M+N+5 s)+2)+3 M s+3 N s-28 s^2-13\right)   \right. \nonumber \\ && \left.  +\cA_h^* \cA_v^{*2} \left(12 \zeta (3) (s (M+N)+3)+M N+4 M s+4 N s-24 s^2-44\right)+2 s \cA_h^{*3} (2 \zeta (3) (M+N+18 s)+M+N-16
   s) \right. \nonumber \\ && \left. + \cA_v^{*3} (2 \zeta (3) (M N+7)+M N-18) \right] + \cO \left(\cA_{v,h}^{*4} \right) \,.
\end{eqnarray}
From the above results, we can also extract the full $1$-loop scaling dimension, which we rewrite as a power series in the couplings
\begin{align}
\Delta_{\mathcal{Q}_{J,s}}^{1-loop} =\bar Q \left(1- \frac{\epsilon}{2}\right) + \frac{4}{N} \left(A_h^* \left(\bar Q-2 s^2\right)+(\bar Q-1) A_v^*\right)  = \bar Q \left(1- \frac{\epsilon}{2}\right) + \frac{\bar Q \left(\bar Q-2 s^2\right)}{ (4 \pi)^2 s} u^* +\frac{2 \bar Q (\bar Q- 1)}{(4 \pi)^2} v^* \,,
\end{align}
and depends neither on $N$ nor $M$ when rewritten in terms of the original couplings $u$ and $v$. We conclude that for the considered family of charge configurations, there is no scaling dimension degeneracy in the perturbative regime and thus the corresponding operators transform in different irreducible representations. These are accessed by varying $s$ at fixed CSD $\bar Q$. In the next section, we will study few concrete examples by setting $\bar Q = 2, 4, 8$. To this end, it is useful to consider one more charge configuration given by
\begin{align} \label{seconda}
\mathcal{Q}_{L}=\rm{diag}\left\{-2J,J,J,0,\cdots,0\right\} \,.
\end{align}
The $M_E$ and $B$ matrices can be parametrized as
\begin{align}
M_E=-i\rm{diag}\left\{\mu_1, \mu_2, \mu_2, 0,\cdots,0\right\} \,, \qquad \qquad B=\rm{diag}\left\{b_1, b_2, b_2, 0,\cdots,0\right\} \,.
\end{align}
The EOM and the charge conditions read
\begin{align}
  &  J =  V \mu_1 b_1^2  \,,  & 2 \mu_1^2 =u_0 b_1^2 +v_0 \left(b_1^2 + 2 b_2^2 \right)+ \frac{m^2}{2} \,, \nonumber \\ &  J =  - 2 V \mu_2 b_2^2  \,,  & 2 \mu_2^2 =u_0 b_2^2 +v_0 \left(b_1^2 + 2 b_2^2 \right)+ \frac{m^2}{2} \,.
    \end{align}
The physical solution is
\begin{eqnarray}
\mu_1 = \frac{2 J \mu _2 m^3 v_0}{\mathcal{C}} \,, \quad \qquad b_1 =\sqrt{ \frac{\mathcal{C}}{4 \pi ^2 \mu _2 v_0}} \,, \qquad \quad b_2 = \sqrt{-\frac{J m^3}{4 \pi ^2 \mu _2}} \,,
\end{eqnarray}
where $\mathcal{C} = J m^3 (u_0+2 v_0)+8 \pi ^2 \mu _2^3-2 \pi ^2 \mu _2 m^2$ and $\mu_2$ solves the following equation
\begin{equation}
  - \frac{16 J^3 \mu _2^2 m^9 v_0^3}{\mathcal{C}^3}+\frac{J m^3 \left(-\frac{2 J m^3 v_0^2}{\mathcal{C}}+u_0+v_0\right)}{2 \pi ^2 \mu _2}+\frac{J m^5 v_0}{\mathcal{C}} = 0 \,.
\end{equation}
We choose the solution of the above equation such that it reproduces the $O(2 N M )$ limit when $u_0 = 0$. The perturbative expansion of this solution reads
\begin{eqnarray}
&& \mu_2 =-\frac{m}{2}-\frac{J m (u_0+4 v_0)}{4 \pi ^2}+\frac{J^2 m \left(3 u_0^2+28 u_0 v_0+48 v_0^2\right)}{16 \pi ^4}+\frac{J^3 \left(-4 m u_0^3-67 m u_0^2 v_0-240 m u_0 v_0^2-256 m v_0^3\right)}{16 \pi ^6}+\cO\left((u_0 J)^4, (v_0 J)^4\right) \,. \nonumber \\
\end{eqnarray}
The range of validity of this solution is determined by the constraint $\mathcal{C}< 0$, which we found out to be always satisfied in the perturbative regime.
The leading contribution in the semiclassical approximation is given by the classical energy \eqref{classE} evaluated on this solution:
\begin{equation}
  J \Delta_{-1} = \bar Q\left[ 1+\frac{ \bar Q (3 u^*+8 v^*)}{64
   \pi ^2} - \frac{\bar Q^2 \left(5 u^{*2}+24 u^* v^* + 32 v^{*2}\right)}{1024 \pi ^4}+ \frac{\bar Q^3 \left(9 u^{*3}+59 u^{*2} v^*+144 u^* v^{*2}+128 v^{*3}\right)}{8192 \pi ^6}+\cO\left((u^* \bar Q)^5, (v^* \bar Q)^5\right) \right] \,.
\end{equation}
 where we set the couplings to their FP values and $\bar Q = 8 J$ is the classical scaling dimension, as expected from Eq.\eqref{MCSD}. We checked that we recover the $O(2 N M)$ case when $u = 0$.
 To compute the fluctuation spectrum, we expand around the classical trajectory as in Eq.\eqref{flutt}. The result for the dispersion relations reads
 \begin{equation}
\label{newdisp}
\begin{split}
\Tilde{\omega}_1=&\sqrt{J^2_\ell+4\mu_1^2}\qquad 2 (N-3)~\rm{d.o.f.}\\
\Tilde{\omega}_2=&\sqrt{J^2_\ell+4\mu_2^2}\qquad 4 (N-3)~\rm{d.o.f.}\\
\Tilde{\omega}_{3,4}=&\sqrt{J^2_\ell+4\mu_1^2}\pm 2\mu_1 \qquad 2 \times (M-3)~\rm{d.o.f.}\\
\Tilde{\omega}_{5,6}=&\sqrt{J^2_\ell+4\mu_2^2}\pm 2\mu_2 \qquad 2 \times 2 (M-3)~\rm{d.o.f.}\\
\Tilde{\omega}_7=&\sqrt{J^2_\ell+m_3^2} \qquad 2 (N-3) (M-3)~\rm{d.o.f.}\\
\Tilde{\omega}_{8,9}=& \sqrt{ J^2_\ell + 2 u b_2^2 + 8 \mu_2^2 \pm \sqrt{\left( J^2_\ell + 2 u b_2^2 + 8 \mu_2^2 \right)^2- J^2_\ell \left(J^2_\ell+ 4 u b_2^2 \right) }} \qquad  2 \times 3 ~\rm{d.o.f.}\\
\Tilde{\omega}_{10,11,12,13} :&\left(J^2_\ell - \omega^2 \pm 2 \omega (\mu_1 - \mu_2) \right) \left[ J^2_\ell - \omega^2 \pm 2 \omega (\mu_1 - \mu_2) + 2 u \left(b_1^2 + b_2^2 \right)\right] \pm 4 u \omega \left(b_1^2 - b_2^2 \right) (\mu_1 + \mu_2) - 4\omega^2  (\mu_1 + \mu_2)^2 = 0 \quad 2 \times 4~\rm{d.o.f.}\\
\Tilde{\omega}_{14,15,16,17} :& \rm{det} \mathcal{D}_A \left(\omega, J^2_\ell \right) = 0  \qquad 4~\rm{d.o.f.} \,,
\end{split}
\end{equation}
 where $m_3^2 = 2 v_0 (b_1^2 +2 b_2^2) + m^2 $ and
\begin{equation}
   \mathcal{D}_A \left(\omega, J^2_\ell \right) =\left( \begin{array}{cccc}
       \omega^2 - J^2_\ell + z_{00}   & - \frac{4}{3} i (\mu_1 + 2 \mu_2) \omega & z_{02} & - \frac{4}{3} \sqrt{2} i (\mu_1 - \mu_2) \omega \\ \\
    \frac{4}{3} i (\mu_1 + 2 \mu_2) \omega & \omega^2 - J^2_\ell &   \frac{4}{3} \sqrt{2} i (\mu_1 - \mu_2) \omega & 0 \\ \\
    z_{02} &  - \frac{4}{3} \sqrt{2} i (\mu_1 - \mu_2) \omega & \omega^2 - J^2_\ell + z_{22}  &  - \frac{4}{3} i (2 \mu_1 +  \mu_2) \omega \\ \\
     \frac{4}{3} \sqrt{2} i (\mu_1 - \mu_2) \omega & 0 & \frac{4}{3} i (2 \mu_1 +  \mu_2) \omega & \omega^2 - J^2_\ell
    \end{array}\right) \,,
        \end{equation}
     \begin{eqnarray}
    && z_{00} = - \frac{4}{3} \left[(u_0+v_0) b_1^2 +4 v_0 b_1 b_2 +2 (u_0+ 2 v_0) b_2^2 \right] \,,  \qquad \qquad z_{02} =  - \frac{4}{3} \sqrt{2} (b_1 - b_2) \left[(u_0 +v_0)  b_1 +  (u_0+ 2v_0) b_2 \right] \,,  \nonumber \\
    && z_{22} =  - \frac{2}{3} \left[4 (u_0+v_0) b_1^2 -8 v_0 b_1 b_2 +2 (u_0+ 2v_0) b_2^2 \right] \ .
 \end{eqnarray}
 Although not obvious, for $u = 0$ we recover the fluctuation spectrum of the $O(2 N M)$ model discussed in Sec.\ref{ONsec} when $k = 3 M$ charges have been fixed. In particular one of the last four d.o.f. reduces to the $U(1)$ conformal mode.

We weren't able to find an analytical expression for $\Delta_0$ in this case. Instead, we computed it numerically at fixed values of parameters and as a function of $\epsilon$. The results are given in the next section.

 \subsection{On how to identify the fixed-charge operators}\label{Identify}

In this section, we focus on identifying the fixed charge operators associated with a certain charge configuration. In particular, we propose a practical identification procedure which we outline by performing detailed examples in the $U(N)\times U(M)$ model. Since the MSCD assumption can be violated at large coupling, the procedure is valid only when the anomalous dimensions of the operators involved are much smaller than one, i.e.~in the perturbative regime. Since in weakly coupled theories it is easy to find the explicit form of the MSCD operator once the irrep in which it transforms is known, we focus on identifying the latter. We first list the conclusions below before conducting a more detailed discussion.
\begin{itemize}
\item The representation of the fixed charge operators can be uniquely determined when a charge configuration is chosen. This conclusion is based on the three conditions summarized in Propositions $1-3$ in \autoref{groupanalysis}.
\item To determine the representation of the fixed charge operators, both group theory and the actual semiclassical computations of the scaling dimensions should be implemented. Group theory alone is not sufficient.
\end{itemize}

We start our analysis by considering operators with $\bar Q=2$; Eq.\eqref{MCSD} implies that we can build only one $N \times N$ charge matrix
\begin{align} \label{firstconfig}
\mathcal{Q}_{L,1/2}=\rm{diag}\left\{-1/2, 1/2,0,\cdots,0\right\} \,.
\end{align}

This charge configuration is of the special type considered in Eq.\eqref{special}. Then we can immediately identify the irreducible representation in which the corresponding fixed charge operator sits as the bi-adjoint representation $(\text{\textbf{Adj}},\text{\textbf{Adj}}) = (N^2-1, M^2-1)$. The underlying reason is that, at the level of operators with $\bar Q = 2$ fields, only this representation contains the weight \eqref{firstconfig}. In the perturbative regime, the corresponding operator can be written as $\Tr \left[ T^a H T^b H^\dagger \right]$. In \cite{Antipin:2020rdw}, the anomalous dimension of this operator has been computed in the semiclassical expansion for $N=M$ and the result has been validated via a diagrammatic calculation at the $1$-loop level. The result for general $N$, $M$ has been given in the previous section, being a special case ($s = 1$, $J=1/2$) of the charge configuration \eqref{complete}.

The simplest nontrivial example is obtained by fixing $N= M =3$ and considering operators with CSD $\bar Q=4$. It has been proved in the previous section (Proposition $1$) that the fixed charge operators belong to the irreducible representations $\left(\Gamma_L,\Gamma_R\right)$ of $SU(N)_L\times\ SU(N)_R$ where $\Gamma_L=\Gamma_R=\left(\textbf{Adj}\right)^{Q/2}$. Thus, in the case at hand, the operators live in the decomposition of the tensor product $\mathbf{8}\otimes \mathbf{8}$, which reads
\begin{equation} \label{decompose}
 \mathbf{8}\otimes\mathbf{8}=\mathbf{1} \oplus 2(\mathbf{8}) \oplus \mathbf{10} \oplus\overline{\mathbf{10}} \oplus\mathbf{27}\,.
 \end{equation}



To construct all the relevant charge configurations, it needs to satisfy the following three requirements:
\begin{enumerate}
\item The matrix of charge configuration is diagonal and traceless i.e.~$\rm{tr}\mathbf{\mathcal{Q}}=0$;

\item The diagonal elements of the charge configuration matrix can only be integer or half-integer (Proposition $3$);

\item The sum of the absolute value of the diagonal elements is equal to $\bar Q/2=2$ (Proposition $4$).
\end{enumerate}
By following the above three constraints, we can only construct two different charge matrices:

\begin{equation}
\begin{split}
&\mathcal{Q}_{3A}^{(4)}=
\left(\begin{matrix}
1 &0&0\\
0&-1&0\\
0&0&0
\end{matrix}\right) \,, \quad
\mathcal{Q}_{3B}^{(4)}=
\left(\begin{matrix}
1 &0&0\\
0&-1/2&0\\
0&0&-1/2
\end{matrix}\right) \,.
\label{charge_configuration}
\end{split}
\end{equation}
It was shown in the previous section that the weight and the charge satisfy: $\mu=2\mathbf{\mathcal{Q}}$. The nonzero roots of $SU(3)$ are
\begin{equation}
\begin{split}
&\alpha_1=
\left(\begin{matrix}
1 &0&0\\
0&-1&0\\
0&0&0
\end{matrix}\right) \,, \quad
\alpha_2=
\left(\begin{matrix}
0 &0&0\\
0&1 &0\\
0&0&-1
\end{matrix}\right) \,.
\quad
\alpha_3= \alpha_1 + \alpha_2 \,,
\end{split}
\end{equation}
and $-\alpha_1$, $-\alpha_2$, $-\alpha_3$. We, therefore, obtain ${\mu}_{3A}^{(4)} = 2 \alpha_1$ and ${\mu}_{3B}^{(4)} = 2 \alpha_1 + \alpha_2$. By using the Cartan matrix of the $SU(3)$ algebra
\begin{equation}
    A_{SU(3)} = \left( \begin{array}{cc}
       2  & -1  \\
       -1  & 2
    \end{array} \right) \,,
\end{equation}
we obtain 
\begin{equation}
    \alpha_1 = 2 w_1 - w_2 \,, \qquad \alpha_2 = - w_1 + 2 w_2
\end{equation}
where $w_1$ and $w_2$ are the fundamental weights of $SU(3)$. Then we can decompose the weights $\mathbf{\mu}_{3A}^{(4)},\,\mathbf{\mu}_{3B}^{(4)}$ as
\begin{equation}
\begin{split}
&\mathbf{\mu}_{3A}^{(4)}=4w_1-2w_2=\left(4,-2\right) \,,\\
&\mathbf{\mu}_{3B}^{(4)}=3w_1=\left(3,0\right) \,.\\
\end{split}
\end{equation}
The next step is to determine the representations containing the above weights.
By analyzing the weight diagrams of the irreducible representation appearing in the RHS of \eqref{decompose}, we see that $\left(4,-2\right)$ only appears in $\mathbf{27}$ while $\left(3,0\right)$ appears in both $\mathbf{27}$ and $\mathbf{10}$. Thus, we can set the following correspondence
\begin{equation} \label{disting}
\mathcal{Q}_{3A}^{(4)}: \left(\mathbf{27},\mathbf{27}\right),~~
\mathcal{Q}_{3B}^{(4)}:
\left\{\begin{matrix}
\left(\mathbf{27},\mathbf{27}\right)\\
\left(\mathbf{10},\mathbf{\overline{10}}\right)\\
\end{matrix}\right.\,,
\end{equation}
where we have already excluded asymmetric representations such as $\left(\mathbf{27},\mathbf{\overline{10}}\right)$ since they do not appear in the decomposition of the four indices traceless symmetric $O(2NM)$ tensor (Proposition $2$). Notice that $\mathcal{Q}_{3A}^{(4)}$ is again of the type \eqref{special} and, indeed, it can be directly uniquely associated with $\left(\mathbf{27},\mathbf{27}\right)$.

\begin{figure}
\includegraphics[width=0.6\textwidth]{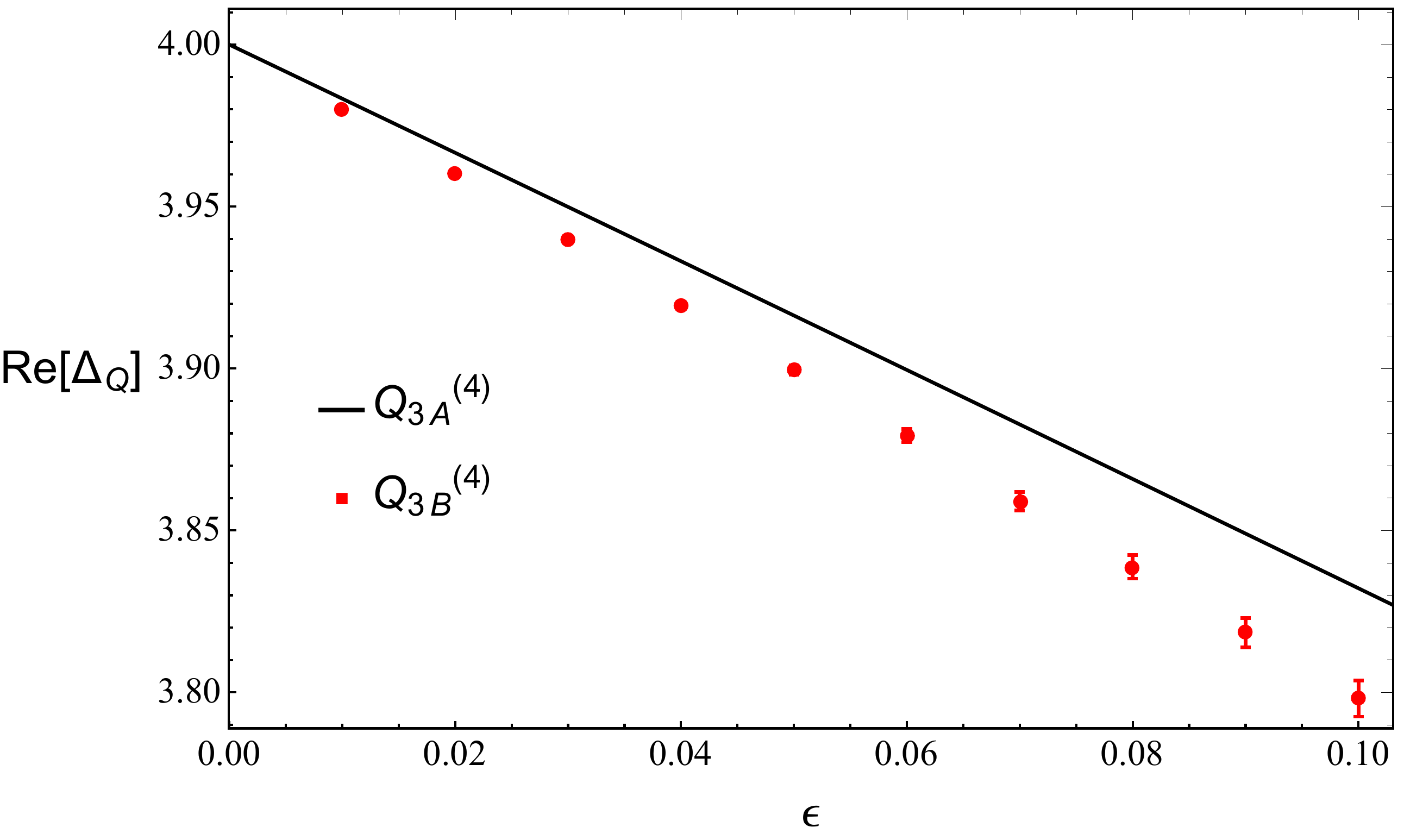}
\caption{\label{fig:diagrams}The results for the real part of the scaling dimension at the fixed point for the $U(3)\times U(3)$ operators with CSD $\Bar Q =4$, carrying the charges $\mathcal{Q}_{3A}^{(4)}$ (black line) and $\mathcal{Q}_{3B}^{(4)}$ (red dots) as a function of $\epsilon$. The error bars encode the numerical error in evaluating $\Delta_0$ for $\mathcal{Q}_{3B}^{(4)}$.}
\label{Config1}
\end{figure}

Using group theory only, \eqref{disting} is the best we can achieve. To further disentangle the representations, we need to employ the fixed-charge semiclassical method, computing at least the first two orders in the semiclassical expansion for both charge matrices. Then, if the corresponding scaling dimensions at the fixed point are different functions of $\epsilon$\footnote{In the small 't Hooft coupling regime we also expect that the scaling dimension corresponds to $\mathcal{Q}_{3B}^{(4)}$ is smaller than that of $\mathcal{Q}_{3A}^{(4)}$. Since the fixed point values are complex, bigger/smaller refers to the real part of the scaling dimensions.} we have that $\mathcal{Q}_{3B}^{(4)}$ corresponds to $\left(\mathbf{10},\mathbf{\overline{10}}\right)$. The scaling dimension of $\mathcal{Q}_{3A}^{(4)}$ at NLO in the semiclassical expansion has been computed analytically in the previous section ($s=J=1$, $N=M=3$ case of Eq.\eqref{complete}) while $\mathcal{Q}_{3B}^{(4)}$ corresponds to the $J=1/2$, $N=M=3$ case of Eq.\eqref{seconda} and thus the associated scaling dimension has been obtained numerically.
The results for the scaling dimensions at NLO are shown in Fig.~\ref{Config1} where the black line and red dots denote, respectively, the cases $\mathcal{Q}_{3A}^{(4)}$ and $\mathcal{Q}_{3B}^{(4)}$.
The error bar on the red dots takes into account all the potential numerical errors\footnote{
The main source of error comes from performing a numerical Taylor expansion in $\epsilon$ of the one-loop functional determinant during the renormalization procedure.
We estimate the numerical error as the difference of the scaling dimensions of $\mathcal{Q}_{3B}^{(4)}$ with the one of $\mathcal{Q}_{3A}^{(4)}$ in the limiting case of $u\rightarrow 0$. As we know, in this case, the scaling dimensions for $\mathcal{Q}_{3B}^{(4)}$ and $\mathcal{Q}_{3A}^{(4)}$ should be equal and coincide with the $O(18)$ result.}.
Clearly, the two scaling dimensions are different functions of $\epsilon$, and thus we can unequivocally associate representations and charges as
\begin{equation}
\mathcal{Q}_{3A}^{(4)}: \left(\mathbf{27},\mathbf{27}\right),~~
\mathcal{Q}_{3B}^{(4)}:
\left(\mathbf{10},\mathbf{\overline{10}}\right) \,.
\end{equation}
Notice that the scaling dimension corresponding to $\mathcal{Q}_{3B}^{(4)}$ is smaller than the one associated with $\mathcal{Q}_{3A}^{(4)}$, consistently with the minimal scaling dimension criteria selecting the fixed-charge operators.

\begin{figure}
\includegraphics[width=0.6\textwidth]{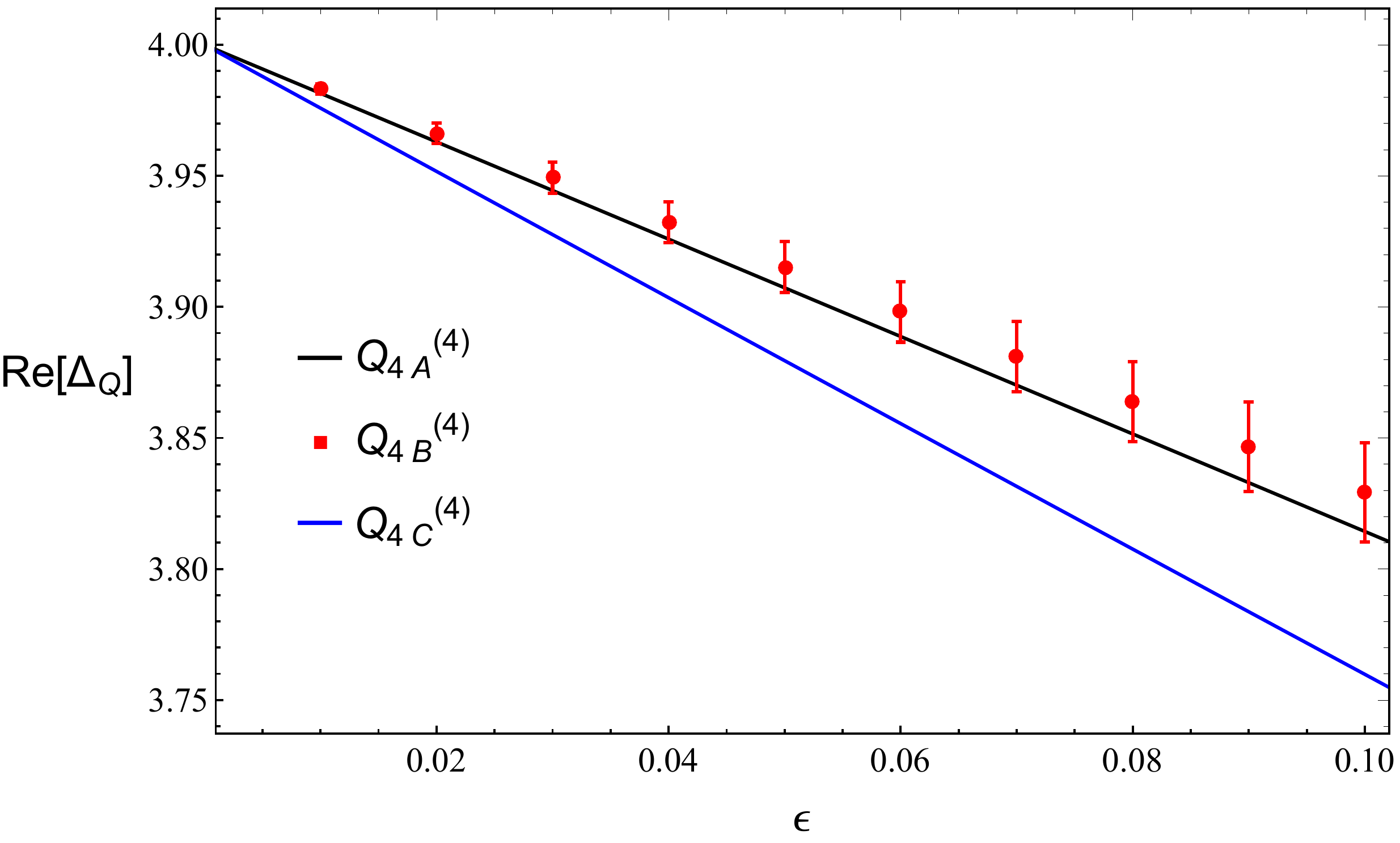}
\caption{\label{fig:diagrams}The results for the real part of scaling dimension for the $U(4)\times U(4)$ operators with CSD $\Bar Q =4$, carrying the charges $\mathcal{Q}_{4A}^{(4)}$ (black line), $\mathcal{Q}_{4B}^{(4)}$ (red dots) and $\mathcal{Q}_{4C}^{(4)}$ (blue lines) as a function of $\epsilon$. The error bars encode the numerical error in evaluating $\Delta_0$ for $\mathcal{Q}_{4B}^{(4)}$.}
\label{Config2}
\end{figure}

\begin{figure}
\includegraphics[width=0.6\textwidth]{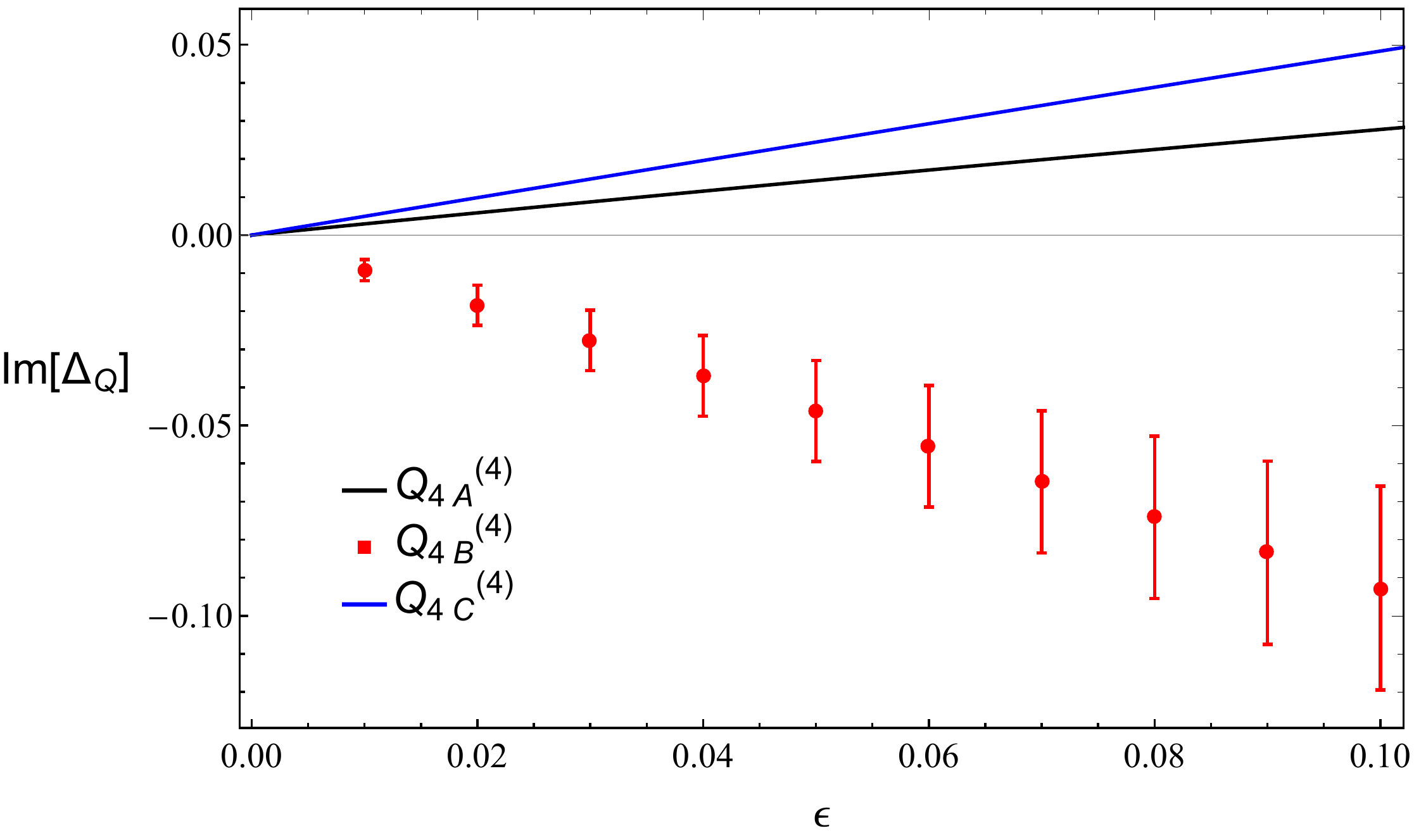}
\caption{\label{fig:diagrams}The imaginary part of the scaling dimension for the $U(4)\times U(4)$ operators with CSD $\Bar Q =4$, carrying the charges $\mathcal{Q}_{4A}^{(4)}$ (black line), $\mathcal{Q}_{4B}^{(4)}$ (red dots) and $\mathcal{Q}_{4C}^{(4)}$ (blue lines )at the fixed point values as a function of $\epsilon$. The error bars encode the numerical error in evaluating $\Delta_0$ for $\mathcal{Q}_{4B}^{(4)}$.}
\label{Config3}
\end{figure}

As a third example, we keep $\bar Q=4$ and we set $N=M=4$. We can now build three independent charge matrices
\begin{equation}
\begin{split}
&\mathcal{Q}_{4A}^{(4)}=
\left(\begin{matrix}
1 &0&0&0\\
0&-1&0&0\\
0&0&0&0\\
0&0&0&0\\
\end{matrix}\right) \,,\quad
\mathcal{Q}_{4B}^{(4)}=
\left(\begin{matrix}
1 &0&0&0\\
0&-1/2&0&0\\
0&0&-1/2&0\\
0&0&0&0\\
\end{matrix}\right) \,, \\
&\mathcal{Q}_{4C}^{(4)}=
\left(\begin{matrix}
1/2 &0&0&0\\
0&-1/2&0&0\\
0&0&1/2&0\\
0&0&0&-1/2\\
\end{matrix}\right) \,.\\
\label{charge_configuration2}
\end{split}
\end{equation}
To connect this example with our semiclassical calculations, we note that Eq.\eqref{complete} encompasses  $\mathcal{Q}_{4A}^{(4)}$ ($s=J=1$, $N=M=4$) and $\mathcal{Q}_{4C}^{(4)}$ ($s=2$, $J=1/2$, $N=M=4$) while Eq.\eqref{seconda} reduces to $\mathcal{Q}_{4B}^{(4)}$ when $J=1/2$ and $N=M=4$. By denoting the three fundamental weights of $SU(4)$  as $W_1$, $W_2$, and $W_3$, we can express the weights associated with the above charge matrices as
\begin{equation}
\begin{split}
&\mathbf{\mu}_{4A}^{(4)}=4 W_1-2 W_2=\left(4,-2,0\right) \,,\\
&\mathbf{\mu}_{4B}^{(4)}=3 W_1-W_3=\left(3,0,-1\right) \,,\\
&\mathbf{\mu}_{4C}^{(4)}=2 W_1-2 W_2+2 W_3=\left(2,-2,2\right)\,.
\end{split}
\end{equation}
The tensor product of two adjoint representations of $SU(4)$ decomposes as
\begin{equation} \label{decompose2}
 \mathbf{15}\otimes\mathbf{15}=\mathbf{1} \oplus2(\mathbf{15})\oplus \mathbf{20}^\prime \oplus \mathbf{45} \oplus \overline{\mathbf{45}} \oplus \mathbf{84}\,.
 \end{equation}
Inspecting the weight diagram of the above representations, we obtain the correspondence summarized below:
\begin{equation} \label{inspect}
\mathcal{Q}_{4A}^{(4)}: \left(\mathbf{84},\mathbf{84}\right),~~
\mathcal{Q}_{4B}^{(4)}:
\left\{\begin{matrix}
\left(\mathbf{84},\mathbf{84}\right)\\
\left(\mathbf{45},\mathbf{\overline{45}}\right)\\
\end{matrix}\right.\,,~~
\mathcal{Q}_{4C}^{(4)}:
\left\{\begin{matrix}
\left(\mathbf{84},\mathbf{84}\right)\\
\left(\mathbf{45},\mathbf{\overline{45}}\right)\\
\left(\mathbf{20'},\mathbf{20'}\right)\\
\end{matrix}\right.\,.
\end{equation}
Again there is one charge matrix, $\mathcal{Q}_{4A}^{(4)}$, which is of the type considered in \eqref{special} and thus corresponds to a unique representation, $\left(\mathbf{84},\mathbf{84}\right)$. Then one can proceed by computing the scaling dimensions associated with $\mathcal{Q}_{4A}^{(4)}$ and $\mathcal{Q}_{4B}^{(4)}$ in the semiclassical expansion. If they are different then $\mathcal{Q}_{4B}^{(4)}$ corresponds to $\left(\mathbf{45},\mathbf{\overline{45}}\right)$ and we can proceed by computing the scaling dimension corresponding to $\mathcal{Q}_{4C}^{(4)}$. If the latter is also different from the previous ones, then we can further conclude that the operator with scaling dimension $\Delta_{\mathcal{Q}_{4C}^{(4)}}$ is in the $\left(\mathbf{20'},\mathbf{20'}\right)$ representation. This is actually the case, as can be seen from the results for the real\footnote{From Fig.\ref{Config2}, one could deduce that $Re\left[\Delta_{\mathcal{Q}_{4B}^{(4)}}\right] > Re\left[\Delta_{\mathcal{Q}_{4A}^{(4)}}\right]$, in violation of the minimal scaling dimension criteria. This apparent puzzle can be solved by taking into account the numerical errors. In addition, the fact that the comparison of the magnitude of the scaling dimensions at small $\epsilon$ should in principle be performed in the expansion of conventional perturbation theory rather than in the semiclassical expansion. The difference is of order $\cO\left(\bar Q \epsilon^2\right)$, but prefactors may magnify it and invert the scaling dimension hierarchy. By making use of the full $2$-loop result of \cite{Antipin:2020rdw, Antipin:2014mga}, we estimated a difference between $0.04\%$ and $0.5\%$ for $\Delta_{Q_{4A}^{(4)}}$ at $\epsilon=0.1$. However, the difference in $\Delta_{Q_{4B}^{(4)}}$ for the same value of epsilon may be much larger.} and imaginary part of scaling dimensions at NLO, which are shown in Figs.~\ref{Config2} and \ref{Config3}, respectively. Due to the numerical error, in this case it is necessary to look also at the imaginary part to disentangle the results. If the results for $\Delta_{\mathcal{Q}_{4A}^{(4)}}$ were the same, we wouldn't have been able to identify the representation associated with $\mathcal{Q}_{4B}^{(4)}$, and would have been necessary to compute higher orders in the semiclassical expansion to check whether they broke the degeneracy or not. However, we could always deduce the irrep related to $\mathcal{Q}_{4C}^{(4)}$ as $\left(\mathbf{20'},\mathbf{20'}\right)$\footnote{If $Re \left[\Delta_{\mathcal{Q}_{4A}^{(4)}}\right] = Re\left[\Delta_{\mathcal{Q}_{4B}^{(4)}}\right]$, it follows that $Re\left[\Delta_{\left(\mathbf{45},\mathbf{\overline{45}}\right)}\right] \ge Re\left[\Delta_{\left(\mathbf{84},\mathbf{84}\right)}\right]$, but since $Re\left[\Delta_{\mathcal{Q}_{4C}^{(4)}}\right] <Re\left[\Delta_{\left(\mathbf{84},\mathbf{84}\right)}\right]$, then $Re\left[\Delta_{\mathcal{Q}_{4C}^{(4)}}\right] = Re\left[\Delta_{\left(\mathbf{20'},\mathbf{20'}\right)}\right]$}.

In conclusion, we have
\begin{equation} 
\mathcal{Q}_{4A}^{(4)}: \left(\mathbf{84},\mathbf{84}\right),~~
\mathcal{Q}_{4B}^{(4)}:
\begin{matrix}
\left(\mathbf{45},\mathbf{\overline{45}}\right)
\end{matrix}\,,~~
\mathcal{Q}_{4C}^{(4)}:
\begin{matrix}
\left(\mathbf{20'},\mathbf{20'}\right)
\end{matrix}\,.
\end{equation}

As the last example, we keep $N=M=4$ and consider $\bar Q=8$. The fourth tensor power of the adjoint representation decomposes as
\begin{eqnarray}
     \mathbf{15}^{\otimes 4} = && 9 (\mathbf{1}) \oplus 43 (\mathbf{15})\oplus 30 (\mathbf{20}^\prime)  \oplus 9(\mathbf{35}) \oplus 9(\overline{\mathbf{35}}) \oplus 39(\mathbf{45}) \oplus 39(\overline{\mathbf{45}}) \oplus 42(\mathbf{84}) \oplus  4(\mathbf{105}) \oplus  39(\mathbf{175}) \oplus 3(\mathbf{189}) \oplus 3(\overline{\mathbf{189}}) \nonumber \\ &&  \oplus 24(\mathbf{256}) \oplus 24(\overline{\mathbf{256}})  \oplus 6(\mathbf{280}) \oplus 6(\overline{\mathbf{280}})\oplus 12(\mathbf{300}^\prime)  \oplus 2(\mathbf{360}^\prime) \oplus 2(\overline{\mathbf{360}}^\prime) \oplus 9 (\mathbf{729}) \oplus (\mathbf{825}) \oplus 3(\mathbf{875}) \oplus 3(\overline{\mathbf{875}}) \ .
\end{eqnarray}

We can build seven charge matrices
   \begin{align}
& \mathcal{Q}_{4A}^{(8)}=\rm{diag}\left\{2, -2, 0, 0, \cdots, 0\right\} \,,  \qquad \qquad \qquad \mathcal{Q}_{4B}^{(8)}=\rm{diag}\left\{-2, 1, 1, 0, \cdots, 0\right\} \,, \nonumber \\ & \mathcal{Q}_{4C}^{(8)}=\rm{diag}\left\{1, -1, 1, -1, 0, \cdots, 0\right\} \,,  \ \  \qquad \qquad
 \mathcal{Q}_{4D}^{(8)}=\rm{diag}\left\{2, -3/2, -1/2, 0, \cdots, 0\right\} \,, \nonumber \\ & \mathcal{Q}_{4E}^{(8)}=\rm{diag}\left\{2, -1, -1/2, -1/2, 0, \cdots, 0\right\} \,, \quad \ \ \   \mathcal{Q}_{4F}^{(8)}=\rm{diag}\left\{3/2, 1/2, -3/2, -1/2, 0, \cdots, 0\right\}\,,\nonumber \\
& \mathcal{Q}_{4G}^{(8)}=\rm{diag}\left\{3/2, 1/2, -1, -1, 0, \cdots, 0\right\} \ .
\label{charge_configuration3}
\end{align}

The corresponding weights, expressed in the fundamental weight basis, read
   \begin{align}
& \mu_{4A}^{(8)}= (8, -4, 0) \,, \qquad  \mu_{4B}^{(8)}= (6,0,-2) \,, \qquad \mu_{4C}^{(8)}= (4, -4, 4) \,, \qquad \mu_{4D}^{(8)}= (7,2,-1) \,,  \nonumber \\ & \mu_{4E}^{(8)}= (6, -1 , 0) \,, \qquad \mu_{4F}^{(8)}= (2, 4, -2) \,, \qquad \mu_{4G}^{(8)}= (2, 3, 0)  \ .
\end{align}
By analyzing the weight diagrams and checking the decomposition of the $O(2 N M)$ traceless symmetric tensor of rank $Q=8$, we obtain the correspondence below
\begin{eqnarray} \label{inspectdeck}
&& \mathcal{Q}_{4A}^{(8)}: {\red{\left(\mathbf{825},\mathbf{825}\right)}} \,,~
\mathcal{Q}_{4D}^{(8)}:
\left\{\begin{matrix}
\left(\mathbf{825},\mathbf{825}\right)\\
\red{\left(\mathbf{875},\mathbf{\overline{875}}\right)}\\
\end{matrix}\right.\,,~~
\mathcal{Q}_{4B}^{(8)}:
\left\{\begin{matrix}
\left(\mathbf{825},\mathbf{825}\right)\\
\left(\mathbf{875},\mathbf{\overline{875}}\right)\\
\red{\left(\mathbf{360'},\mathbf{\overline{360}'}\right)}\\
\end{matrix}\right.\,, ~~
\mathcal{Q}_{4F}^{(8)}:
\left\{\begin{matrix}
\left(\mathbf{825},\mathbf{825}\right)\\
\left(\mathbf{875},\mathbf{\overline{875}}\right)\\
\red{\left(\mathbf{729},\mathbf{729}\right)}\\
\end{matrix}\right.\,, \nonumber \\
&& \mathcal{Q}_{4E}^{(8)}:
\left\{\begin{matrix}
\left(\mathbf{825},\mathbf{825}\right)\\
\left(\mathbf{875},\mathbf{\overline{875}}\right)\\
\left(\mathbf{360'},\mathbf{\overline{360}'}\right)\\
\red{\left(\mathbf{189},\mathbf{\overline{189}}\right)}\\
\end{matrix}\right.
 \mathcal{Q}_{4G}^{(8)}:
\left\{\begin{matrix}
\left(\mathbf{825},\mathbf{825}\right)\\
\left(\mathbf{875},\mathbf{\overline{875}}\right)\\
\left(\mathbf{729},\mathbf{729}\right)\\
\left(\mathbf{360'},\mathbf{\overline{360}'}\right)\\
\red{\left(\mathbf{280},\mathbf{\overline{280}}\right)}\\
\end{matrix}\right.\,,~~
\mathcal{Q}_{4C}^{(8)}:
\left\{\begin{matrix}
\left(\mathbf{825},\mathbf{825}\right)\\
\left(\mathbf{875},\mathbf{\overline{875}}\right)\\
\left(\mathbf{729},\mathbf{729}\right)\\
\left(\mathbf{360'},\mathbf{\overline{360}'}\right)\\
\left(\mathbf{280},\mathbf{\overline{280}}\right)\\
\red{\left(\mathbf{105},\mathbf{105}\right)}\\
\end{matrix}\right.\,.
\end{eqnarray}

If all the corresponding semiclassical computations give different functions of $\epsilon$ as result, then we can uniquely identify all the irreps where the fixed charge operators sit. These are the representations outlined in red above. To be more precise, to allow a complete identification it is not needed that all the scaling dimensions differ; for instance, the results for $\mathcal{Q}_{4B}^{(4)}$ and $\mathcal{Q}_{4F}^{(4)}$ or for $\mathcal{Q}_{4E}^{(4)}$ and $\mathcal{Q}_{4F}^{(4)}$ can be equal.

To summarize, a general identification procedure consists in inspecting the weight diagrams of the various representations in order to obtain correspondences as Eq.\eqref{inspectdeck} and then looking if the corresponding scaling dimensions are degenerate or not, following a sort of "identification chain" that starts from "highest weight" charge matrices of the form \eqref{special} for which there is only one candidate for the corresponding irrep. In the second step, the choice will be between this irrep plus one new candidate, and so on.
We analyzed many other examples and our findings strongly suggest that this identification procedure can always be implemented and fails only in presence of particular degeneracies in the semiclassical results.

\section{Conclusions} \label{conclusion}

We introduced a general strategy apt at determining the relation between a given large charge configuration and the associated operators. In fact, we demonstrated how, varying charge configurations, we could determine the specific anomalous dimensions of distinct operators transforming according to a variety of irreducible representations of the non-abelian symmetry group  going beyond traditional diagrammatical computations.

To demonstrate the usefulness of our methodology, which fuses semiclassical methods with group theoretical considerations, we determined the anomalous dimensions of several composite operators to the next-to-leading order in the semiclassical expansion of the $U(N) \times U(M)$ model in $4-\epsilon$ dimensions.

Our work brings us one step closer to investigating the dynamics of theories similar in structure to the standard model of particle interactions which, in many respects, can be seen as a slight deformation of a CFT \cite{Antipin:2013sga,Shaposhnikov:2009pv}.  Additionally one can envision computing processes involving large number of SM Higgses useful for the next generation of colliders \cite{Benedikt:2018csr,CEPCStudyGroup:2018rmc,CEPCStudyGroup:2018ghi} that have attracted past \cite{Rubakov:1995hq,Son:1995wz} and recent attention \cite{Khoze:2018mey}.

 In fact, critical phenomena play an important role also for applications to social and health sciences. For example, it has been recently shown that (near) fixed points are a useful way to organise the dynamics and diffusion of infectious diseases. The approach, known as the epidemiological Renormalization Group approach \cite{DellaMorte:2020wlc} has been shown to emerge in \cite{cacciapaglia2021field} from either stochastic (percolation models, random walks, diffusion models) or deterministic (compartmental-type) models that themselves can be viewed as mean field theories at criticality \cite{Cardy_1985}.

\section*{Acknowledgements}
The work of O.A. and J.B. is partially supported by the Croatian Science Foundation project number 4418 as well as European Union through the European Regional Development Fund - the Competitiveness and Cohesion Operational Programme (KK.01.1.1.06). F.S. and Z.W. acknowledge the partial support by Danish National Research Foundation grant DNRF:90. We thank Johan Henriksson for his very helpful comments. For the group theoretical analyses, we made use of the Mathematica package LieART \cite{Feger:2019tvk, Feger:2012bs}.

\appendix

\section{The functions $\rho (x^*,M, N, s, {\cA_h}^*, \cA_v^*)$ and $\sigma(\ell,x^*, M, N, s,  {\cA_h}^*,  \cA_v^*)$} \label{funzioni}
In this appendix, we provide explicit expression for the functions appearing in Eq.\eqref{NLO}. Recalling that $x^*=  \frac{72}{N}(\mathcal{A}_h^* + \mathcal{A}_v^*)  + \sqrt{-3+ \left(\frac{72}{N}(\mathcal{A}_h^* + \mathcal{A}_v^*) \right)^2 }$, we have
\begin{eqnarray}
 && \rho (x^*,M, N, s, {\cA_h}^*,  {\cA_v^*}) = \frac{1}{144 {x^*}^{4/3} \left({\cA_h}+{\cA_v^*}\right)^2}\left [-2 {\cA_h^*} {\cA_v^*} \left(2 M \left(102 N {x^*}^{4/3}+24 \sqrt[3]{3} N {x^*}^{2/3}+8\ 3^{2/3} N {x^*}^2+s \left(16 \sqrt[3]{3} {x^*}^{8/3}+291 {x^*}^{4/3}  \right. \right. \right. \right. \nonumber \\  && \left. \left. \left. \left. +216 \sqrt[3]{3} {x^*}^{2/3} +72\ 3^{2/3} {x^*}^2+48\ 3^{2/3}\right)\right)+2 N s \left(16
   \sqrt[3]{3} {x^*}^{8/3}+291 {x^*}^{4/3}+216 \sqrt[3]{3} {x^*}^{2/3}+72\ 3^{2/3} {x^*}^2+48\ 3^{2/3}\right)+48 \sqrt[3]{3} {x^*}^{8/3}+753 {x^*}^{4/3}  \right. \right. \nonumber \\  && \left. \left. +552 \sqrt[3]{3} {x^*}^{2/3} +184\ 3^{2/3} {x^*}^2+144\ 3^{2/3}\right)+{\cA_h^*}^2 \left(-2 M \left(72 N
   {x^*}^{4/3}+s \left(16 \sqrt[3]{3} {x^*}^{8/3}+351 {x^*}^{4/3}+264 \sqrt[3]{3} {x^*}^{2/3}+88\ 3^{2/3} {x^*}^2+48\ 3^{2/3}\right)\right) \right. \right. \nonumber \\  &&  \left. \left. -2 s \left(N \left(16 \sqrt[3]{3} {x^*}^{8/3}  +351 {x^*}^{4/3}+264 \sqrt[3]{3} {x^*}^{2/3}+88\ 3^{2/3} {x^*}^2+48\
   3^{2/3}\right)+6 s \left(16 \sqrt[3]{3} {x^*}^{8/3}+231 {x^*}^{4/3}+168 \sqrt[3]{3} {x^*}^{2/3}+56\ 3^{2/3} {x^*}^2+48\ 3^{2/3}\right)\right)\right) \right.  \nonumber \\  &&  \left.   -{\cA_v^*}^2 \left(M N \left(16 \sqrt[3]{3} {x^*}^{8/3} +495 {x^*}^{4/3}+264 \sqrt[3]{3}
   {x^*}^{2/3}+88\ 3^{2/3} {x^*}^2+48\ 3^{2/3}\right)+4 \left(16 \sqrt[3]{3} {x^*}^{8/3}+261 {x^*}^{4/3}+192 \sqrt[3]{3} {x^*}^{2/3}+64\ 3^{2/3} {x^*}^2  \right. \right. \right. \nonumber \\  &&  \left.   \left.    \left. +48\ 3^{2/3}\right)\right)\right] +   (M-2 s) (N-2 s)  \sqrt{\frac{{\cA_h^*}+\frac{4 \left({x^*}^{2/3}+\sqrt[3]{3}\right)^2 {\cA_v^*}}{3 \sqrt[3]{3} {x^*}^{2/3}}}{{\cA_h^*}+{\cA_v^*}}}+\frac{s^2}{3}  \sqrt{\frac{1}{{x^*}^{2/3} \left({\cA_h^*}+{\cA_v^*}\right)}}\left( 6 \left(2\ 3^{2/3} {x^*}^{4/3}+9 {x^*}^{2/3}+6
   \sqrt[3]{3}\right) {\cA_h^*}  \right. \nonumber \\  &&  \left. +4 \left(3^{2/3} {x^*}^{4/3}+6 {x^*}^{2/3}+3 \sqrt[3]{3}\right) {\cA_v^*}\right)^{1/2}+\frac{2 s^2-1}{2}   \left(- \frac{3 \left(4\ 3^{2/3}
   {x^*}^{4/3}+21 {x^*}^{2/3}+12 \sqrt[3]{3}\right) {\cA_h^*}+8 \left(3^{2/3} {x^*}^{4/3}+6 {x^*}^{2/3}+3 \sqrt[3]{3}\right) {\cA_v^*}}{9 {x^*}^{4/6} \left({\cA_h^*}+{\cA_v^*}\right)} \right. \nonumber \\  &&  \left. + \frac{\left(\frac{4
   \left({x^*}^{2/3}+\sqrt[3]{3}\right)^2}{3 \sqrt[3]{3} {x^*}^{2/3}}-1\right) {\cA_h^*}}{{\cA_h^*}+{\cA_v^*}}+\frac{8 \left({x^*}^{2/3}+\sqrt[3]{3}\right)^2}{3 \sqrt[3]{3} {x^*}^{2/3}} \right)^{1/2}+\frac{2 s^2-1}{2} \left(
   \frac{3 \left(4\ 3^{2/3} {x^*}^{4/3}+21 {x^*}^{2/3}+12 \sqrt[3]{3}\right) {\cA_h^*}+8 \left(3^{2/3} {x^*}^{4/3}+6 {x^*}^{2/3}+3 \sqrt[3]{3}\right) {\cA_v^*}}{9{x^*}^{4/6} \left({\cA_h^*}+\cA
   _v^*\right)}  \right. \nonumber \\  &&  \left. +\frac{\left(\frac{4 \left({x^*}^{2/3}+\sqrt[3]{3}\right)^2}{3 \sqrt[3]{3} {x^*}^{2/3}}-1\right) {\cA_h^*}}{{\cA_h^*}+{\cA_v^*}}+\frac{8 \left({x^*}^{2/3}+\sqrt[3]{3}\right)^2}{3 \sqrt[3]{3} {x^*}^{2/3}}\right)^{1/2}+\frac{4 s
   \left({x^*}^{2/3}+\sqrt[3]{3}\right) (M-N)}{3^{2/3} \sqrt[3]{{x^*}}}+\frac{8 s \left({x^*}^{2/3}+\sqrt[3]{3}\right) (N-2 s)}{3^{2/3} \sqrt[3]{{x^*}}}+\frac{2 s^2 \left({x^*}^{2/3}+\sqrt[3]{3}\right)}{3^{2/3} \sqrt[3]{{x^*}}}
  \end{eqnarray}
 and
 \begin{eqnarray}
 && \sigma(\ell,{x^*}, M, N, s,  {\cA_h^*},  \cA_v^*) = -2 \ell^3 M N-6 \ell^2 M N + \frac{\left(4\ 3^{2/3} {x^*}^{4/3}+15 {x^*}^{2/3}+12 \sqrt[3]{3}\right)^2}{324 \ell {x^*}^{4/3} \left({\cA_h^*}+{\cA_v^*}\right){}^2}
 \left(2 s {\cA_h^*}^2 (M+N+6 s)+2 {\cA_h^*} {\cA_v^*} (2 M s+2 N s+3)  \right. \nonumber \\  && \left. +(M N+4) {\cA_v^*}^2\right) + \frac{\ell}{9 {x^*}^{2/3} (\cA_h^* +\cA_v^*)} \left[ {\cA_h^*} \left(54 M N {x^*}^{2/3}+M s \left(8\ 3^{2/3} {x^*}^{4/3}+30 {x^*}^{2/3}+24 \sqrt[3]{3}\right)+2 N s \left(4\ 3^{2/3} {x^*}^{4/3}+15 {x^*}^{2/3}+12 \sqrt[3]{3}\right)\right) \right. \nonumber \\  && \left. + {\cA_v^*} \left(M N \left(4\ 3^{2/3} {x^*}^{4/3}+69 {x^*}^{2/3}+12
   \sqrt[3]{3}\right)+4\ 3^{2/3} {x^*}^{4/3}+15 {x^*}^{2/3}+12 \sqrt[3]{3}\right)\right] -\frac{1}{9 {x^*}^{2/3} (\cA_h^* +\cA_v^*)} \left[-2 {\cA_h^*} \left(9 M N {x^*}^{2/3}+M s \left(4\ 3^{2/3} {x^*}^{4/3} \right.  \right.  \right. \nonumber \\  && \left.  \left.  \left. +15 {x^*}^{2/3}   +12 \sqrt[3]{3}\right)+N s \left(4\ 3^{2/3} {x^*}^{4/3}+15 {x^*}^{2/3}  +12 \sqrt[3]{3}\right)\right)-{\cA_v^*} \left(M N \left(4\ 3^{2/3} {x^*}^{4/3}+33 {x^*}^{2/3}+12
   \sqrt[3]{3}\right)+4\ 3^{2/3} {x^*}^{4/3}+15 {x^*}^{2/3}+12 \sqrt[3]{3}\right) \right] \nonumber \\
 \end{eqnarray}


\begin{thebibliography}{99}

\bibitem{Hellerman:2015nra}
S.~Hellerman, D.~Orlando, S.~Reffert and M.~Watanabe,
JHEP \textbf{12} (2015), 071
doi:10.1007/JHEP12(2015)071
[arXiv:1505.01537 [hep-th]].

\bibitem{Alvarez-Gaume:2016vff}
  L.~Alvarez-Gaume, O.~Loukas, D.~Orlando and S.~Reffert,
  JHEP {\bf 1704}, 059 (2017)
  doi:10.1007/JHEP04(2017)059
  [arXiv:1610.04495 [hep-th]].

\bibitem{Jafferis:2017zna}
D.~Jafferis, B.~Mukhametzhanov and A.~Zhiboedov,
JHEP \textbf{05} (2018), 043
doi:10.1007/JHEP05(2018)043
[arXiv:1710.11161 [hep-th]].

\bibitem{Hellerman:2017sur}
S.~Hellerman and S.~Maeda,
JHEP \textbf{12} (2017), 135
doi:10.1007/JHEP12(2017)135
[arXiv:1710.07336 [hep-th]].


\bibitem{Gaume:2020bmp}
L.~\`A.~Gaum\'e, D.~Orlando and S.~Reffert,
[arXiv:2008.03308 [hep-th]].

\bibitem{Orlando:2019skh}
D.~Orlando, S.~Reffert and F.~Sannino,
Phys. Rev. D \textbf{101} (2020) no.6, 065018
doi:10.1103/PhysRevD.101.065018
[arXiv:1909.08642 [hep-th]].

\bibitem{Orlando:2020yii}
D.~Orlando, S.~Reffert and F.~Sannino,
[arXiv:2003.08396 [hep-th]].

\bibitem{Sannino:2009za}
F.~Sannino,
Acta Phys. Polon. B \textbf{40} (2009), 3533-3743
[arXiv:0911.0931 [hep-ph]].

\bibitem{Cacciapaglia:2020kgq}
G.~Cacciapaglia, C.~Pica and F.~Sannino,
Phys. Rept. \textbf{877} (2020), 1-70
doi:10.1016/j.physrep.2020.07.002
[arXiv:2002.04914 [hep-ph]].

\bibitem{Litim:2014uca}
D.~F.~Litim and F.~Sannino,
JHEP \textbf{12} (2014), 178
doi:10.1007/JHEP12(2014)178
[arXiv:1406.2337 [hep-th]].

\bibitem{Banks:1981nn}
T.~Banks and A.~Zaks,
Nucl. Phys. B \textbf{196} (1982), 189-204
doi:10.1016/0550-3213(82)90035-9


\bibitem{Orlando:2019hte}
  D.~Orlando, S.~Reffert and F.~Sannino,
  JHEP {\bf 1908}, 164 (2019)
  doi:10.1007/JHEP08(2019)164
  [arXiv:1905.00026 [hep-th]].

\bibitem{Badel:2019oxl}
G.~Badel, G.~Cuomo, A.~Monin and R.~Rattazzi,
JHEP \textbf{11} (2019), 110
doi:10.1007/JHEP11(2019)110
[arXiv:1909.01269 [hep-th]].

\bibitem{Arias-Tamargo:2019xld}
G.~Arias-Tamargo, D.~Rodriguez-Gomez and J.~Russo,
JHEP \textbf{10} (2019), 201
doi:10.1007/JHEP10(2019)201
[arXiv:1908.11347 [hep-th]].

\bibitem{Antipin:2020abu}
O.~Antipin, J.~Bersini, F.~Sannino, Z.~W.~Wang and C.~Zhang,
Phys. Rev. D \textbf{102} (2020) no.4, 045011
doi:10.1103/PhysRevD.102.045011
[arXiv:2003.13121 [hep-th]].

\bibitem{Antipin:2020rdw}
O.~Antipin, J.~Bersini, F.~Sannino, Z.~W.~Wang and C.~Zhang,
Phys. Rev. D \textbf{102} (2020) no.12, 125033
doi:10.1103/PhysRevD.102.125033
[arXiv:2006.10078 [hep-th]].

\bibitem{Jack:2020wvs}
I.~Jack and D.~R.~T.~Jones,
Phys. Rev. D \textbf{102} (2020) no.8, 085012
doi:10.1103/PhysRevD.102.085012
[arXiv:2007.07190 [hep-th]].

\bibitem{Jack:2021ypd}
I.~Jack and D.~R.~T.~Jones,
[arXiv:2101.09820 [hep-th]].


\bibitem{Giombi:2020enj}
S.~Giombi and J.~Hyman,
[arXiv:2011.11622 [hep-th]].


\bibitem{Arias-Tamargo:2020fow}
G.~Arias-Tamargo, D.~Rodriguez-Gomez and J.~G.~Russo,
[arXiv:2003.13772 [hep-th]].

\bibitem{Codello:2016muj}
A.~Codello, K.~Lang\ae{}ble, D.~F.~Litim and F.~Sannino,
JHEP \textbf{07} (2016), 118
doi:10.1007/JHEP07(2016)118
[arXiv:1603.03462 [hep-th]].

\bibitem{Watanabe:2019pdh}
M.~Watanabe,
[arXiv:1909.01337 [hep-th]].

\bibitem{Arias-Tamargo:2019kfr}
G.~Arias-Tamargo, D.~Rodriguez-Gomez and J.~G.~Russo,
JHEP \textbf{01} (2020), 171
doi:10.1007/JHEP01(2020)171
[arXiv:1912.01623 [hep-th]].

\bibitem{Badel:2019khk}
G.~Badel, G.~Cuomo, A.~Monin and R.~Rattazzi,
Phys. Lett. B \textbf{802} (2020), 135202
doi:10.1016/j.physletb.2020.135202
[arXiv:1911.08505 [hep-th]].

  \bibitem{Wilson}
  K.~G.~Wilson and M.~E.~Fisher,
  Phys. Rev. Lett. 28, 240 (1972).

\bibitem{Rychkov:2016iqz}
S.~Rychkov,
doi:10.1007/978-3-319-43626-5
[arXiv:1601.05000 [hep-th]].

\bibitem{Simmons-Duffin:2016gjk}
D.~Simmons-Duffin,
doi:10.1142/9789813149441\_0001
[arXiv:1602.07982 [hep-th]].

\bibitem{Farnsworth:2017tbz}
K.~Farnsworth, M.~A.~Luty and V.~Prilepina,
JHEP \textbf{10} (2017), 170
doi:10.1007/JHEP10(2017)170
[arXiv:1702.07079 [hep-th]].

\bibitem{Cardy:1984rp}
J.~L.~Cardy,
J. Phys. A \textbf{17} (1984), L385-L387

\bibitem{Cardy:1985lth}
J.~Cardy,
J. Phys. A \textbf{18} (1985) no.13, L757-L760
doi:10.1088/0305-4470/18/13/005

\bibitem{Brown:1980qq}
L.~S.~Brown and J.~C.~Collins,
Annals Phys. \textbf{130} (1980), 215
doi:10.1016/0003-4916(80)90232-8

\bibitem{PalanquesMestre:1983zy}
A.~Palanques-Mestre and P.~Pascual,
Commun. Math. Phys. \textbf{95} (1984), 277
doi:10.1007/BF01212398

\bibitem{Gracey:1996he}
J.~A.~Gracey,
Phys. Lett. B \textbf{373} (1996), 178-184
[arXiv:hep-ph/9602214 [hep-ph]].

\bibitem{Holdom:2010qs}
B.~Holdom,
Phys. Lett. B \textbf{694} (2011), 74-79
doi:10.1016/j.physletb.2010.09.037
[arXiv:1006.2119 [hep-ph]].

\bibitem{Antipin:2018zdg}
O.~Antipin, N.~A.~Dondi, F.~Sannino, A.~E.~Thomsen and Z.~W.~Wang,
Phys. Rev. D \textbf{98} (2018) no.1, 016003
[arXiv:1803.09770 [hep-ph]].

\bibitem{Mann:2017wzh}
R.~Mann, J.~Meffe, F.~Sannino, T.~Steele, Z.~W.~Wang and C.~Zhang,
Phys. Rev. Lett. \textbf{119} (2017) no.26, 261802
[arXiv:1707.02942 [hep-ph]].

\bibitem{Pelaggi:2017abg}
G.~M.~Pelaggi, A.~D.~Plascencia, A.~Salvio, F.~Sannino, J.~Smirnov and A.~Strumia,
Phys. Rev. D \textbf{97} (2018) no.9, 095013
[arXiv:1708.00437 [hep-ph]].

\bibitem{Kowalska:2017pkt}
K.~Kowalska and E.~M.~Sessolo,
JHEP \textbf{04} (2018), 027
doi:10.1007/JHEP04(2018)027
[arXiv:1712.06859 [hep-ph]].

\bibitem{Molinaro:2018kjz}
E.~Molinaro, F.~Sannino and Z.~W.~Wang,
Phys. Rev. D \textbf{98} (2018) no.11, 115007
[arXiv:1807.03669 [hep-ph]].



\bibitem{Alanne:2018csn}
T.~Alanne and S.~Blasi,
Phys. Rev. D \textbf{98} (2018) no.11, 116004
doi:10.1103/PhysRevD.98.116004
[arXiv:1808.03252 [hep-ph]].

\bibitem{Wang:2018yer}
Z.~W.~Wang, A.~Al Balushi, R.~Mann and H.~M.~Jiang,
Phys. Rev. D \textbf{99} (2019) no.11, 115017
[arXiv:1812.11085 [hep-ph]].

\bibitem{Sannino:2019sch}
F.~Sannino, J.~Smirnov and Z.~W.~Wang,
Phys. Rev. D \textbf{100} (2019) no.7, 075009
[arXiv:1902.05958 [hep-ph]].

\bibitem{Dondi:2020qfj}
N.~A.~Dondi, G.~V.~Dunne, M.~Reichert and F.~Sannino,
Phys. Rev. D \textbf{102} (2020) no.3, 035005
doi:10.1103/PhysRevD.102.035005
[arXiv:2003.08397 [hep-th]].


\bibitem{Huang:2020bbe}
W.~C.~Huang, F.~Sannino and Z.~W.~Wang,
Phys. Rev. D \textbf{102} (2020) no.9, 095025
[arXiv:2004.02332 [hep-ph]].

\bibitem{Cacciapaglia:2020tzd}
G.~Cacciapaglia and S.~Vatani,
[arXiv:2005.07540 [hep-ph]].



\bibitem{Banerjee:2017fcx}
D.~Banerjee, S.~Chandrasekharan and D.~Orlando,
Phys. Rev. Lett. \textbf{120} (2018) no.6, 061603
doi:10.1103/PhysRevLett.120.061603
[arXiv:1707.00711 [hep-lat]].

\bibitem{Monin:2016jmo}
A.~Monin, D.~Pirtskhalava, R.~Rattazzi and F.~K.~Seibold,
JHEP \textbf{06} (2017), 011
doi:10.1007/JHEP06(2017)011
[arXiv:1611.02912 [hep-th]].



\bibitem{Cuomo:2020rgt}
G.~Cuomo,
Phys. Lett. B \textbf{812} (2021), 136014
doi:10.1016/j.physletb.2020.136014
[arXiv:2010.00407 [hep-th]].

\bibitem{Banerjee:2019jpw}
D.~Banerjee, S.~Chandrasekharan, D.~Orlando and S.~Reffert,
Phys. Rev. Lett. \textbf{123} (2019) no.5, 051603
doi:10.1103/PhysRevLett.123.051603
[arXiv:1902.09542 [hep-lat]].

\bibitem{Hellerman:2017efx}
S.~Hellerman, N.~Kobayashi, S.~Maeda and M.~Watanabe,
JHEP \textbf{10} (2019), 038
doi:10.1007/JHEP10(2019)038
[arXiv:1705.05825 [hep-th]].

\bibitem{Hellerman:2018sjf}
S.~Hellerman, N.~Kobayashi, S.~Maeda and M.~Watanabe,
[arXiv:1804.06495 [hep-th]].

\bibitem{Nielsen:1975hm}
  H.~B.~Nielsen and S.~Chadha,
  Nucl.\ Phys.\ B {\bf 105} (1976) 445.
  doi:10.1016/0550-3213(76)90025-0


\bibitem{Kehrein:1995ia}
  S.~K.~Kehrein,
  Nucl.\ Phys.\ B {\bf 453} (1995) 777
  doi:10.1016/0550-3213(95)00375-3
  [hep-th/9507044].

\bibitem{Kleinert:1991rg}
H.~Kleinert, J.~Neu, V.~Schulte-Frohlinde, K.~G.~Chetyrkin and S.~A.~Larin,
Phys. Lett. B \textbf{272} (1991), 39-44
[erratum: Phys. Lett. B \textbf{319} (1993), 545]
doi:10.1016/0370-2693(91)91009-K
[arXiv:hep-th/9503230 [hep-th]].

\bibitem{Braun:2013tva}
V.~M.~Braun and A.~N.~Manashov,
Eur. Phys. J. C \textbf{73} (2013), 2544
doi:10.1140/epjc/s10052-013-2544-1
[arXiv:1306.5644 [hep-th]].

\bibitem{Kompaniets:2019zes}
M.~Kompaniets and K.~J.~Wiese,
Phys. Rev. E \textbf{101} (2020) no.1, 012104
doi:10.1103/PhysRevE.101.012104
[arXiv:1908.07502 [cond-mat.stat-mech]].

\bibitem{Calabrese:2002bm}
P.~Calabrese, A.~Pelissetto and E.~Vicari,
Phys. Rev. B \textbf{67} (2003), 054505
doi:10.1103/PhysRevB.67.054505
[arXiv:cond-mat/0209580 [cond-mat]].

\bibitem{Wallace:1974nu}
  D.~J.~Wallace and R.~K.~P.~Zia,
   J. Phys. C: Solid State Phys. 8 839 (1975) 10.1088/0022-3719/8/6/014.

\bibitem{Calabrese:2002qi}
P.~Calabrese, A.~Pelissetto and E.~Vicari,
Phys. Rev. E \textbf{65} (2002), 046115
doi:10.1103/PhysRevE.65.046115
[arXiv:cond-mat/0111160 [cond-mat.stat-mech]].

\bibitem{Cornwell:1985xs}
   J.~F.~Cornwell, “Group Theory in Physics, Vol. 1”, (Academic Press, 1984).

\bibitem{Cornwell:1985xt}
   J.~F.~Cornwell, “Group Theory in Physics, Vol. 2”, (Academic Press, 1984).

\bibitem{Hall:2015tb}
   B.~C.~Hall, “Lie Groups, Lie Algebras, and Representations”, (Springer, Second Edition, 2015).

\bibitem{Calabrese:2004uk}
P.~Calabrese and P.~Parruccini,
JHEP \textbf{05} (2004), 018
doi:10.1088/1126-6708/2004/05/018
[arXiv:hep-ph/0403140 [hep-ph]].

\bibitem{Butti:2003nu}
A.~Butti, A.~Pelissetto and E.~Vicari,
JHEP \textbf{08} (2003), 029
doi:10.1088/1126-6708/2003/08/029
[arXiv:hep-ph/0307036 [hep-ph]].

\bibitem{Pisarski:1980ix}
R.~Pisarski and D.~Stein,
Phys. Rev. B \textbf{23} (1981), 3549-3552
doi:10.1103/PhysRevB.23.3549

\bibitem{Kaplan:2009kr}
D.~B.~Kaplan, J.~W.~Lee, D.~T.~Son and M.~A.~Stephanov,
Phys. Rev. D \textbf{80} (2009), 125005
doi:10.1103/PhysRevD.80.125005
[arXiv:0905.4752 [hep-th]].

\bibitem{Gorbenko:2018ncu}
V.~Gorbenko, S.~Rychkov and B.~Zan,
JHEP \textbf{10} (2018), 108
doi:10.1007/JHEP10(2018)108
[arXiv:1807.11512 [hep-th]].


\bibitem{Antipin:2014mga}
O.~Antipin, E.~Mølgaard and F.~Sannino,
JHEP \textbf{06} (2015), 030
doi:10.1007/JHEP06(2015)030
[arXiv:1406.6166 [hep-th]].

\bibitem{Antipin:2013sga}
O.~Antipin, M.~Gillioz, J.~Krog, E.~M\o{}lgaard and F.~Sannino,
JHEP \textbf{08} (2013), 034
doi:10.1007/JHEP08(2013)034
[arXiv:1306.3234 [hep-ph]].

\bibitem{Shaposhnikov:2009pv}
M.~Shaposhnikov and C.~Wetterich,
Phys. Lett. B \textbf{683} (2010), 196-200
doi:10.1016/j.physletb.2009.12.022
[arXiv:0912.0208 [hep-th]].

\bibitem{Benedikt:2018csr}
A.~Abada \textit{et al.} [FCC],
Eur. Phys. J. ST \textbf{228} (2019) no.4, 755-1107
doi:10.1140/epjst/e2019-900087-0



\bibitem{CEPCStudyGroup:2018rmc}
 [CEPC Study Group],
[arXiv:1809.00285 [physics.acc-ph]].

\bibitem{CEPCStudyGroup:2018ghi}
J.~B.~Guimar\~aes da Costa \textit{et al.} [CEPC Study Group],
[arXiv:1811.10545 [hep-ex]].

\bibitem{Rubakov:1995hq}
V.~A.~Rubakov,
``Nonperturbative aspects of multiparticle production,''
[arXiv:hep-ph/9511236].




\bibitem{Son:1995wz}
D.~T.~Son,
Nucl. Phys. B \textbf{477} (1996), 378-406
doi:10.1016/0550-3213(96)00386-0
[arXiv:hep-ph/9505338 [hep-ph]].

\bibitem{Khoze:2018mey}
V.~V.~Khoze and J.~Reiness,
Phys. Rept. C \textbf{822} (2019), 1-52
doi:10.1016/j.physrep.2019.06.004
[arXiv:1810.01722 [hep-ph]].
LaTeX (EU)


\bibitem{DellaMorte:2020wlc}
M.~Della Morte, D.~Orlando and F.~Sannino,
Front. in Phys. \textbf{8} (2020), 144
doi:10.3389/fphy.2020.00144
%



\bibitem{cacciapaglia2021field}
G.~Cacciapaglia, C.~Cot, M.~Della~Morte, S.~Hohenegger, F.~Sannino and S.~Vatani,
q-bio, PE, arXiv:2101.11399
https://arxiv.org/abs/2101.11399



\bibitem{Cardy_1985}
J. L.~ Cardy and P.~ Grassberger
Journal of Physics A: Mathematical and General, {Vol.18, n.6} (1985), L267-L271
doi:10.1088/0305-4470/18/6/001



\bibitem{Feger:2012bs}
R.~Feger and T.~W.~Kephart,
Comput. Phys. Commun. \textbf{192} (2015), 166-195
doi:10.1016/j.cpc.2014.12.023
[arXiv:1206.6379 [math-ph]].

\bibitem{Feger:2019tvk}
R.~Feger, T.~W.~Kephart and R.~J.~Saskowski,
Comput. Phys. Commun. \textbf{257} (2020), 107490
doi:10.1016/j.cpc.2020.107490
[arXiv:1912.10969 [hep-th]].























































  \end{thebibliography}
  \end{document}